\documentclass[aps,preprintnumbers,amsmath,amssymb,nofootinbib,eqsecnum,showkeys,preprintnumbers,10pt,onecolumn,floatfix,superscriptaddress]{revtex4}
\usepackage{graphicx}
\usepackage{epsf}
\usepackage{amsmath}
\usepackage{epstopdf}
\usepackage{bm}
\usepackage{color}
\usepackage{tabularx}
\usepackage{enumitem}
\usepackage{float}
\usepackage{array,booktabs}
\usepackage{footnote}
\usepackage{threeparttable}
\usepackage{graphicx}
\usepackage{hyperref}
\usepackage{amssymb,epsf}
\usepackage{latexsym}
\usepackage{epstopdf}
\usepackage{epsfig}
\usepackage{morefloats}
\usepackage{multirow} 

\begin{document}
\title{Holographic $s$- and $p$-wave superconductors from the $4D$ regularization\\of Einstein-Lovelock theory}


\author{Soodeh Zarepour}
\email{szarepour@phys.usb.ac.ir (Corresponding author)}
\affiliation{Department of Physics, University of Sistan and Baluchestan, Zahedan, Iran}

\author{Ali Dehghani}
\email{ali.dehghani.phys@gmail.com}
\affiliation{School of Physics, Damghan University, Damghan 3671645667, Iran}



\begin{abstract}
We investigate holographically dual descriptions of $(2+1)$-dimensional $s$-wave and $p$-wave superconductors in the framework of four-dimensional 4$D$ regularization of  Einstein-Lovelock gravity theories,  incorporating higher curvature corrections beyond the Gauss-Bonnet sector.  We first implement the 4$D$ regularization of Einstein-Lovelock gravity with finely tuned coupling constants to include corrections up to any $K$th order in curvature. The bulk geometry is constructed from exact black-brane solutions characterized by the fine-tuned Lovelock coupling $\alpha$ and the highest power of curvature in the Lagrangian $K$, allowing systematic control over higher curvature effects. We then analyze the condensation of scalar and vector operators dual to minimally coupled matter fields, focusing on two possible prescriptions for fixing the bulk-field masses. We show that these choices play a nontrivial role in determining both quantitative and qualitative features of the superconducting phase. Our results demonstrate that higher curvature terms significantly modify the phase structure of both $s$-wave and $p$-wave systems and enhance the sensitivity of the condensates and critical temperatures to the gravitational couplings. In both cases, the critical temperature generally increases with the maximal curvature order $K$, leading to higher-$T_c$ phases compared to Einstein gravity, particularly for negative values of the fine-tuned Lovelock coupling. This enhancement is typically accompanied by a suppression of the condensate at low temperatures. The fine-tuned coupling $\alpha$ effectively governs the strength of higher curvature interactions in the bulk, acting as a geometric control parameter for superconductivity: positive $\alpha$ suppresses condensation and lowers $T_c$, whereas negative $\alpha$ enhances superconducting order and promotes higher-$T_c$ phases relative to Einstein gravity. We further study the optical conductivity and find that both the gap frequency and the ratio $\omega_g/T_c$ exhibit a strong dependence on the Lovelock coupling $\alpha$ and the curvature order $K$, deviating from the universal Einstein-gravity result. In particular, higher curvature effects enhance the superconducting gap scale, while preserving the expected large-frequency asymptotics and the divergent DC conductivity. Despite these modifications, all condensates display mean-field scaling near the critical temperature, consistent with a second-order phase transition. Notably, the $p$-wave system shows a stronger sensitivity to the mass-fixing prescription compared to the $s$-wave case, indicating a richer dependence on the effective AdS scale. Overall, higher curvature corrections in the 4$D$ regularization of Einstein-Lovelock gravity theories provide a robust mechanism for controlling superconducting observables and significantly amplify the imprint of the bulk gravitational sector on the dual strongly coupled system.
\end{abstract}

\keywords{AdS/CFT Correspondence, Holography and Condensed Matter Physics (AdS/CMT), Classical Theories of Gravity}

\maketitle

\tableofcontents

\section{Introduction} \label{sect:1}
Gauge/gravity duality, also known as AdS/CFT correspondence, claims that quantum field theories in lower dimensions could have a dual description in terms of asymptotically anti-de Sitter (AdS) gravitational field theories in one higher dimension \cite{Maldacena1998,MAGOO2000}. This claim can mathematically be formulated by the GKPW relation as \cite{Witten1998a,GKP1998}
\begin{equation} \label{GKPW}
{\left\langle {{e^{i\int {{d^d}x} {\phi _{(0)}}(\vec x){\cal O}(\vec x)}}} \right\rangle _{{\rm{QFT}}}} = {{\cal Z}_{{\rm{AdS - gravity}}}} \approx {\left. {{e^{ - {{\bar {S}}_E}[{{\left. {\phi (\vec x,r)} \right|}_{r = \infty }} = {\phi _{(0)}}(\vec x)]}}} \right|_{{\text{AdS boundary}}}},
\end{equation}
in which the left hand side is the generating functional (partition function) of the quantum field theory living on the AdS boundary while the right hand side is that of the gravitational bulk theory in one higher dimension. Usually, the gravitational theory is weakly-coupled, for which gravity is classical and saddle-point approximation is valid. As a result, by applying gauge/gravity duality, a weakly-coupled gravitational theory in AdS black hole/brane background translates into a strongly-coupled gauge field theory at finite temperature, where the perturbation theory fails. This technique is a powerful tool whose applications in different areas of physics are becoming more and more, e.g., a number of novel applications have been discovered in high-energy and elementary particle physics, nuclear physics, hydrodynamics, condensed matter physics etc. \cite{Witten1998b,Rey2001Yee,Policastro2000PRL,Policastro2002a,Policastro2002b,Hartnoll2009Lectures,Zaanen2015Book,ErdmengerBook2015,Nastase2015Book,NatsuumeBookAdS/CFT}. Particularly, by implementing gauge/gravity duality, it has been shown that gravity models minimally coupled with matter fields in AdS background are useful in building
holographic models of superconductors \cite{Gubser2008,HHH2008PRL,HHH2008JHEP}. This is a completely different approach compared to the framework of the BCS microscopic theory of superconductivity \cite{BCS1957} by which a number of important experimental phenomena are explained including the diverging conductivity, the superconducting transition temperature, and the existence of an energy gap at the Fermi level \cite{Zaanen2015Book,Annett2004Book}. In what follows, we shall restrict ourselves to the subject of holographic superconductors. \vspace{1mm}

The holographic applications of the gauge/gravity duality to superconductors have been extensively studied in a number of research papers after the seminal works of Gubser, Hartnol, Herzog, and Horowitz \cite{Gubser2008,HHH2008PRL}. In order to construct a dual holographic description of superconductors with the associated superconductivity on the boundary, gravity as the bulk theory should be coupled with some matter fields (with spin $0$, $1/2$, or $1$), which makes a second-order phase transition possible below a certain critical temperature for the holographic background \cite{HHH2008PRL,HHH2008JHEP,Hartnoll2009Lectures,Herzog2009Lectures,Roberts2008PRD,BravoGaete2026CQG}. As a result, a new ground state below the critical temperature forms which is dual to a condensate. So far, various kinds of holographic superconductors have been constructed and studied in the literature which provide a better understanding of this subject including $s$-wave \cite{HHH2008PRL,HHH2008JHEP}, $p$-wave \cite{Gubser2008pWave,Aprile2011pWave,Cai2013pWave,Cai2014pWave,Paul,Chae2026}, and $d$-wave \cite{d-Wave2010a,d-Wave2010b} holographic superconductors using different couplings of matter fields with Einstein gravity (see \cite{Cai-2015Review} and references therein for a nice review on holographic models of superconductors). In this regard, implementing modified gravity theories rather than Einstein gravity has been found many applications in building holographic superconducting systems since they incorporate modifications in a number of various ways, which enrich the holographic model \cite{HS2010Cai,HS2011Meyer,HS2013Banerjee,HS2014PRD,HS2018GRG,HS2019GRG,HS2020EPJP,HS2020Nam}. Among them, modified gravity theories with higher curvature corrections are of central importance for several reasons in the view of both sides of the gravitational bulk theory \cite{SG-Stelle1977,SG-Donoghue1994,SG-Smolin2003,SG-BurgessLivingReviewQG2004,SG-Woodard2009,SG-Donoghue2012,SG-Donoghue2015,SG-Donoghue2018,SG-Hartman,SG1,SG2,SG3,SG4,SG5,SG6} and the dual boundary field theory  \cite{HCC-Holography2002,KSS2008violation,HolographicLovelocka,HolographicLovelockb,HolographicLovelockc,CamanhoThesis,Sinamuli2017}. Higher-order curvature terms are essential for describing strong gravity's effects, appear in different approaches to quantum gravity, and it strongly seems that such corrections are necessary for renormalizing the gravitational field \cite{SG-Stelle1977,SG-Donoghue1994,SG-Smolin2003,SG-BurgessLivingReviewQG2004,SG-Woodard2009,SG-Donoghue2012,SG-Donoghue2015,SG-Donoghue2018,SG-Hartman,SG1,SG2,SG3,SG4,SG5,SG6}. Indeed, they can no longer be ignored when quantum effects, strong-field regime, or loop corrections are taken into account. In the view of dual boundary field theory, modified gravity theories with higher curvature corrections could result in dual boundary field theories which violate the universal bound (KSS bound \cite{KSS2005PRL}) on the ratio of the shear viscosity over entropy density \cite{KSS2008violation}. Furthermore, the investigation of holographic Smarr relation beyond the large $N$ limit has also disclosed that the bulk correlates of sub-leading $1/N$ corrections to the Smarr relation are related to the couplings in Lovelock gravity theories \cite{Sinamuli2017}. Inspired by string theory, coupling higher curvature corrections to gravity also gives a wide range of holographic superconducting systems \cite{HS-GB-2009Gregory,HS-GB-2010Gregory,HS-GB-2010Cai,HS-GB-2010Kuang,HS-HCC-2010Siani,HS-HCC-2021,HS-GB-2021PLB}. \vspace{1mm}

Probably, the most interesting, fruitful theory involving higher curvature corrections is the Lovelock's theory of gravity \cite{Lovelock1971,Lovelock1972} which is regarded as a natural generalization of Einstein's general relativity to higher dimensions and has been the subject of many researches so far. In the present paper, we employ this theory in the limit $D\to 4$ \cite{Tomozawa2011,Zerbini2013,Glavan2020,Casalino2021,Zhidenko2020PRDa,Zhidenko2020PRDb,Zhidenko2020PLB,Zhidenko2020DarkUniv,Mann2020,Kobayashi2020,LuPang2020,Mukohyama2020,EGB4Dreview} as the framework to build a new class of 2+1 dimensional holographic superconductors up to any order of higher curvature corrections beyond the quadratic, ghost-free Gauss-Bonnet term. One important observation, which motivates researchers to holographically explore the lower-dimensional superconductivity, is the absence of Mermin-Wagner-Coleman theorem \cite{MerminWagner1966,Coleman1973} in 2+1 dimensional holographic superconductors \cite{HHH2008PRL}. This theorem states that spontaneous breaking of a continuous symmetry is impossible in 2+1 or fewer dimensions \cite{Zee2010QFT}, but spatially two-dimensional versions of the superconducting phase transition occur in holographic models of superconductors \cite{HHH2008PRL}. On the other hand, the phenomenon of superconductivity in high-$T_C$ oxide materials with $\text{Cu-O}$ planes \cite{SC1986,SC1989}, $\text{CoO}_2$ layers \cite{SC2003}, and $\text{LaAlO}_3/\text{SrTiO}_3$ interfaces \cite{SC2011} plays an important role in condensed matter physics which gives the study of this subject in 2+1 dimensions through gauge/gravity duality (better known as AdS/CMT correspondence) a special significance since, in terms of spatial dimensions, the physics is essentially two dimensional. (We refer the interested reader to refs. \cite{SC-2D-review1,SC-2D-review2,SC-2D-review3,SC-2D-review4,SC-2D-review5,SC-2D-2017,SC-2D-2021,SC-2D-2022} for more details, further directions, and relevant references for spatially two-dimensional superconductors.) For these reasons, there are several analyses investigating the phenomena of holographic superconductivity in lower dimensions starting from the simplest one (Einstein-Maxwell-scalar field system \cite{HHH2008PRL}) to more complicated models involving higher curvature corrections \cite{HS-4GB-2020JHEP,HS-4GB-2021PLBa,HS-4GB-2021PLBb,HS-4GB-2022EPJP,Edelstein2022}.\vspace{1mm}

Now, let us focus on the holographic model we are dealing with in the present paper, namely the $D \to 4$ limit of Lovelock gravity theories \cite{Tomozawa2011,Zerbini2013,Glavan2020,Casalino2021,Zhidenko2020PRDa,Zhidenko2020PRDb,Zhidenko2020PLB,Zhidenko2020DarkUniv,Mann2020,Kobayashi2020,LuPang2020,Mukohyama2020}. By dropping the requirement of linearity in the second derivatives of metric ($g_{\mu \nu}$) and preserving the rest of Einstein's assumptions in general relativity, Lovelock found that the resultant gravitational field (Lovelock-Lanczos) tensors are nonlinear in the Riemann tensor and differ from the Einstein tensor only if the spacetime possesses more than four dimensions \cite{Lovelock1971,Lovelock1972,Lanczos1938}. As a result, the field equations in Lovelock gravity contain metric derivatives no higher than second order which lead to ghost-free nontrivial gravitational interactions in higher dimensions. (Besides $f(R)$ gravity \cite{Sotiriou2010}, it is the only family of relativistic gravitational theory with higher curvature corrections that is ghost-free in general \cite{Zwiebach1985,Boulware1985Deser,Zumino1986}.) The degrees of polarizations for the massless graviton are also the same as Einstein gravity. It is remarkable that this theory is in strong connection with different models of string theory. In fact, the classical gravity approximation of string theory can be obtained in the low-energy limit where the string coupling constant is too small \cite{StringRef1,StringRef2,StringRef3,StringRef4,StringRef5} and the resultant Lagrangian consists of an infinite number of terms with powers of curvature tensors such as $R^2$, $R^3$, ${R^{\mu \nu}}{R_{\mu \nu}}$ etc. Remarkably, a ghost-free combination of such higher derivative terms leads to the Lovelock Lagrangian \cite{Zwiebach1985,Boulware1985Deser,Zumino1986} which is given by a sum of dimensionally extended Euler densities as \cite{Lovelock1971,Lovelock1972}
\begin{gather}
{\cal L} = \frac{1}{{16\pi {G_N}}}\sum\limits_{i = 0}^K {{{\tilde \alpha }_i}{{\cal L}^{(i)}}}, \label{Lovelock Lagrangian1}\\
{{\cal L}^{(i)}} = \frac{1}{{{2^i}}}\delta _{{\rho _1}\,{\sigma _1}\,...\,{\rho _i}\,{\sigma _i}}^{{\mu _1}\,{\nu _1}\,...\,{\mu _i}\,{\nu _i}}{R_{{\mu _1}\,{\nu _1}}}^{{\rho _1}\,{\sigma _1}}...\,{R_{{\mu _i}\,{\nu _i}}}^{{\rho _i}\,{\sigma _i}}, \label{Lovelock Lagrangian2}
\end{gather}
where $\delta _{{\rho _1}\,{\sigma _1}\,...\,{\rho _i}\,{\sigma
		_i}}^{{\mu _1}\,{\nu _1}\,...\,{\mu _i}\,{\nu _i}}$ and ${R_{{\mu_i}\,{\nu _i}}}^{{\rho _i}\,{\sigma _i}}$ are the generalized
totally antisymmetric Kronecker delta and the Riemann tensor, respectively, and ${{\tilde \alpha }_i}$'s are the so-called Lovelock coupling constants. $K = \left[ {\frac{{D - 1}}{2}} \right]$ determines the maximal order of possible higher curvature corrections (the highest power of curvature in the Lagrangian) in a $D$-dimensional spacetime, in which the closed brackets $[...]$ denote taking the integer part. Those terms with $2i = D$ are purely topological and do not contribute to the field equations and the terms with $2i > D$ vanish identically. The trouble is that such corrections appear in higher dimensions; e.g., the quadratic terms (better known as Gauss-Bonnet term) contribute in $D \ge 5$ dimensions, while the cubic Lovelock curvature terms will show up in the spacetimes having $D \ge 7$, and so on. This means the consequences of higher curvature corrections of Lovelock gravity theories can not be examined in the physical (three- or four-) dimensions. However, according to gauge/gravity duality, such corrections can in principle affect the physics of lower-dimensional boundary field theory. In this regard, the five-dimensional Gauss-Bonnet gravity can be translated into a four-dimensional dual boundary gauge theory \cite{HS-GB-2009Gregory,HS-GB-2010Cai,HS-GB-2010Gregory}, but the rest of higher-order corrections such as cubic ($K=3$) and quartic ($K=4$) terms remain to be understood in the physical dimensions. There exist some dimensional-reduction techniques for regularizing Lovelock gravity by which the
effects of ghost-free, higher-order curvatures can be extended to three- and four-dimensional spacetimes. Regardless of the approach employed, the Lovelock coupling constants must ultimately be rescaled at the level of the action or the gravitational field equations as \cite{Tomozawa2011,Zerbini2013,Glavan2020,Casalino2021,Zhidenko2020PRDa,Zhidenko2020PRDb,Zhidenko2020PLB,Zhidenko2020DarkUniv,Mann2020,Kobayashi2020,LuPang2020,Mukohyama2020,EGB4Dreview} 
\begin{equation} \label{rescalings}
{\tilde{\alpha} _i} \to {\alpha _i}\prod\limits_{j = 3}^{2i} {\frac{1}{{D - j}}},
\end{equation}
which leads to higher curvature corrections also contributing to the four-dimensional gravitational field equations. Due to the this regularization trick, all the higher-order curvature terms of Lovelock Lagrangian can affect the physics of boundary ($2+1$) gauge field system. \vspace{1mm}

The effect of higher curvature corrections up to quadratic (Gauss-Bonnet) terms on the physics of holographic superconductors have been intensively studied in $3+1$ (using the conventional Gauss-Bonnet gravity in five-dimensions \cite{HS-GB-2009Gregory,HS-GB-2010Cai,HS-GB-2010Gregory}) and more recently in $2+1$ dimensions (using the regularized $4D$ Einstein-Gauss-Bonnet theory \cite{HS-4GB-2020JHEP,HS-4GB-2021PLBa,HS-4GB-2021PLBb}). The recent studies have primarily focused on the consequences of the $4D$ regularization of Gauss-Bonnet gravity, in which only quadratic curvature corrections are included. However, holographic applications of regularized 4$D$ Einstein-Lovelock theories beyond the Gauss-Bonnet sector remain considerably less explored, and their implications for holographic superconductivity at higher curvature orders have not been systematically investigated. In this paper we elaborate on such studies by considering more higher-order corrections of spacetime curvature. The purpose of the present work is to investigate how higher-order (quadratic, cubic, quartic and etc.) curvature corrections affect the physics of superconductivity in comparison to Einstein's general relativity and also the $4D$ Einstein-Gauss-Bonnet gravity. Motivated by the fact that a compelling bound on negative higher curvature couplings of $4D$ Einstein-Lovelock theories can be justified by demanding that atomic nuclei should not be shielded by a horizon \cite{Charmousis2022}, we also pay special attention to negative couplings, which are usually less addressed in the literature. We utilize the $4D$ regularization of Einstein-Lovelock gravity to build $s$-wave and $p$-wave models of holographic superconductors with the associated superconductivity. The black brane background needed for this purpose has already been found in ref. \cite{DS2022}, where they were constructed and studied in detail. Using this background, one can also take the $K \to \infty$ limit of the resultant black brane solution to include the infinite number of higher curvature corrections, but, here due to the numerical nature of the study, we focus our attention to the cases with $K=1$ (Einstein gravity), $K=2$ (Gauss-Bonnet case), and beyond Gauss-Bonnet, i.e., $K=3, 4, ...$ up to $K=100$. Adding matter (scalar and vector) fields as probes is our assumption in the present paper and it would be interesting to compare the outcomes of this work with fully backreacting holographic superconductors of this model in future. \vspace{1mm}

We emphasize that the $D \to 4$ limit of Einstein-Lovelock gravity theories based on the original Glavan-Lin prescription \cite{Glavan2020} has well-known subtleties \cite{iss1-Tekin2020,iss2-Mahapatra,iss3-Ai2020,iss4-Hinterbichler2020,iss5-Hohmann2021GB4D}, which have motivated several alternative regularization procedures, including conformal regularization \cite{Mann2020}, regularized Kaluza-Klein reduction \cite{Kobayashi2020,LuPang2020}, temporal diffeomorphism breaking \cite{Mukohyama2020}, and regularization with the dimensional derivative \cite{EGB4Dreview}. These constructions generally introduce additional degrees of freedom and provide well-defined four-dimensional descriptions of the corresponding higher curvature theories \cite{EGB4Dreview}. For the planar black brane sector relevant to the present holographic construction, the agreement of the corresponding backgrounds across different regularization procedures is particularly important, since the higher curvature effects enter the holographic matter equations through the background geometry. In fact, for our black brane background (introduced in Section \ref{sect2:setup}), the alternative regularized theories reproduce the same black brane backgrounds as the original construction, as already shown for the Gauss-Bonnet and cubic sectors \cite{EGB4Dreview,Alkac2022}. Accordingly, setting $K=2$ in the present work, one can recover the results of Ref. \cite{HS-4GB-2020JHEP}, in which the authors studied $s$-wave and $p$-wave holographic superconductors in the $4D$ Einstein-Gauss-Bonnet theory based on the regularization procedure of Aoki \textit{et al.} \cite{Mukohyama2020} by breaking temporal diffeomorphism invariance. More generally, a well-defined dimensional conformal regularization of the Lovelock invariants leads to four-dimensional scalar-tensor theories belonging to the Horndeski class and is applicable at arbitrary Lovelock order \cite{Fernandes2025,Fernandes2026}. It also provides an independent framework for the planar backgrounds considered here beyond the Gauss-Bonnet and cubic sectors \cite{ZD2026}. A complete analytical treatment of this correspondence, including its relation to other regularization procedures and higher-curvature solutions, will be presented separately \cite{ZD2026}. \vspace{1mm}

To achieve our goals, we organize the rest of this paper as follows. In the next section (\ref{sect2:setup}), the regularized $4D$ bulk theory and the corresponding black brane solutions as the background for building 2+1 holographic superconductors are discussed in detail. In section \ref{sect3:sWave}, in the probe limit, we then present the holographic description of $s$-wave superconductors in the black brane background with an arbitrary number of higher curvature corrections. We mainly concentrate on the effects of the Lovelock coupling ($\alpha$) and the order of higher curvature terms ($K$) on the scalar condensate, critical temperature, and the optical conductivity of
the $s$-wave superconductors. Next, in section \ref{sect4:pWave}, the holographic model of $p$-wave superconductors in the probe limit are explored in the regularized model under consideration. We investigate the effects of varying Lovelock coupling $\alpha$ as well as adding higher curvature corrections (by increasing $K$) on the holographic phenomenon of $p$-wave superconductivity. Finally, in section \ref{sect5:final}, we summarize the main results and finish our paper with some concluding remarks.

 \section{Holographic background: Einstein-Lovelock black branes} \label{sect2:setup}

 The action of Einstein-Lovelock gravity is considered as our bulk theory, where matter fields like scalar and vector fields can propagate. The Einstein-Lovelock action in the presence of a negative cosmological constant in $D$-dimensions may be written as
 \begin{equation} \label{bulk action}
S_{bulk} = \frac{1}{{16\pi {G_N}}}\int {{d^D}x\sqrt { - g} } \sum\limits_{i = 0}^K {{{\tilde \alpha }_i}{{\cal L}^{(i)}}},
 \end{equation}
 where the elements of this action has already been introduced in eqs. (\ref{Lovelock Lagrangian1}) and (\ref{Lovelock Lagrangian2}). The first term in the Lovelock Lagrangian (\ref{Lovelock Lagrangian1}), i.e. $\alpha_0 {\cal L}_0$, is a constant and , in principle, it can be interpreted as the cosmological constant term, $- 2 \Lambda$, where $\Lambda=-3/L^2$ ($L$ is the AdS scale). Setting $\alpha_1 =1$, the second term gives the Ricci scalar, $\alpha_1 {\cal L}_1=R$, and this choice recovers Einstein gravity in the absence of higher curvature corrections. The leading order of higher curvature corrections is obtained by considering the case with $i=2$ which is the well-known (quadratic) Gauss-Bonnet term with $K$ specifies the maximal order of higher curvature corrections beyond Gauss-Bonnet. For conventional Lovelock gravity in $D$-dimensions, there exist always a finite number of dimensionful parameters, $\alpha_i$'s, but the rescaling (\ref{rescalings}) allows us to have any number of dimensionful parameters in four-dimensions, which most likely enriches the thermodynamic phase space as well as the holographic system in the gauge/gravity applications. The action (\ref{bulk action}) in a $D$-dimensional spacetime has exactly $K-1 = \left[ {\frac{{D - 3}}{2}} \right]$ new dimensionful parameters in addition to those of Einstein gravity with a cosmological constant. \vspace{1mm}
 
In the present work, we use the simple mathematical prescription of Glavan and Lin \cite{Glavan2020} as a convenient shortcut for obtaining the relevant black brane backgrounds, since our main focus is their holographic applications in the probe limit. Importantly, these backgrounds also arise from conformal regularization \cite{ZD2026}. Considering the bulk action (\ref{bulk action}), the Einstein-Lovelock field equations are found to be as
 \begin{equation} \label{field eq gravity}
 {G_{\mu \nu }} + \Lambda {g_{\mu \nu }} + \sum\limits_{i = 2}^K {\frac{{{\tilde \alpha _i}}}{{{2^{i + 1}}}}{g_{\mu \lambda }}\delta _{\nu {\rho _1}...{\rho _i}{\sigma _1}...{\sigma _i}}^{\lambda {\eta _1}...{\eta _i}{\tau _1}...{\tau _i}}{R_{{\eta _1}{\tau _1}}}^{{\rho _1}{\sigma _1}}...{R_{{\eta _i}{\tau _i}}}^{{\rho _i}{\sigma _i}}}  = 0.
 \end{equation}
Implementing the coupling rescalings (\ref{rescalings}) and then taking the $D \to 4$ limit, the gravitational field equations for a four-dimensional static line element ansatz with a planar horizon reduce to a polynomial equation of degree $K$, given by \cite{Zhidenko2020PRDa,Casalino2021}
 \begin{equation} \label{reduced field eq}
 \sum\limits_{i = 1}^K {{\alpha _i}{{\left( {\frac{{f(r)}}{{{r^2}}}} \right)}^i}}  = \frac{\Lambda }{3} + \frac{M}{{{r^3}}},
 \end{equation}
where $M$ is an integration constant proportional to the ADM mass of the spacetime (see the details in refs. \cite{Zhidenko2020PRDa,Casalino2021}). Now, following \cite{ICP2014}, it can be shown that, in four dimensions, an analytic black hole solution up to any $K$th order in curvature can be found by assuming that the Lovelock couplings ($\alpha_i$'s with $i\ge 2$ in our notation) obey the following special relationship \cite{DS2022}
 \begin{equation} \label{fine tuning}
 {\alpha _i} = {\alpha _K}{\left( {\frac{1}{{K{\alpha _K}}}} \right)^{\frac{{K - i}}{{K - 1}}}}\left( {\begin{array}{*{20}{c}}
 	K\\
 	i
 	\end{array}} \right).
 \end{equation}
Such fine-tunings in standard Lovelock gravity has led to analytic black hole solutions of order $K$ which have been the subject of many studies, providing the first example of non-standard critical exponents along with a number of novel phase transitions in black hole physics \cite{ICP2014,Frassino2014,2021}. It should also be noted that, from the low-energy limit of the heterotic string theory, the Lovelock couplings are proportional to the inverse string tension, which means they are positive definite \cite{Boulware1985Deser}. However, following \cite{Charmousis2022}, a compelling bound on negative values of $\alpha$ in regularized $4D$ Einstein-Lovelock theories can always be found by demanding that atomic nuclei should not be shielded by a horizon. Hence, for the sake of completeness, besides the positive sign of the Lovelock coefficients, we will consider the cases with negative couplings. Assuming the fine-tunings (\ref{fine tuning}), the background we consider is the planar $\text{AdS}_4$ black hole -\textit{black brane}- solutions of the field equations (\ref{field eq gravity}) as \cite{DS2022}
\begin{equation} \label{metric}
d{s^2} =  - f(r)d{t^2} + \frac{{d{r^2}}}{{f(r)}} + {r^2}\left( {d{x^2} + d{y^2}} \right),
\end{equation}
where 
\begin{equation} \label{metric function}
 f(r) = \frac{{{r^2}}}{{{{\left( {K{\alpha _K}} \right)}^{\frac{1}{{K - 1}}}}}}\left \{{1 - {{\left[ {1 + {{\left( {{K^K}{\alpha _K}} \right)}^{\frac{1}{{K - 1}}}}\left( {\frac{M}{{{r^3}}} + \frac{\Lambda }{3}} \right)} \right]}^{\frac{1}{K}}}} \right \}.
 \end{equation}

Up to quadratic curvature corrections ($K=2$), alternative regularization procedures have been shown to share the relevant black hole/brane solutions with the original Glavan-Lin construction \cite{Glavan2020,Mann2020,Kobayashi2020,LuPang2020,Mukohyama2020,EGB4Dreview}. For cubic corrections ($K=3$) the corresponding agreement has also been established for maximally symmetric black hole geometries with planar horizons (black branes) through a regularized Kaluza-Klein reduction. It is conceivable that this agreement can also  be obtained from the conformal regularization approach because the same Lagrangian is obtained from both approaches \cite{Alkac2022}. More generally, a well-defined dimensional conformal regularization of the invariants in Lovelock Lagrangian leads to four-dimensional scalar-tensor theories belonging to the Horndeski class and is applicable at arbitrary Lovelock order \cite{Fernandes2025,Fernandes2026}. For the planar black hole ansatz, the logarithmic scalar profile
$\phi(r)=\log(r/r_0)$ satisfies the scalar-field equation, while the remaining gravitational equations reduce to a single algebraic equation for the metric function. In this sense, the planar backgrounds considered here admit an independent scalar-tensor realization beyond the original $D\to4$ prescription. In particular, the reduced equation obtained in this framework has the same planar Lovelock structure as the reduced field equation used above, Eq. (\ref{reduced field eq}), after the appropriate identification of coupling and mass normalizations \cite{ZD2026}. This observation also provides a useful connection between the black brane background of the $4D$ regularized construction employed here and the planar black brane solutions of the conformal regularization of Einstein-Lovelock gravity. The latter have reduced equations with the same algebraic curvature structure.\footnote{Note that this statement assumes a static metric ansatz with planar (Ricci flat) horizon. For maximally symmetric (static) ansatz with horizon geometries of constant curvature, solutions with curvature corrections beyond Gauss-Bonnet from the naive limit generally do not match the other regularization procedures, as indicated in ref. \cite{Alkac2022} for cubic Lovelock gravity.} The relevance of this connection for the present holographic analysis is that the background geometry, rather than the particular $4D$ regularization used to obtain it, enters the probe-limit matter equations. A detailed derivation of this connection, together with another higher-curvature solution and a systematic analytical study of their geometric, thermodynamic, and holographic-superconducting properties, will be presented in a separate work currently in preparation \cite{ZD2026}. We therefore restrict the present paper to the holographic consequences of these planar backgrounds. \vspace{1mm}

\begin{figure}[!htbp]
	\begin{center}
		\epsfxsize = 4 cm
		\includegraphics[keepaspectratio=true,scale=0.75]{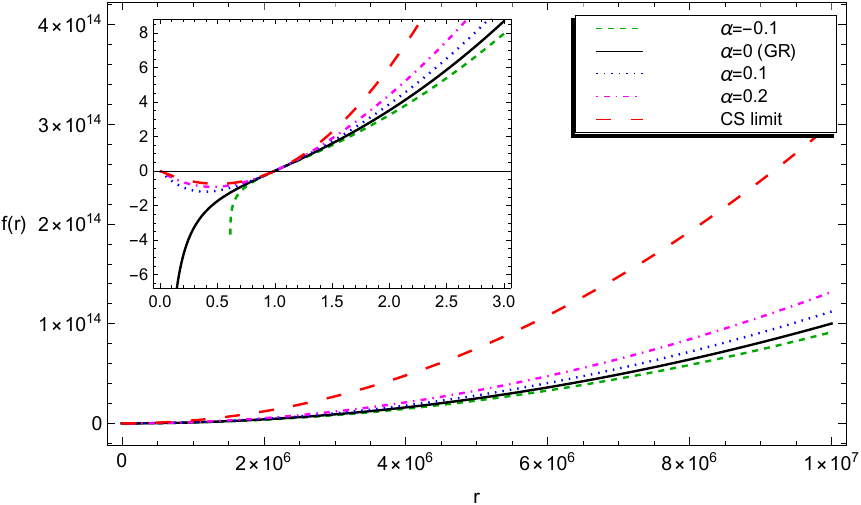}
	\end{center}
	\caption{\textit{Numerical solutions of $f(r)$ versus $r$ for several values of $\alpha$ with $M=L=1$. This plot is specific to the $4D$ third-order (cubic) Einstein-Lovelock gravity, i.e. the case with $K=3$ in our notation, but qualitatively the same behavior is observed for the theory up to any $K$th order of higher curvature corrections. As seen from the long-dashed (red) curve, in the CS limit ($\alpha=1/3$), the asymptotic behavior of $f(r)$ changes drastically and, from Eq. (\ref{f(r)_CS limit}), it is seen that the correction to the AdS quadratic gravitational potential shows a ``${r^{\frac{{2K - 3}}{K}}}$" dependence.}}
	\label{f_behaviors}
\end{figure}

Now, let us investigate some required features of the $\text{AdS}_4$ black brane spacetimes (first presented in \cite{DS2022}), defined by eqs. (\ref{metric}) and (\ref{metric function}). First, it is found that the obtained black hole spacetime using the fine-tunings (\ref{fine tuning}), given by eqs. (\ref{metric}) and (\ref{metric function}), can be recast in a more compact and practical form. To do so, using the fine-tunings (\ref{fine tuning}), one can write $\alpha_K$ in terms of the leading order of corrections, $\alpha_2=\alpha$, as ${\alpha _K} = \frac{1}{K}{\big( {\frac{{2\alpha }}{{K - 1}}} \big)^{K - 1}}$. Thus, the metric function (\ref{metric function}) can be rewritten as
\begin{equation} \label{metric_function_newform}
f(r) = \frac{{(K - 1){r^2}}}{{2\alpha }}\left \{ {1 - {{\left[ {1 + \frac{{2K\alpha }}{{K - 1}}\left( {\frac{M}{{{r^3}}} + \frac{\Lambda }{3}} \right)} \right]}^{\frac{1}{K}}}} \right \}.
\end{equation}
This new expression for the metric function is very practical to straightforwardly access the $K \to \infty$ limit of the spacetime (i.e., black branes with the entire number of higher-order curvature corrections) in analytical studies. At large distance, we have the asymptotic Schwarzschild-AdS metric with planar symmetry by expanding the metric function about $\alpha=0$, yielding ${\left. {f(r)} \right|_{\alpha  \to 0}} = \frac{{{r^2}}}{{{L^2}}} - \frac{M}{r} + {\cal O}(\alpha )$. Expanding $f(r)$ around $K = 1$ leads to the same result and recovers the planar Schwarzschild black brane solution again. The location of the event horizon is found by use of the largest root of the metric function (\ref{metric function}), i.e. ${{{\left. {{g^{rr}}} \right|}_{r = {r_ + }}}}=f(r_+)=0$, yielding ${r_ + } = {\left( {M{L^2}} \right)^{1/3}}$. To determine the domain in which the metric function $f(r)$ is real, we need to investigate the asymptotic behavior of the spacetime. The metric function $f(r)$ at large distance, $r \to \infty$, behaves as 
\begin{equation}  \label{asymp f(r)}
{\left. {f(r)} \right|_{r \to \infty }} = \frac{{(K - 1){r^2}}}{{2\alpha }}\left( {1 - {{\left[ {1 - \frac{{2K\alpha }}{{(K - 1){L^2}}}} \right]}^{\frac{1}{K}}}} \right) + {\cal O} \big(r^{-1} \big),
\end{equation}
which implies the following restriction on the parameters for ensuring an asymptotically AdS background and avoiding naked singularities 
\begin{equation} \label{restriction}
{\alpha _K} \leqslant \frac{{{L^{2(K - 1)}}}}{{{K^K}}} \quad \Leftrightarrow \quad \alpha  \leqslant \frac{{(K - 1){L^2}}}{{2K}}.
\end{equation}
As a result, $\alpha$ approaches $L^2/2$ when the infinite series of higher curvature terms ($K \to \infty$) are included. Eq. \ref{asymp f(r)} shows that the correction to the AdS quadratic gravitational potential near the AdS boundary has a $1/r$ dependence (expanding $f(r)$ about $\alpha = 0$ also reveals that the leading order of corrections has a $1/r$ dependence everywhere). Nevertheless, the metric function behaves quite different near the upper bound of $\alpha$ (\ref{restriction}), i.e. $\alpha_{\rm{CS}} = \frac{{(K - 1){L^2}}}{{2K}}$, a limit that is similar to a class of Chern-Simons (CS) theories in odd higher dimensions (see refs. \cite{Zanelli2000,Zanelli2005} for more details). For convenience, throughout the paper this bound will be referred to as the CS limit in keeping with convention \cite{HS-GB-2009Gregory,HS-GB-2010Gregory,HS-4GB-2020JHEP,ICP2014} (in this nomenclature, the reader should be careful about the different contexts of this naming). In the CS limit, $f(r)$ behaves as
\begin{equation}\label{f(r)_CS limit}
{\left. {f(r)} \right|_{\alpha  \to {\alpha _{{\rm{CS}}}}}} = \frac{{K{r^2}}}{{{L^2}}}\left( {1 - {{\left[ {\frac{{M{L^2}}}{{{r^3}}}} \right]}^{1/K}}} \right),
\end{equation}
indicating a ``${r^{\frac{{2K - 3}}{K}}}$" dependence instead of ``$1/r$", which subsequently results in a much stronger effect in this limit at the AdS boundary.\footnote{For the five-dimensional Gauss-Bonnet gravity, the correction to the AdS quadratic gravitational potential changes from the $1/r$ dependence to a constant in the CS limit \cite{HS-GB-2009Gregory}.} In Fig. \ref{f_behaviors}, the numerical solutions for several values of $\alpha$ are plotted, which clearly shows the impact of varying the coupling constant on the asymptotic behavior of the spacetime. Consequently, since the holographic condensation is a model-dependent phenomenon, one should expect dramatic behavior when approaching the CS limit, as already observed for Einstein-Gauss-Bonnet theories in lower and higher dimensions \cite{HS-GB-2009Gregory,HS-4GB-2020JHEP}; Note that this conclusion is not affected by considering the gravitational backreaction of the matter fields on the spacetime geometry. However, we anticipate that the bulk causal structure analysis further constrain the physical parameter space of the theory and, especially, make the upper limit of the physical range of $\alpha$ in Eq. (\ref{restriction}) smaller, the same as the analysis performed in refs. \cite{KSS2008violation,deBoer2010} for the Gauss-Bonnet and the cubic Lovelock gravity theories in higher dimensions. To our knowledge, there has been no such analysis in the literature regarding this issue for the regularized $4D$ Einstein-Lovelock theory, but, for the case of $4D$ Einstein-Gauss-Bonnet gravity, it has been shown that the Gauss-Bonnet coupling constant ($\alpha_2 =\alpha$, in our notation) is restricted to the negative-$\alpha$ regime \cite{causality4dGH2020}. This gives consideration to the study of holographic superconductors in the regime with negative coupling constant $\alpha$ a special significance. For the sake of completeness, we will consider the whole range of (\ref{restriction}) in order to gain more insights into the parameter space of the holographic model. \vspace{1mm}

One can also rewrite Eq. (\ref{asymp f(r)}) as ${\left. {f(r)} \right|_{r \to \infty }} = \frac{{{r^2}}}{{L_{{\text{eff}}}^2}} + {\cal O}({r^{ - 1}})$, which implies that the effective asymptotic AdS scale, $L_\text{eff}$, up to any order of higher curvature corrections can be defined as\footnote{Eq. (\ref{Leff}) in terms of $\alpha_K$ is rewritten as
	\begin{equation}
	L_{{\rm{eff}}}^2 \equiv \frac{{{{\left( {K{\alpha _K}} \right)}^{\frac{1}{{K - 1}}}}}}{{1 - {{\left[ {1 + \frac{{{{\left( {{K^K}{\alpha _K}} \right)}^{\frac{1}{{K - 1}}}}}}{{{L^2}}}} \right]}^{\frac{1}{K}}}}} = \left\{ \begin{array}{ll}
	{L^2}\,,\quad &{\text{for}}\,\alpha  \to {\rm{0}}\,\\
	\,  \\
	\frac{{{L^2}}}{K}\,\,, \quad &{\text{for}}\,\alpha  \to \alpha_{\rm{CS}}= \frac{{{L^{2(K - 1)}}}}{{{K^K}}}\,
	\end{array} \right. \, . \nonumber
	\end{equation}}
\begin{equation} \label{Leff}
L_{{\text{eff}}}^2 \equiv \frac{{2\alpha }}{{(K - 1)\left[ {1 - {{\left( {1 - \frac{{2K\alpha }}{{(K - 1){L^2}}}} \right)}^{\frac{1}{K}}}} \right]}} = \left\{ \begin{array}{ll}
{L^2}\,, \quad &{\text{for}}\,\alpha  \to {\text{0}}\, \\
\, \\
\frac{{{L^2}}}{K}\,, \quad & {\text{for}}\,\alpha  \to \alpha_{\rm{CS}}= \frac{{(K - 1){L^2}}}{{2K}}
\end{array}  \right. \,.
\end{equation}
Furthermore, using the definition of surface gravity \cite{Hawking1975Radiance,BardeenCarterHawking1973} or the Euclidean trick \cite{Gibbons1977Hawking,Hawking1983Page}, the Hawking temperature of the spacetime as the temperature of the dual boundary field theory is computed as 
 \begin{equation}
T = \frac{1}{{4\pi }}{\left. {\frac{{\partial f(r)}}{{\partial r}}} \right|_{r = {r_ + }}} = \frac{{3{r_ + }}}{{4\pi {L^2}}} = \frac{{3{M^{1/3}}}}{{4\pi {L^{4/3}}}}.
 \end{equation}
 
 It should be noted that the fine-tunings (\ref{fine tuning}) imply 
 \begin{equation}
 {\alpha _2} \propto \alpha \,,\,\,\,{\alpha _3} \propto {\alpha ^2}\,,\,\,\,{\alpha _4} \propto {\alpha ^3}\,,\,\,\,...\,,\,\,{\alpha _K} \propto {\alpha ^{K - 1}},
 \end{equation}
 meaning that, the Gauss-Bonnet term dominates in the weak-gravity regime ($\alpha \ll 1$), while the role of higher-order curvatures such as cubic and quartic terms becomes more and more important in the strong gravitational field regime, as expected from string theory. The two important cases of physical interest beyond Gauss-Bonnet term are corrections up to cubic and quartic curvature terms, for which the fine-tuned coupling constants of the theory are as follows 
\begin{eqnarray}
\text{when} && K = 3: \quad {\alpha _1} = 1,\,{\alpha _2} = \alpha ,\,{\alpha _3} = \frac{\alpha ^2}{3}, \\
\text{when} && K = 4: \quad {\alpha _1} = 1,\,{\alpha _2} = \alpha ,\,{\alpha _3} = \frac{{4{\alpha ^2}}}{9}\,,\,{\alpha _4} = \frac{{2{\alpha ^3}}}{{27}}.
\end{eqnarray}
Note that once the sign of the first (Gauss-Bonnet) coupling is established, Eq. (\ref{fine tuning}) dictates that the remaining couplings will follow suit. Specifically, if the first coupling ($\alpha_2=\alpha$) is negative, then the other couplings that represent corrections of the even order will also be negative. From now on, we call $\alpha$ the fine-tuned Lovelock coupling, for convenience.
 

\section{Holographic $s$-wave superconductors} \label{sect3:sWave}
In order to have a holographically dual description of $s$-wave superconductors with an isotropic order parameter, a Maxwell vector field ($A_{\mu}$)  and a charged (complex) scalar field ($\Psi$, with the charge $q$ and the mass $m$) as the minimum ingredients should be considered in the black brane background (\ref{metric}) and \ref{metric function}. Hence, the appropriate matter action may be written as \cite{HHH2008PRL}
\begin{equation} \label{s-wave matter}
{S_{matter}} =  - \frac{1}{{16\pi {G_N}}}\int {{d^4}x\sqrt { - g} \left[ {\frac{1}{4}{F_{\mu \nu }}{F^{\mu \nu }} + {{\left| {{\nabla _\mu }\Psi  - iq{A_\mu }\Psi } \right|}^2} + {m^2}{{\left| \Psi  \right|}^2}} \right]},
\end{equation}
where ${F _{\mu \nu }} = {\nabla _\mu }{A _\nu } - {\nabla _\nu }{A _\mu }$ is the usual strength field tensor in Maxwell's electrodynamics. Considering static (isotropic) ans\"{a}tze for both the gauge and the scalar fields as
\begin{equation} \label{ansatz_sWave}
{A_\mu } = \Phi (r)\delta _\mu ^0\,, \quad \Psi  = \Psi (r),
\end{equation}
and then varying the matter action (\ref{s-wave matter}) in the black brane background (reviewed in section \ref{sect2:setup}), the equations of motion for the Maxwell scalar potential $\Phi$ and the scalar field $\Psi$ in the probe limit turn out to be as
\begin{equation} \label{scalarEq_s-wave}
\Phi '' + \frac{2}{r}\Phi ' - \frac{{2{q^2}{\Psi ^2}}}{f}\Phi  = 0,
\end{equation}
\begin{equation} \label{vectorEq_s-wave}
\Psi '' + \left( {\frac{{f'}}{f} + \frac{2}{r}} \right)\Psi ' + \left( {\frac{{{q^2}{\Phi ^2}}}{{{f^2}}} - \frac{{{m^2}}}{f}} \right)\Psi  = 0,
\end{equation}
respectively, where the prime denotes the derivative with respect to $r$ and $f \equiv f(r)$. \vspace{1mm}

In the next subsections, we first determine the regularity conditions at both the event horizon and the AdS boundary which are necessary for solving the Maxwell and scalar field equations, and hence obtaining the corresponding dual boundary operators. Note that we consider matter fields as probes, the so-called probe limit, where they decouple from gravity (Effects of the backreaction of matter fields need to be extensively studied in an another study). We then discuss the scalar condensate, the critical temperature, and the (super)conductivity of dual boundary gauge theory. The results clearly indicate the $s$-wave holographic superconductivity (with an isotropic order parameter) in the presence of any number of higher curvature corrections. We find that both the fine-tuned Lovelock coupling ($\alpha$) and the maximal order of higher curvature corrections ($K$) have dramatic effects on the holographic phenomenon of superconductivity.

\subsection{Regularity at the horizon and the AdS boundary} \label{sect3-1}

The differential equations (\ref{scalarEq_s-wave}) and (\ref{vectorEq_s-wave}) have to be solved by imposing regularity at both the horizon ($r=r_+$) and the AdS boundary ($r \to \infty$). As usual, at the horizon one finds
\begin{equation} 
A_t(r_+)=\Phi ({r_ + }) = 0, \quad \Psi ({r_ + }) = \frac{{f'({r_ + })\Psi '({r_ + })}}{{{m^2}}}, 
\end{equation}
which follow from the fact that the norm of the gauge field ($A_\mu A^\mu$) must not diverge and also the corresponding current $j_\nu$ in the Maxwell equations remains finite at the horizon in any gauge \cite{Hartnoll2009Lectures}. The asymptotic forms of the solutions near the AdS boundary read
\begin{equation} \label{fields behavior at AdS_sWave}
{\left. \Phi  \right|_{r \to \infty }} = \mu  - \frac{\rho }{r}, \quad {\left. \Psi  \right|_{r \to \infty }} = \frac{{{\Psi _ - }}}{{{r^{{\Delta _ - }}}}} + \frac{{{\Psi _ + }}}{{{r^{{\Delta _ + }}}}},
\end{equation}
where ${\Delta _ \pm }$ may be expressed in terms of ${L_{{\rm{eff}}}^2}$ or in terms of $L$, $\alpha$ and $K$ as
\begin{eqnarray} \label{lamdapm_Swave}
{\Delta _ \pm } &=& \frac{3}{2} \pm \frac{1}{2}\sqrt {9 + 4{m^2}L_{{\text{eff}}}^2} \nonumber \\
&=& \frac{3}{2} \pm \frac{1}{2}\sqrt {9 + \frac{{8{m^2}\alpha }}{{K - 1}}{{\left( {1 - {{\left[ {1 - \frac{{2\alpha K}}{{(K - 1){L^2}}}} \right]}^{1/K}}} \right)}^{ - 1}}} \, \, . 
\end{eqnarray}
In the view of AdS/CFT correspondence, $\mu$ and $\rho$ are interpreted as the chemical potential and charge density of the boundary gauge field theory, respectively. As far as the boundary field theory is stable, it is permissible to set a negative squared mass for the scalar field, i.e. ${m^2} < 0$. Hence, the squared mass obeys the corresponding Breitenlohner-Freedman (BF) bound \cite{BFbound1982} as
\begin{equation} \label{BF bound}
{m^2} \ge m_{{\rm{BF}}}^2 = \frac{{ - 9}}{{4L_{{\rm{eff}}}^2}},
\end{equation}
 where the effects of higher curvature terms on the BF bound are directly understood from the fine-tuned coupling ($\alpha$) and the maximal order of higher curvature corrections ($K$) in Eq. (\ref{lamdapm_Swave}). On the other hand, a positive squared mass (${m^2}>0$) is not allowed whenever the slow falloff sector ($\Psi_-/r^{ \Delta_-}$) of the scalar field (\ref{fields behavior at AdS_sWave}) is taken into account since it diverges at the AdS boundary (thereby leading to a diverging energy-momentum tensor and hence being non-renormalizable). Putting these together, the scalar field mass should remain in the specific range of $m_{{\rm{BF}}}^2< m^2 <0$ in order for both falloffs to be renormalizable. On the other hand, stability of AdS requires that one of the falloffs, $\Psi_-$ or $\Psi_+$, must always be zero \cite{Hertog2004}. So, regarding the superconductivity phenomenon, we can take either $\Psi_-$ or $\Psi_+$ as zero, often $\Psi_-=0$, to play the role of a turned-off source in the spontaneous symmetry breaking of $U(1)$ gauge symmetry. This ensures a normalizable solution always. Therefore, obeying the BF bound (\ref{BF bound}), one can set either $m^2 {L_{{\rm{eff}}}^2}$ or $m^2 {L^2}$ to be fixed; for example, $m^2 {L_{{\rm{eff}}}^2} = -2$ everywhere (of course, the positive values for the squared mass are also allowed when $\Psi_-$ is dual to the source for the boundary operator $\cal O$). We will examine both cases for fixing the squared mass of the scalar field.

\subsection{Scalar condensate}
The black brane solution, (\ref{metric}) and (\ref{metric function}) with $\Psi=0$, corresponds to the normal high-temperature phase ($T > T_c$). The scalar condensate appears at low enough temperatures, lower than a critical temperature $T_c$, where the normal phase becomes unstable. Obviously, in this region, the solution is replaced by a black brane in the presence of non-zero $\Psi$ (so the system undergoes a second-order phase transition), which allows a spontaneous symmetry breaking. Therefore, $\Psi$ in principle controls the transition from a normal phase to a superconducting phase, thereby leading to interpreting it as the isotropic order parameter in the boundary gauge theory. However, in the view of AdS/CFT correspondence, either $\Psi_+$ or $\Psi_-$ can be dual to the expectation value of the scalar (condensation) operator $\cal O$ with the conformal dimension $\Delta_\pm$ at strong coupling in the boundary, i.e. ${\Psi _ + } = \left\langle {{{\cal O}_ + }} \right\rangle $ or ${\Psi _ - } = \left\langle {{\cal O}_ - } \right\rangle $, due to the two possible quantization schemes \cite{MAGOO2000,GKP1998,Witten1998a,Nastase2015Book,NatsuumeBookAdS/CFT,ErdmengerBook2015,HHH2008PRL,Cai-2015Review}. The general consensus, in line with physical intuition, is that the slow falloff of bulk field represents the external source of the corresponding operator in the boundary field theory while the fast falloff being its response. We refer to the quantization with ``\textit{fast falloff} $ \leftrightarrow$ \textit{response}'' as the standard picture while addressing the other quantization, ``\textit{slow falloff} $ \leftrightarrow$ \textit{response}'', as the alternative picture \cite{NatsuumeBookAdS/CFT}. \vspace{1mm}

To fully study scalar condensate, we first concentrate on the case with $m^2 {L^2} = \textit{fixed}$ (which deals with the true cosmological constant) for both the standard and the alternative quantization pictures. As will become clear later, the same qualitative description of superconductivity is also observed for the case of $m^2 {L_{\text{eff}}^2} = \textit{fixed}$ (dealing with the effective cosmological constant), although the variations of the relevant basic quantities (like the condensate and critical temperature) with respect to the theory's parameter space ($\alpha$ and $K$) are quite different in the alternative picture.

\subsubsection{${\Psi _ + } = \left\langle {{{\cal O}_ + }} \right\rangle $ and ${\Psi _ - }=0$ with $m^2 {L}^2 = \textit{fixed}$} \label{subsection3:2:1}

\begin{figure}[!htbp] 
	\begin{center}
		\epsfxsize = 4 cm
		\includegraphics[keepaspectratio=true,scale=0.78]{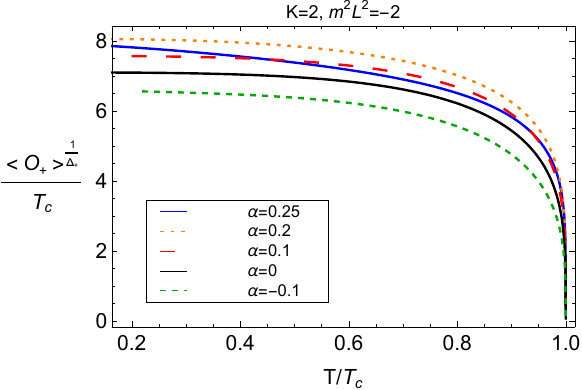}
		\hskip .1 cm
		\epsfxsize = 4 cm
		\includegraphics[keepaspectratio=true,scale=0.78]{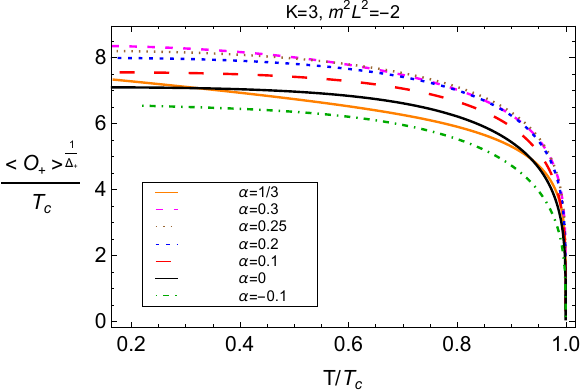}
		\vskip .1 cm
		\epsfxsize = 4 cm
		\includegraphics[keepaspectratio=true,scale=0.78]{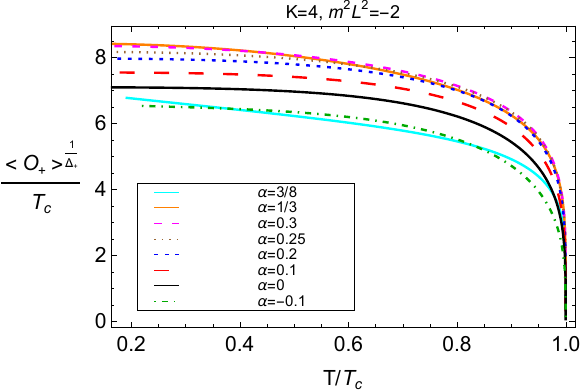}
		\hskip .1 cm
		\epsfxsize = 4 cm
		\includegraphics[keepaspectratio=true,scale=0.78]{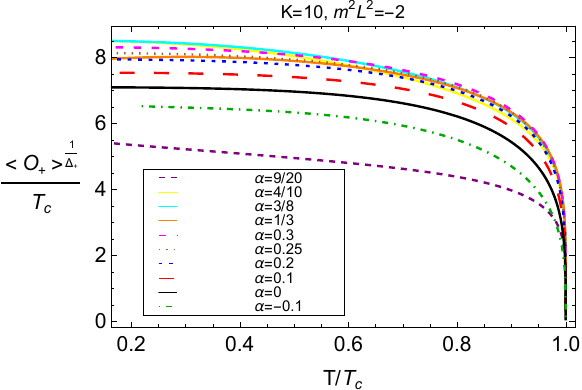}
		\vskip .1 cm
		\epsfxsize = 4 cm
		\includegraphics[keepaspectratio=true,scale=0.78]{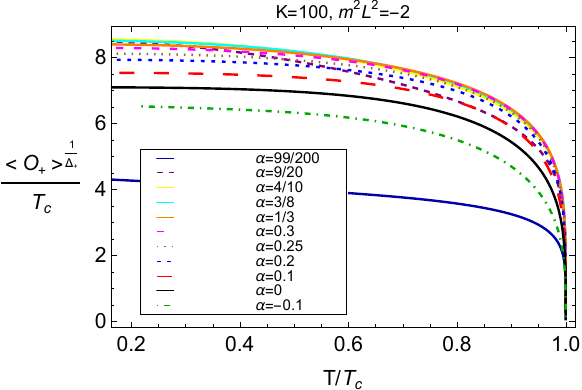}
	\end{center}
	\caption{\textit{Condensation of the scalar operator ${\cal O}_ + $ as a function of temperature for $4D$ Einstein-Lovelock gravity theories at different orders in curvature, labeled by $K=2, 3,4,10$ and $100$, with the mass of the scalar field fixed by $m^2 L^2=-2$. For each case, different values of $\alpha$ in the allowed ranges are considered. The curves with $\alpha =0 $ are the same in all panels and correspond to the Einstein gravity background.}} \label{fig:O+_scalar}
\end{figure}


Considering $\Psi _ -$ as the source and thereby imposing the boundary condition ${\Psi _ - }=0$, then ${\Psi _ + }$ is the spontaneous condensation of the operator $\cal O$. Let now calculate the condensate associated with this case, ${\Psi _ + } = \left\langle {{{\cal O}_ + }} \right\rangle $, that is the standard picture. In order to find $\Psi_+$, $\mu$ and $\rho$, the second order coupled differential equations (\ref{scalarEq_s-wave}) and (\ref{vectorEq_s-wave}) have been solved numerically by employing the shooting method with the boundary conditions discussed before. The resulting scalar condensates for regularized $4D$ Einstein-Lovelock theories at different orders in curvature ($K$) are depicted in Fig. \ref{fig:O+_scalar}, which all the condensation curves satisfy the BF bound (\ref{BF bound}). This figure exhibits the dimensionless quantity ${\left\langle {{{\cal O}_ + }} \right\rangle ^{1/{\Delta _ + }}}/{T_c}$ versus the scale invariant temperature $T/T_c$ for various values of the fine-tuned Lovelock coupling constant $\alpha$ and different $K$'s in the allowed range with the mass of the scalar field fixed by ${m^2}{L^2} =  - 2$. As seen, for all cases, the charged scalar operator $\left\langle {{{\cal O}_ + }} \right\rangle $ has condensed for temperatures in the range $T<T_c$. We also see that the condensate goes to a constant value as $T \to 0$. All these are similar to the results obtained in BCS theory and also Einstein gravity \cite{HHH2008PRL}. The condensate generally tends to increase with respect to $\alpha$, assuming any number of higher-order curvature corrections that may be considered. However, for $\alpha > 0$, the condensate at $T=0$ or at any fixed temperature is larger than that of Einstein gravity, meaning that positive fine-tuned couplings make the condensation harder (the black brane in the bulk has developed scalar hair and, as $\alpha$ increases, the formation of the scalar hair becomes harder). While, for $\alpha <0$, its value is smaller than the one in Einstein gravity and the effect of a negative $\alpha$ is to make it easier for scalar hair to form for any subclass of $4D$ Einstein-Lovelock gravity theories including the Einstein-Gauss-Bonnet class. Numerically, it is also observed that increasing $K$ with respect to the Gauss-Bonnet case ($K=2$) results in a decrease of the condensate provided that the value of fine-tuned coupling constant $\alpha$ is held fixed (unfortunately, this case is difficult to be seen in Fig. \ref{fig:O+_scalar}). Furthermore, for any $K$th order Einstein-Lovelock theory in four-dimensions, an abnormal behavior is observed for the curves associated with the values of $\alpha$ that are close to the upper bound (\ref{restriction}), i.e. the ones approaching the CS limit with $\alpha_{\rm{CS}} = \frac{{(K - 1){L^2}}}{{2K}}$; however, such curves qualitatively behave the ones in BCS theory as well as Einstein gravity, but they do not follow the pattern mentioned above since they exhibit a much stronger or unusual variation of the condensate with temperature (this has also been seen in gravity with Gauss-Bonnet corrections in higher dimensions \cite{HS-GB-2009Gregory,HS-4GB-2020JHEP}. However,  this behavior is intensified when more higher-order curvature corrections are included.). Although assuming a fully backreacting situation may provide a better understanding of this issue, but this behavior will still persist since in the CS limit the behavior of gravity specially near the horizon and the AdS boundary becomes different (see Fig. \ref{f_behaviors}), yielding a slightly different (often stronger) dependence of the condensate on temperature, as such behavior has also been reported in backreacting Gauss-Bonnet superconductors \cite{HS-GB-2010Gregory,HS-4GB-2021PLBa,HS-4GB-2021PLBb}.  Nevertheless, as mentioned in the previous section, since the causality constraint from the boundary CFT can place more restrictions on the physical parameter space of Einstein-Lovelock gravity theories \cite{KSS2008violation,deBoer2010,causality4dGH2020} and make the previous range obtained in Eq. (\ref{restriction}) smaller, we speculate the same happens for the regularized $4D$ Einstein-Lovelock theories, thereby leading to such abnormal behaviors near the upper limit of $\alpha$ to be out of the acceptable range.

\begin{figure}[!htbp]
	\begin{center}
		\epsfxsize = 4 cm
		\includegraphics[keepaspectratio=true,scale=0.75]{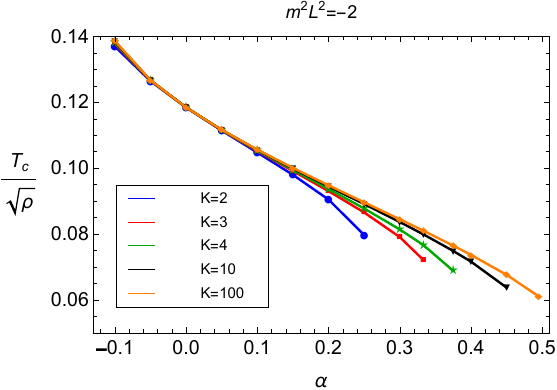}
		\hskip .1 cm
		\epsfxsize = 4 cm
		\includegraphics[keepaspectratio=true,scale=0.75]{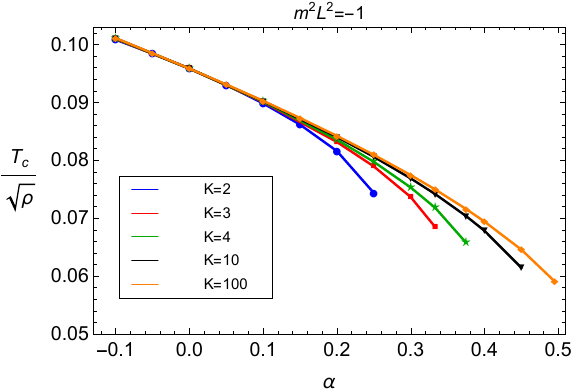}
		\vskip .1 cm
		\epsfxsize = 4 cm
		\includegraphics[keepaspectratio=true,scale=0.75]{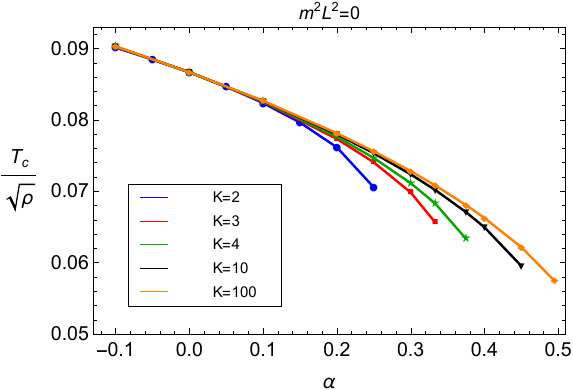}
	\end{center}
	\caption{\textit{Critical temperature for the superconducting transition, associated with the scalar operator ${\cal O}_ +$, as a function of $\alpha$ for $4D$ Einstein-Lovelock gravity theories at different orders in curvature, labeled by $K=2,3,4,10$ and $100$. The mass of the scalar field has been fixed by $m^2 L^2=-2$, $m^2 L^2=-1$ and $m^2 L^2=0$. The points with $\alpha =0$ correspond to the results of the Einstein gravity background. Note that, the upper left panel corresponds to Fig. \ref{fig:O+_scalar}.}} \label{fig:Tc_sWave_O+}
\end{figure}


The condensate is a temperature dependent phenomenon, which occurs below a characteristic critical temperature $T_c$. In Einstein gravity, for the condensate associated with the fast falloff of the bulk scalar field, one obtains $T_c \approx 0.118 \rho^{1/2}$ provided that ${m^2}{L^2} =  - 2$ \cite{HHH2008PRL}, while in the presence of higher curvature corrections, there exist more room for condensation to occur. Returning to our model, Fig. \ref{fig:Tc_sWave_O+} shows the change of the critical temperature for $4D$ Einstein-Lovelock theories of order $K$ as a function of $\alpha$ with different choices for fixing the mass of the scalar field. For every panel of Fig. \ref{fig:Tc_sWave_O+}, all curves meet each other at the point $\alpha =0 $, which corresponds to the results of Einstein's general relativity obtained in \cite{HHH2008PRL}. Generally, the critical temperature decreases with increasing $\alpha$, meaning that the scalar hair ($\propto$ the condensate) is harder to form as $\alpha$ increases. From this figure it is evident that we can also access higher critical temperatures with respect to the one obtained in Einstein gravity, and for this purpose, it is sufficient to consider the model in the regime with negative fine-tuned Lovelock couplings, $\alpha < 0$, where it is easier for the scalar hair to form. Increasing the maximal order of higher curvature corrections, through $K$, naturally leads to more possibilities for condensation. From this figure, it is inferred that increasing $K$ with respect to the Gauss-Bonnet case ($K=2$) results in an increase of the critical temperature $T_c$ (subsequently a decrease of the condensate, as seen in Fig. \ref{fig:O+_scalar}) provided that the value of fine-tuned coupling $\alpha$ is held fixed, making it easier for scalar hair to form for any subclass of $4D$ finely tuned Einstein-Lovelock gravity theories beyond the Einstein-Gauss-Bonnet class. \vspace{1mm}

Furthermore, near the critical temperature for small condensates, the usual square root behavior characteristic of Ginzburg-Landau mean-field theory as $\left\langle {{{\cal O}_ + }} \right\rangle  \propto {\left( {{T_c} - T} \right)^{1/2}}$ is observed by fitting the condensate curves in Fig. \ref{fig:O+_scalar}, in which the exponent $\beta=1/2$ confirms a second-order phase transition, as expected. All these capture the relevant physics of the superconductivity phenomenon. \vspace{1mm}

\subsubsection{${\Psi _ - } = \left\langle {{{\cal O}_ - }} \right\rangle $ and ${\Psi _ + }=0$ with $m^2 {L}^2 = \textit{fixed}$}

\begin{figure}[!htbp]
	\begin{center}
		\epsfxsize = 4 cm
		\includegraphics[keepaspectratio=true,scale=0.78]{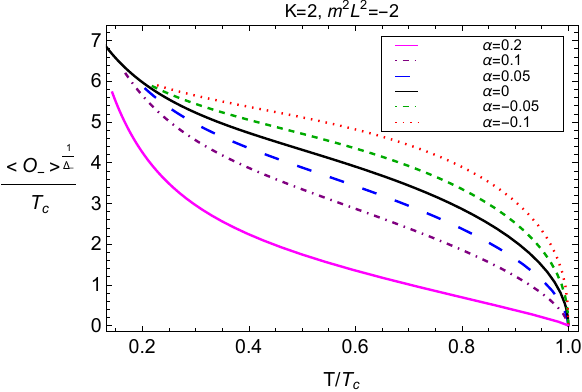}
		\hskip .1 cm
		\epsfxsize = 4 cm
		\includegraphics[keepaspectratio=true,scale=0.78]{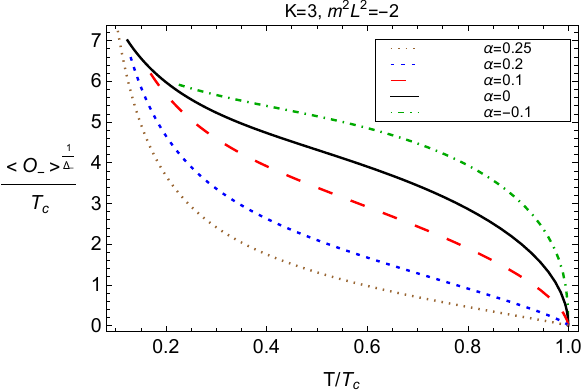}
		\vskip .1 cm
		\epsfxsize = 4 cm
		\includegraphics[keepaspectratio=true,scale=0.78]{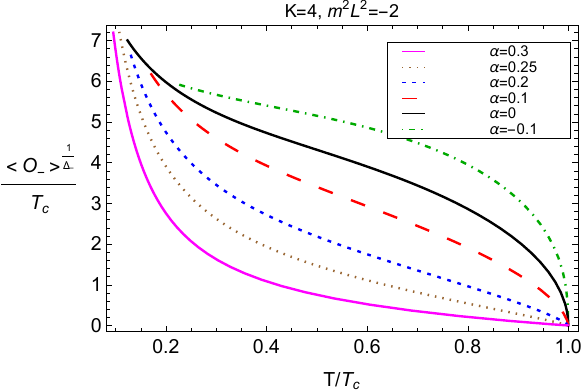}
		\hskip .1 cm
		\epsfxsize = 4 cm
		\includegraphics[keepaspectratio=true,scale=0.78]{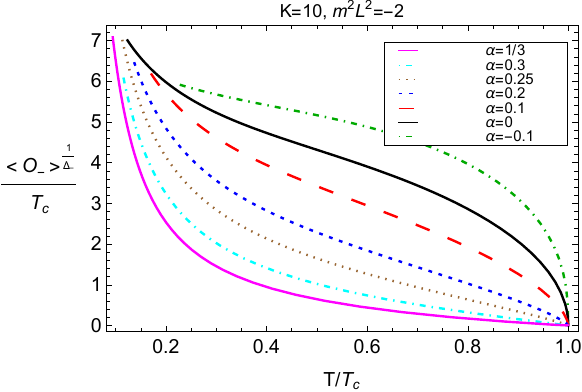}
		\vskip .1 cm
		\epsfxsize = 4 cm
		\includegraphics[keepaspectratio=true,scale=0.78]{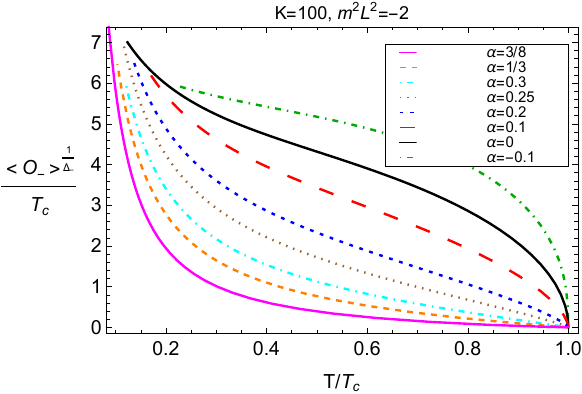}
	\end{center}
	\caption{\textit{Condensation of the scalar operator ${\cal O}_ - $ as a function of temperature for $4D$ Einstein-Lovelock gravity theories at different orders in curvature, labeled by $K=2, 3,4,10$ and $100$, with the mass of the scalar field fixed by $m^2 L^2=-2$. For each case, different values of $\alpha$ in the allowed ranges are considered. The curves with $\alpha =0 $ are the same in all panels and correspond to the Einstein gravity background.}} \label{fig:O-_scalar}
\end{figure}

In the alternative picture, where the condensate is given by ${\Psi _ - } = \left\langle {{{\cal O}_ - }} \right\rangle $ with the boundary condition ${\Psi _ + } =0$, one can holographically describe the phenomenon of superconductivity again, but we observe the opposite behavior for variations of $\left\langle {{{\cal O}_ - }} \right\rangle $ and $T_c$ with respect to $\alpha$ and $K$ in comparison with the standard picture discussed in the previous subsection, (\ref{subsection3:2:1}). Fig. \ref{fig:O-_scalar} shows condensation of the scalar operator ${{\cal O}_ - }$ below $T_c$ as a function of temperature in the presence of various orders of higher curvature corrections, $K$. In this figure, all curves for any $4D$ Einstein-Lovelock theory of order $K$ qualitatively mimic the behavior of the Einstein-Maxwell-scalar system \cite{HHH2008PRL} where the slow falloff sector (${\Psi _ - }$) is identified with the spontaneous condensation of the operator $\cal O$. Unlike the standard picture, the condensate tends to decrease with increasing $\alpha$, which means the scalar hair forms easier as $\alpha$ increases. We also observe that, by adding higher-order curvature corrections, the value of the $\left\langle {{{\cal O}_ - }} \right\rangle$ condensate becomes large provided that $\alpha$ is held fixed.\vspace{1mm}


\begin{figure}[!htbp]
	\begin{center}
		\epsfxsize = 4 cm
		\includegraphics[keepaspectratio=true,scale=0.8]{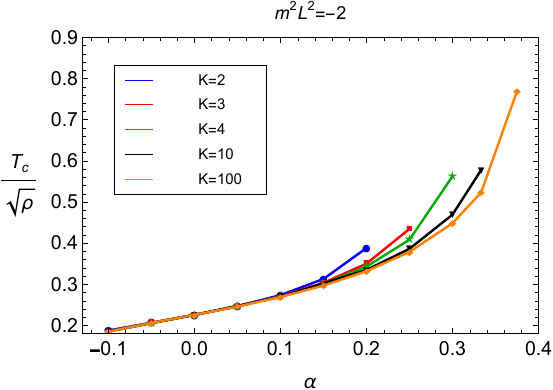}
	\end{center}
	\caption{\textit{Critical temperature for the superconducting transition, associated with the scalar operator ${\cal O}_ -$, as a function of $\alpha$ for $4D$ Einstein-Lovelock gravity theories at different orders in curvature, labeled by $K=2,3,4,10$ and $100$. The mass of the scalar field has been fixed by $m^2 L^2=-2$. The point with $\alpha =0$ corresponds to the result of Einstein gravity. Note that this figure corresponds to Fig. \ref{fig:O-_scalar}}} \label{fig:Tc_sWave_O-}
\end{figure}

The critical temperatures corresponding to the plots in Fig. \ref{fig:O-_scalar} are also shown in Fig. \ref{fig:Tc_sWave_O-}, which clearly indicate that, for any $K$th order of the theory, increasing $\alpha$ leads to higher values for $T_c$, which subsequently makes the formation of the scalar hair easier with respect to the Einstein-Maxwell-scalar system (compare with Fig. \ref{fig:O-_scalar}). In this figure, all the curves meet each other at the point $\alpha = 0$, which corresponds to the result of Einstein gravity, i.e. $T_c \approx 0.226 \rho^{1/2}$ when ${m^2}{L^2} =  - 2$ \cite{HHH2008PRL}. Assuming a fixed fine-tuned Lovelock coupling $\alpha$, we also confirm that increasing the order of higher curvature corrections ($K$) leads to decreasing $T_c$, which is easily inferred from Fig. \ref{fig:Tc_sWave_O-}. Interestingly, we also observe that the previous abnormal behavior associated with the condensation curves near the upper bound of $\alpha$ (the CS limit) for the operator ${\cal O}_+$ in Fig. \ref{fig:O+_scalar} is not present for the case of the $\left\langle {{{\cal O}_ - }} \right\rangle$ condensate, as seen in Fig. \ref{fig:O-_scalar}. On the other hand, at low temperatures, the condensate tends to diverge (developing a large scalar hair), signaling that the backreaction onto the bulk geometry can no longer be neglected, so the probe limit is not suitable in this region. Once again, we observe the usual square root behavior $\left\langle {{{\cal O}_ - }} \right\rangle  \propto {\left( {{T_c} - T} \right)^{1/2}}$ near the critical temperature, confirming a second-order phase transition. In conclusion, contrary to the  the standard picture discussed in Sec. \ref{subsection3:2:1}, the scalar condensate and the corresponding critical temperature behave quite different in the alternative picture and, compared to the Einstein gravity framework, for $\alpha>0$ we see higher critical temperatures (with less condensates) while lower critical temperatures (with larger condensates) for $\alpha <0$. \vspace{1mm}


\subsubsection{Scalar condensate with $m^2 {L_{{\rm{eff}}}^2} = \textit{fixed}$}

\begin{figure}[!htbp]
	\begin{center}
		\epsfxsize = 4 cm
		\includegraphics[keepaspectratio=true,scale=0.78]{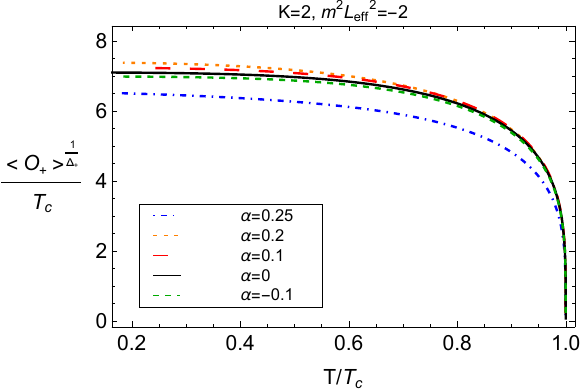}
		\hskip .1 cm
		\epsfxsize = 4 cm
		\includegraphics[keepaspectratio=true,scale=0.78]{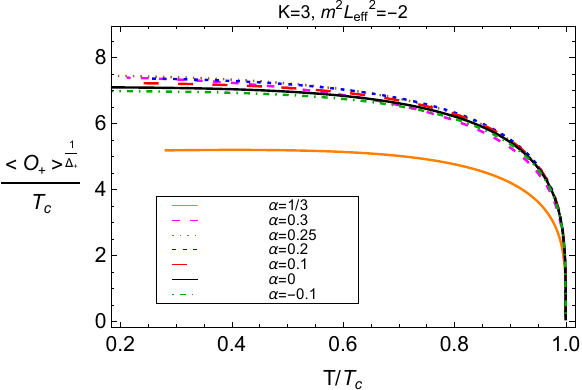}
		\vskip .1 cm
		\epsfxsize = 4 cm
		\includegraphics[keepaspectratio=true,scale=0.78]{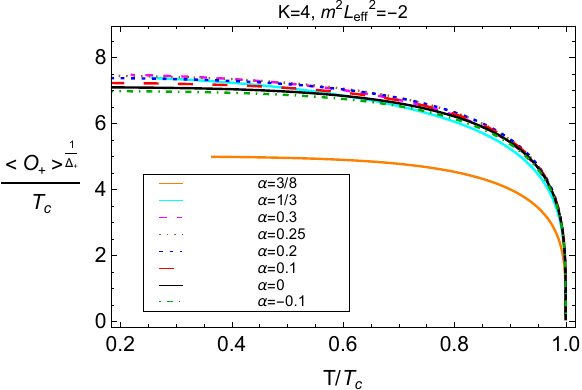}
		\hskip .1 cm
		\epsfxsize = 4 cm
		\includegraphics[keepaspectratio=true,scale=0.78]{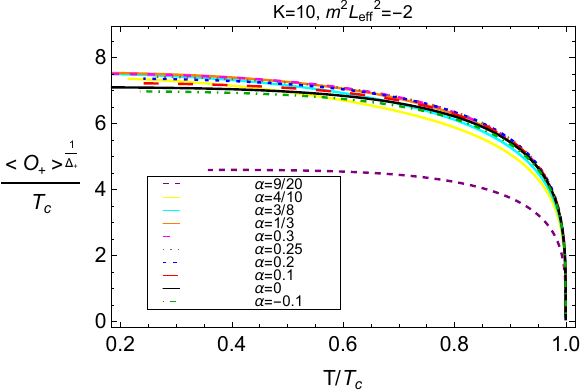}
		\vskip .1 cm
		\epsfxsize = 4 cm
		\includegraphics[keepaspectratio=true,scale=0.78]{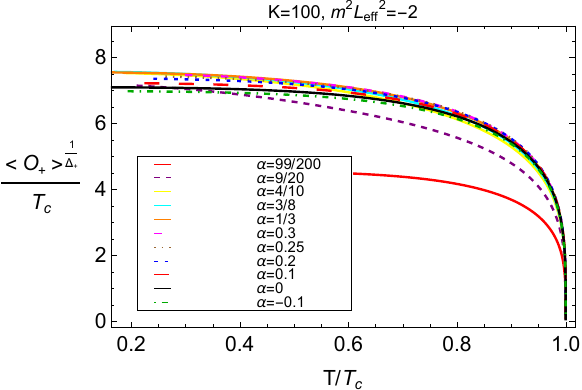}
	\end{center}
	\caption{\textit{Condensation of the scalar operator ${\cal O}_ + $ as a function of temperature for $4D$ Einstein-Lovelock gravity theories at different orders in curvature, labeled by $K=2, 3,4,10$ and $100$, with the mass of the scalar field fixed by ${m^2}L_{{\rm{eff}}}^2 =  - 2$. For each case, different values of $\alpha$ in the allowed ranges are considered. The curves with $\alpha =0 $ are the same in all panels and correspond to the Einstein gravity background where $\mathop {\lim }\limits_{\alpha  \to 0} {L_{{\text{eff}}}} = L$.}} \label{fig:O+_scalar_Leff}
\end{figure}

\begin{figure}[!htbp]
	\begin{center}
		\epsfxsize = 4 cm
		\includegraphics[keepaspectratio=true,scale=0.78]{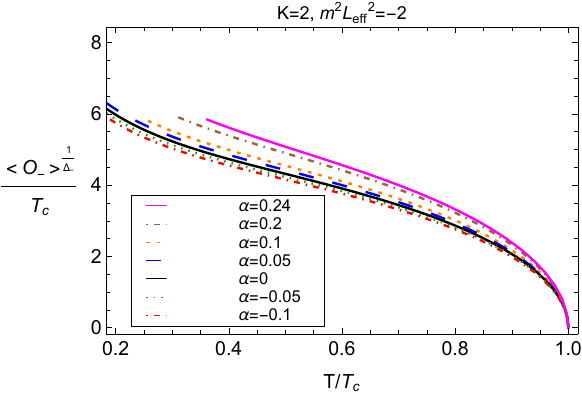}
		\hskip .1 cm
		\epsfxsize = 4 cm
		\includegraphics[keepaspectratio=true,scale=0.78]{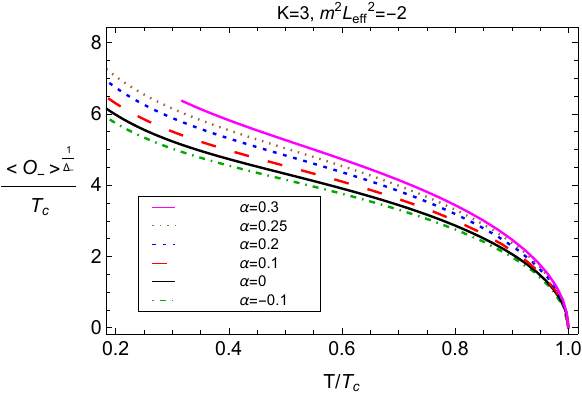}
		\vskip .1 cm
		\epsfxsize = 4 cm
		\includegraphics[keepaspectratio=true,scale=0.78]{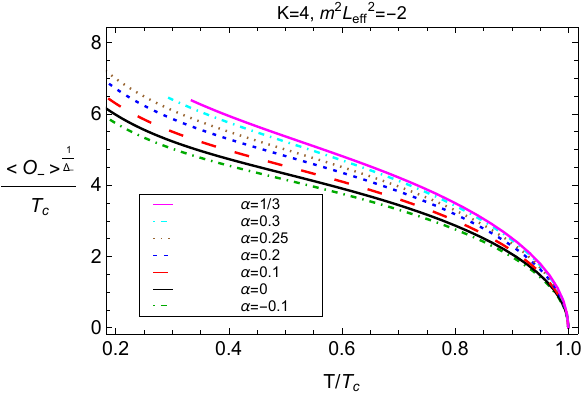}
		\hskip .1 cm
		\epsfxsize = 4 cm
		\includegraphics[keepaspectratio=true,scale=0.78]{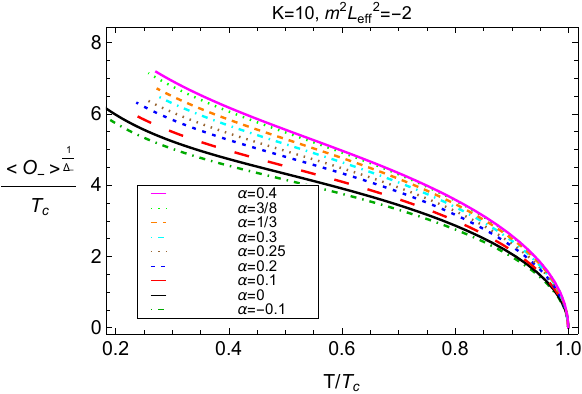}
		\vskip .1 cm
		\epsfxsize = 4 cm
		\includegraphics[keepaspectratio=true,scale=0.78]{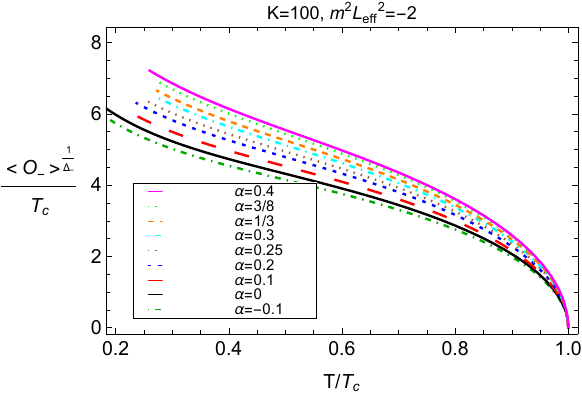}
	\end{center}
	\caption{\textit{Condensation of the scalar operator ${\cal O}_ - $ as a function of temperature for $4D$ Einstein-Lovelock gravity theories at different orders in curvature, labeled by $K=2, 3,4,10$ and $100$, with the mass of the scalar field fixed by ${m^2}L_{{\rm{eff}}}^2 =  - 2$. For each case, different values of $\alpha$ in the allowed ranges are considered. The curves with $\alpha =0 $ are the same in all panels and correspond to the Einstein gravity background where $\mathop {\lim }\limits_{\alpha  \to 0} {L_{{\text{eff}}}} = L$.}}\label{fig:O-_scalar_Leff}
\end{figure}

So far, we have concentrated on the scalar condensates and the corresponding critical temperatures by fixing the mass of scalar field as $m^2 L^2 = \textit{fixed}$, i.e. with respect to the AdS radius $L$, so that the mass remains the same as we vary $\alpha$. Instead, one can think of fixing the scalar field's mass relative to the effective asymptotic AdS scale (\ref{Leff}), i.e. $m^2 {L_{\text{eff}}^2} = \textit{fixed}$ \cite{HS-GB-2009Gregory,HS-GB-2010Gregory}, as discussed in Sec. \ref{sect3-1}. In this case, if one varies the coupling constant $\alpha$, $L_\text{eff}$ will vary accordingly, and in order for $m^2 L_{\text{eff}}^2$ to remain fixed, the mass of the scalar field will necessarily change. From this basic distinction, it is naturally expected that scalar operators within this new mass fixing do not behave similarly to the previous case where $m^2 L^2 = \textit{fixed}$ (see Sec. \ref{subsection3:2:1}). Regarding this new choice of fixing the mass, $m^2 L_{\text{eff}}^2 = \textit{fixed}$, Figs. \ref{fig:O+_scalar_Leff} and \ref{fig:O-_scalar_Leff} illustrate the condensate of the scalar operator, $\left\langle {{{\cal O}_ +}} \right\rangle $ and $\left\langle {{{\cal O}_ - }} \right\rangle $, in the boundary field theory dual to the bulk field $\Psi$. At first glance, compared to the previous condensates with $m^2 L^2=\textit{fixed}$, it is seen that the condensation curves are closer to each other, and as a result, the intensity of the effect of varying $\alpha$ and $K$ is less than before. We observe that almost the same features for the scalar operator in both (the standard and the alternative) pictures are obtained qualitatively, but the variations of the condensate as well as the critical temperature with $\alpha$ and $K$ are different, especially for the condensate associated with the scalar operator ${{\cal O}_ -}$. To be more specific, in the standard picture ($\Psi_+=\left\langle {{{\cal O}_ +}} \right\rangle $ and $\Psi_-=0$), increasing the fine-tuned coupling $\alpha$ results in increasing the condensate. While increasing the order of higher curvature corrections ($K$) with respect to the Gauss-Bonnet case ($K=2$) for a fixed $\alpha$ makes the formation of the scalar hair easier for most of the allowed range of the Lovelock coupling $\alpha$. In comparison with the Einstein gravity (i.e., the Einstein-Maxwell-scalar $s$-wave system), mostly, the formation of the scalar hair is harder for $\alpha > 0$ but easier for $\alpha <0$. \vspace{1mm}

However, for $K$th order Einstein-Lovelock theories beyond the Gauss-Bonnet corrections, we disclose some deviations from the aforementioned pattern for the condensates associated with the values of $\alpha$ approaching the upper bound (the CS limit) of this parameter (\ref{restriction}): such condensation curves intersect the ones belong to Einstein's theory (i.e., the curves with $\alpha = 0$ in Fig. \ref{fig:O+_scalar_Leff}) at a characteristic temperature, say $T_i$. Regarding such cases, the condensate is larger for the range $T<T_i$ but smaller for $T>T_i$, compared to the Einstein gravity framework. It is worth mentioning that this behavior is absent in $4D$ (or higher-dimensional) Einstein-Gauss-Bonnet gravity. Furthermore, we again observe the previous abnormal behavior of the curves close to the upper bound of $\alpha$, i.e. $\alpha = \frac{{(K - 1){L^2}}}{{2K}}$, in every panels of Fig. \ref{fig:O+_scalar_Leff}. We again expect that this behavior persists when the effect of backreaction into the black brane geometry is taken into account. However, this region is less interesting and will be excluded if one further considers the causality constraint to have a well-defined boundary CFT dual \cite{KSS2008violation,deBoer2010,causality4dGH2020}. \vspace{1mm}


Interestingly, in the alternative picture ($\Psi_-=\left\langle {{{\cal O}_ -}} \right\rangle $ and $\Psi_+=0$), we again observe the same qualitative behaviors as the ones for the standard picture (but, of course, a cleaner behavior without those deviations near the maximum value of $\alpha$, like the same scalar operator  ${\cal O}_ -$ in the previous mass fixing where $m^2 L^2 = \textit{fixed}$; see Fig. \ref{fig:O-_scalar}). This can be understood from the variations of ${\left\langle {{{\cal O}_ - }} \right\rangle ^{1/{\Delta _ - }}}/{T_c}$ with the dimensionless quantity $T/T_c$ for different values of $\alpha$ and $K$, as plotted in Fig. \ref{fig:O-_scalar_Leff}. As increasing $\alpha$, the formation of the scalar hair is getting harder while increasing $K$ (with fixing $\alpha$) makes its formation easier. Note that, compared to the case of $m^2 L^2 = \textit{fixed}$, the opposite behavior is observed for variations of $\left\langle {{{\cal O}_ - }} \right\rangle $ and $T_c$ with respect to $\alpha$ and $K$. In addition, in this alternative picture, those abnormal behaviors associated with the values of $\alpha$ approaching the CS limit does not exist for the condensate, like the same scalar operator in the case with $m^2 L^2 = \textit{fixed}$. \vspace{1mm}

\begin{figure}[!htbp]
	\begin{center}
		\epsfxsize = 4 cm
	\includegraphics[keepaspectratio=true,scale=0.85]{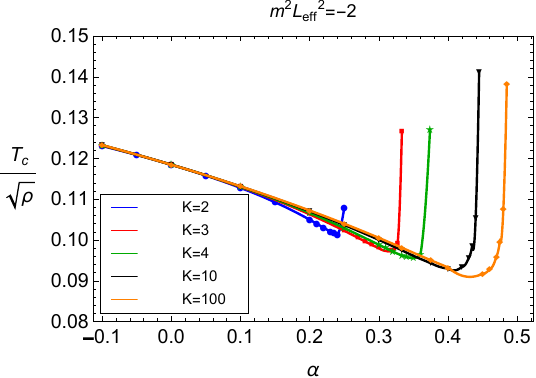}
	\hskip .1 cm
	\epsfxsize = 4 cm
	\includegraphics[keepaspectratio=true,scale=0.85]{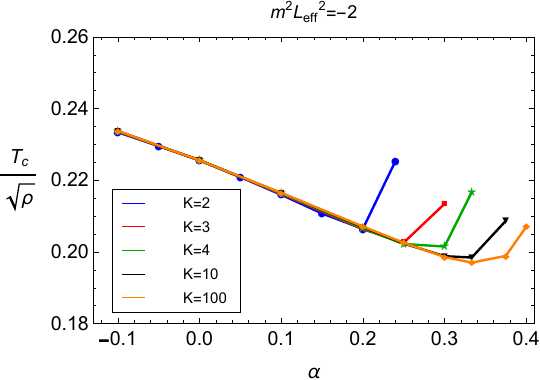}
	\end{center}
	\caption{\textit{Critical temperature of the superconducting transition, associated with the scalar operators ${\cal O}_ +$ (left panel) and ${\cal O}_ -$ (right panel), as a function of $\alpha$ for $4D$ Einstein-Lovelock gravity theories at different orders in curvature, labeled by $K=2,3,4,10$ and $100$. The mass of the scalar field has been fixed relative to the effective asymptotic AdS scale $L_{\rm{eff}}$ as  ${m^2}L_{{\rm{eff}}}^2 =  - 2$. The points with $\alpha =0$ correspond to the results of the Einstein gravity background where $\mathop {\lim }\limits_{\alpha  \to 0} {L_{{\text{eff}}}} = L$. }} \label{fig:Tc_sWave_Leff}
\end{figure}

To complete our discussion regarding the case of $m^2 {L_{\text{eff}}^2} = \textit{fixed}$, we further investigate both the standard and the alternative pictures for this choice of fixing numerically by plotting the corresponding critical temperatures ($T_c$) as a function of fine-tuned Lovelock coupling $\alpha$ in Fig \ref{fig:Tc_sWave_Leff}. The left panel of Fig. \ref{fig:Tc_sWave_Leff}, that is associated with the scalar operator ${\cal O}_ +$, indicates that the critical temperature $T_c$ decreases with the increase of $\alpha$ until it reaches a point after which further increasing of $\alpha$ (until reaching the CS limit) causes its increase. We expect this behavior to still exist even after accounting for the fully backreacting case, as such behavior was previously reported for backreacting superconductors in both the higher and the lower-dimensional Einstein-Gauss-Bonnet gravity theories \cite{HS-GB-2010Gregory,HS-4GB-2021PLBa,HS-4GB-2021PLBb}. For a fixed value of $\alpha$, increasing $K$ also leads to increasing $T_c$. However, for the same scalar operator  ${\cal O}_ +$ in the previous fixing ($m^2 L^2 = \textit{fixed}$) in Fig. \ref{fig:Tc_sWave_O+}, $T_c$ was a decreasing function of $\alpha$ everywhere and an increasing function of $K$ always. Considering the alternative picture ($\Psi_-=\left\langle {{{\cal O}_ -}} \right\rangle $ and $\Psi_+=0$), the critical temperature is a decreasing function of $\alpha$ and, for a fixed $\alpha$, an increasing function of the order of higher curvature corrections, $K$. However, when $\alpha$ approaches the CS limit it reaches a point that after which we observe the opposite behavior: the critical temperature increases with increasing $\alpha$ but decreases with increasing $K$ (as seen, this also happens for the Einstein-Gauss-Bonnet class which is not reported before.). We should emphasize that these behaviors are observed for all $K$th order theories of $4D$ Einstein-Lovelock gravity. \vspace{1mm}


\subsection{Electric conductivity}

\begin{figure}[!htbp]
	\begin{center}
		\epsfxsize = 4 cm
		\includegraphics[keepaspectratio=true,scale=0.8]{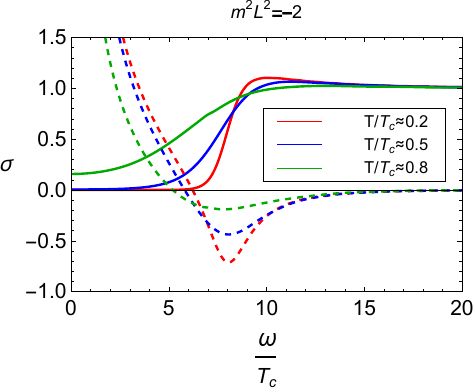}
	\end{center}
	\caption{\textit{Conductivity of the s-wave holographic superconductor versus $\frac{\omega }{{{T_c}}}$  for the Einstein limit (where $\alpha=0$ and $L_{{\rm{eff}}} \to L$) at three different values of $\frac{T}{{{T_c}}} \approx 0.2,0.5$ and $0.8$. The solid and dashed curves show the real and imaginary parts of the conductivity, respectively.}}
	\label{fig:conductivity:Einstein}
\end{figure}

\begin{figure}[!htbp]
	\begin{center}
		\epsfxsize = 4 cm
		\includegraphics[keepaspectratio=true,scale=0.72]{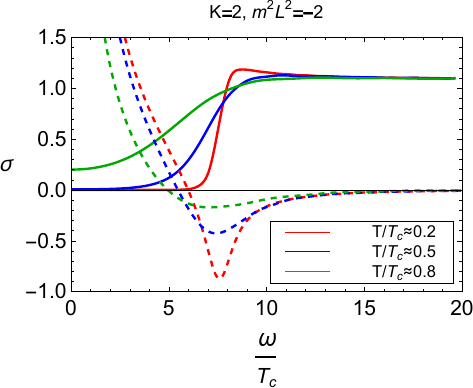}
		\hskip .1 cm
		\epsfxsize = 4 cm
		\includegraphics[keepaspectratio=true,scale=0.72]{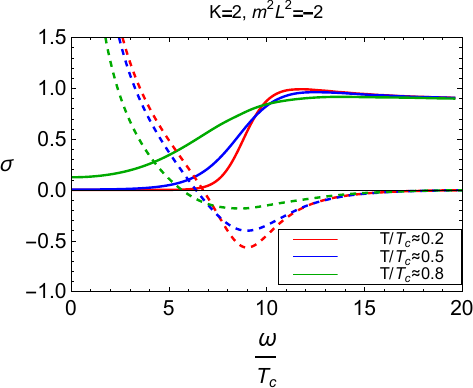}
		\hskip .1 cm
		\epsfxsize = 4 cm
		\includegraphics[keepaspectratio=true,scale=0.72]{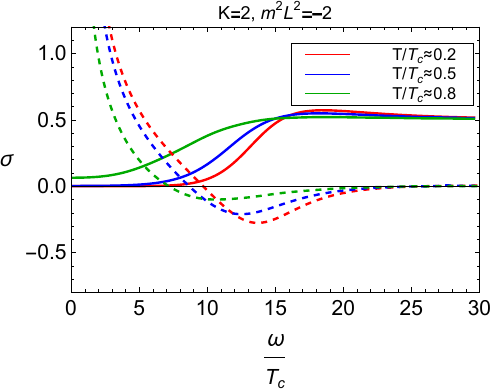}
	\end{center}
	\caption{
		\textit{Conductivity of the s-wave holographic superconductor versus $\frac{\omega }{{{T_c}}}$  for the $4D$ Einstein-Gauss-Bonnet gravity ($K=2$) at three different values of $\frac{T}{{{T_c}}} \approx 0.2,0.5$ and $0.8$. The solid and dashed curves display the real and imaginary parts of the conductivity, respectively. The left, middle and right panels are plotted for $\alpha=-0.1,0.1$ and $0.25$ (the upper bound), respectively and the mass of the scalar field is fixed by $m^2 L^2=-2$. }}
	\label{fig:conductivity:Swave_K2}
\end{figure}

The most striking property of superconductors is the ability to conduct DC electricity without losing energy when they are cooled below a certain critical temperature ($T<T_c$), meaning that the resistance drops abruptly to zero (that is equivalent to an infinite electrical conductivity). The diverging DC conductivity that implies a delta function at zero frequency in the AC conductivity $\sigma(\omega)$, i.e. Re[$\sigma(\omega \to 0)]$ $\propto \delta(\omega)$, is manifested from the $1/\omega$ pole in the imaginary part of the frequency dependent conductivity, Im [$\sigma(\omega)$], with the help of the Kramers-Kronig relation \cite{Hartnoll2009Lectures,Herzog2009Lectures}. To compute the conductivity in the dual gauge field theory as a function of frequency ($\omega$), we should consider electromagnetic responses. Considering the black brane spacetime (\ref{metric}) and \ref{metric function} as the background, we turn on the electromagnetic perturbation of the background $U(1)$ gauge field as $A_\mu = \left( {\Phi (r),{A_x}(r){e^{ - i\omega t}},0,0} \right)$ which gives rise the conductivity in the dual boundary gauge theory \cite{Gubser2008,HHH2008PRL}. Note that $x$ denotes any
direction in the dual CFT because of the isotropic property. The variation of Lagrangian with respect to the perturbative potential field, $\delta {A_x} = {A_x}(r){e^{ - i\omega t}}$, leads to 
\begin{equation} \label{purturbed_EM_Swave}
{A''_x}(r) + \frac{{f'(r)}}{{f(r)}}{A'_x}(r) + \left( {\frac{{{\omega ^2}}}{{{f^2}(r)}} - 2q^2\frac{{{\Psi ^2}(r)}}{{f(r)}}} \right){A_x}(r) = 0.
\end{equation}
 The above differential equation at the AdS boundary ($r \to \infty$) reduces to
\begin{equation}
{A''_x}(r) + \frac{2}{r}{A'_x}(r) + \frac{{{\omega ^2}L_{{\text{eff}}}^4}}{{{r^4}}}{A_x}(r) = 0.
\end{equation}
It is found that the vector potential close to AdS boundary behaves as
\begin{equation}
{A_x}(r) = {A_x^{(0)}} + \frac{{{A_x^{(1)}}}}{r} + ... ,
\end{equation}
where ${A_x^{(0)}}$ and ${{A_x^{(1)}}}$ are arbitrary integration constants, acting as the dual source (${A_x} = A_x^{(0)}$) and the expectaion value of the dual current ($\left\langle {{J_x}} \right\rangle  = A_x^{(1)}$), respectively. The superconducting (low-temperature) phase implies $\Psi \ne 0$ when $T<T_c$, so we shall solve the differential equation \ref{purturbed_EM_Swave} numerically with ingoing wave
boundary condition near the horizon, ${A_x}(r) = S(r){f^{ - i\omega t}} = S(r){f^{ - i\omega /4\pi T}} = S(r){f^{ - i\omega /3{r_ + }}}$ \cite{HHH2008PRL,Son2002Starinets}.\footnote{$S(r) = 1 + a\left( {r - {r_ + }} \right) + b{\left( {r - {r_ + }} \right)^2} + ...$ .} The electrical conductivity is then read off through the Ohm’s law, $\sigma_x  = {\left\langle {{J_x}} \right\rangle }/{E_x}$, leading to the so-called Kubo formula in the linear response theory as
\begin{equation}
\sigma  = \frac{{\left\langle {{J_x}} \right\rangle }}{{{E_x}}} =  - \frac{{\left\langle {{J_x}} \right\rangle }}{{{{\partial_t A}_x}}} = \frac{{A_x^{(1)}}}{{i\omega A_x^{(0)}}}.
\end{equation}
For Einstein-Maxwell-scalar system, there is a universal relation with deviations of less than $8\%$, given by \cite{Roberts2008PRD}
\begin{equation}
\frac{\omega_g}{T_c} \approx 8,
\end{equation}
where the so-called frequency gap, $\omega_g$, is the frequency minimizing Im [$\sigma(\omega)$]. A numeric example in the context of Einstein gravity has been presented in Fig. \ref{fig:conductivity:Einstein}, confirming the above universality and also suitable as a base for comparing our next results. However, this universality no longer holds in the presence of higher curvature corrections \cite{HS-GB-2009Gregory,HS-GB-2010Gregory,HS-4GB-2020JHEP,Edelstein2022}, including $4D$ Einstein-Gauss-Bonnet gravity as shown in Fig. \ref{fig:conductivity:Swave_K2}. We reaffirm this latter result for our model, but further seek to understand the effect of adding higher curvature corrections beyond Gauss-Bonnet and also pay special attention to both options of fixing the mass of the scalar field. \vspace{1mm}


\begin{figure}[!htbp]
	\begin{center}
		\epsfxsize = 4 cm
		\includegraphics[keepaspectratio=true,scale=0.72]{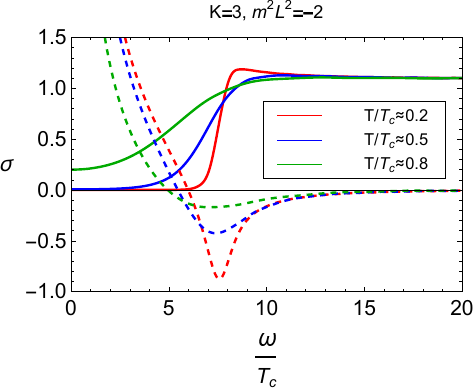}
		\hskip .1 cm
		\epsfxsize = 4 cm
		\includegraphics[keepaspectratio=true,scale=0.72]{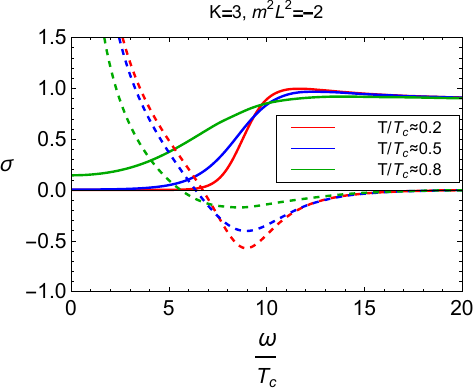}
		\hskip .1 cm
		\epsfxsize = 4 cm
		\includegraphics[keepaspectratio=true,scale=0.71]{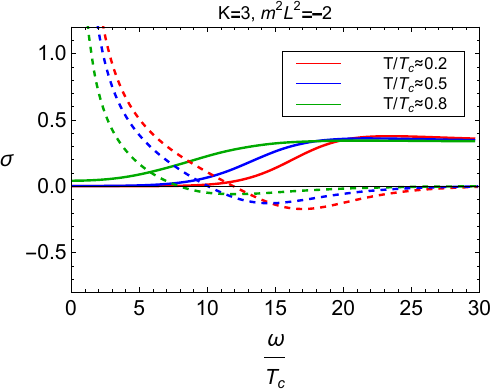}
		\vskip .1 cm
		\epsfxsize = 4 cm
		\includegraphics[keepaspectratio=true,scale=0.72]{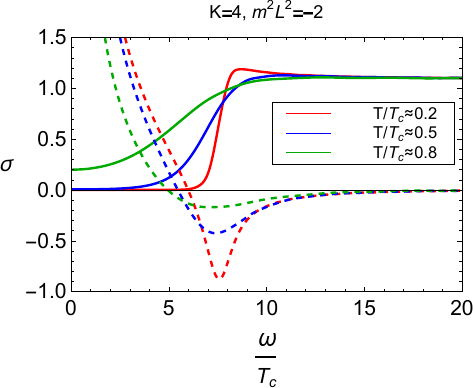}
		\hskip .1 cm
		\epsfxsize = 4 cm
		\includegraphics[keepaspectratio=true,scale=0.72]{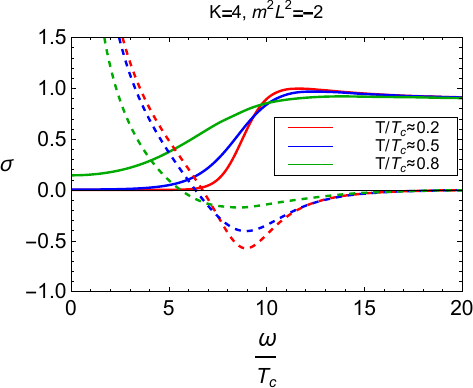}
		\hskip .1 cm
		\epsfxsize = 4 cm
		\includegraphics[keepaspectratio=true,scale=0.72]{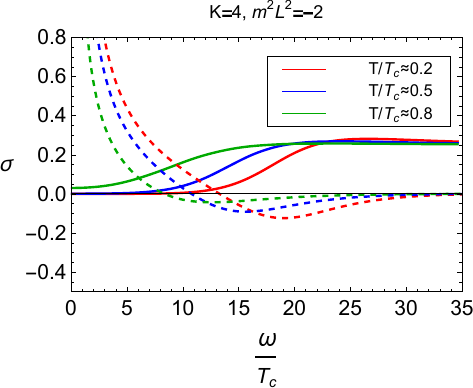}
		\vskip .1 cm
		\epsfxsize = 4 cm
		\includegraphics[keepaspectratio=true,scale=0.72]{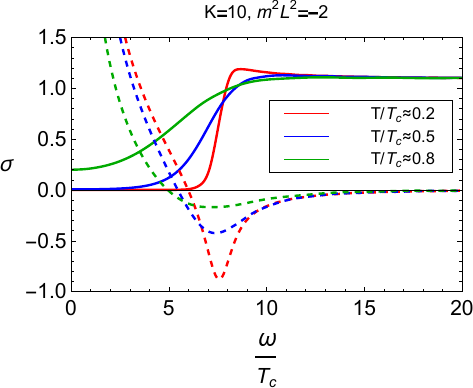}
		\hskip .1 cm
		\epsfxsize = 4 cm
		\includegraphics[keepaspectratio=true,scale=0.72]{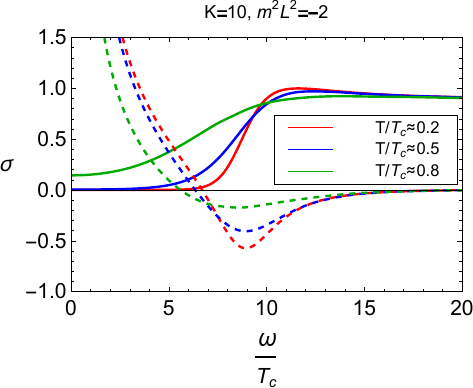}
		\hskip .1 cm
		\epsfxsize = 4 cm
		\includegraphics[keepaspectratio=true,scale=0.72]{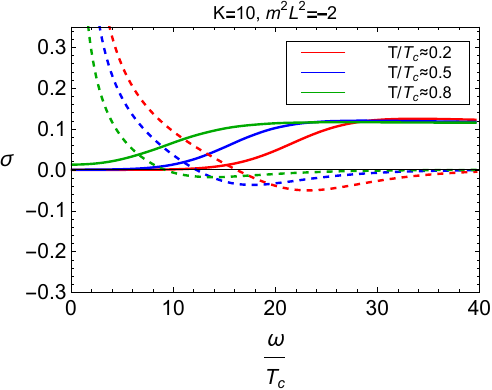}
		\vskip .1 cm
		\epsfxsize = 4 cm
		\includegraphics[keepaspectratio=true,scale=0.72]{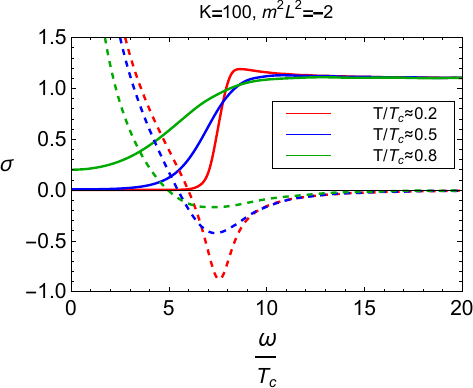}
		\hskip .1 cm
		\epsfxsize = 4 cm
		\includegraphics[keepaspectratio=true,scale=0.72]{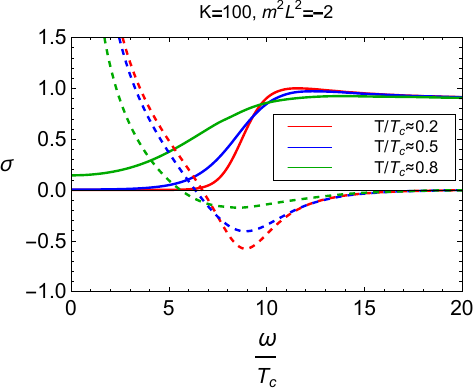}
		\hskip .1 cm
		\epsfxsize = 4 cm
		\includegraphics[keepaspectratio=true,scale=0.7]{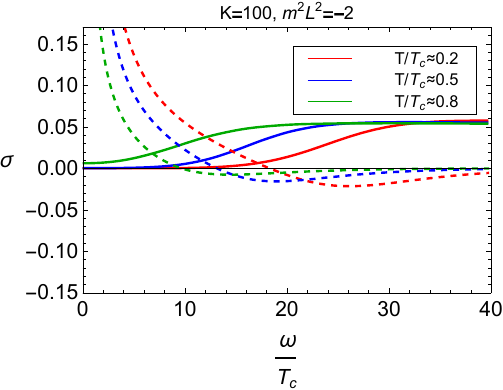}
	\end{center}
	\caption{
		\textit{Conductivity of the $ s $-wave holographic superconductors versus $\frac{\omega }{{{T_c}}}$ in $4D$ Einstein-Lovelock gravity theories (labeled by $K=3, 4, 10$ and $100$), at three different values of $\frac{T}{{{T_c}}} \approx 0.2,0.5$ and $0.8$. The solid and dashed curves display the real and imaginary parts of the conductivity, respectively. The left and middle panels are plotted for $\alpha=-0.1$ and $\alpha=0.1$, respectively. The right panels are plotted for the upper bound of $\alpha$ (the Chern-Simons limits), i.e. $\alpha = \frac{{(K - 1){L^2}}}{{2K}}$, resulting in $\alpha=1/3, 3/8, 9/20$ and $99/200$ for $K=3,4,10$ and $100$, respectively. The mass of the scalar field is also fixed by $m^2 L^2=-2$.}}
	   \label{fig:conductivity:Swave_K}
\end{figure}

\begin{figure}[!htbp]
	\begin{center}
		\epsfxsize = 4 cm
		\includegraphics[keepaspectratio=true,scale=0.72]{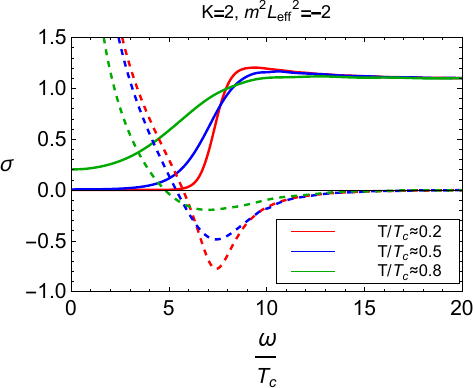}
		\hskip .1 cm
		\epsfxsize = 4 cm
		\includegraphics[keepaspectratio=true,scale=0.72]{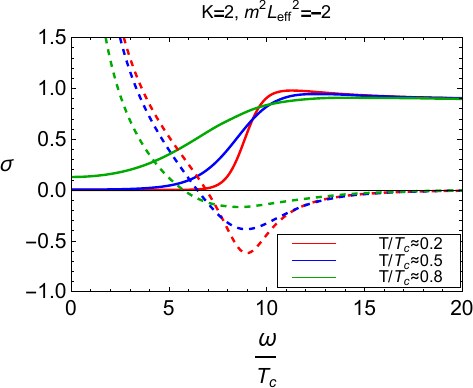}
		\hskip .1 cm
		\epsfxsize = 4 cm
		\includegraphics[keepaspectratio=true,scale=0.72]{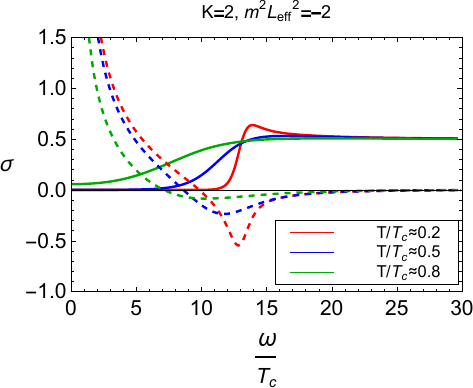}
		\vskip .1 cm
		\epsfxsize = 4 cm
		\includegraphics[keepaspectratio=true,scale=0.72]{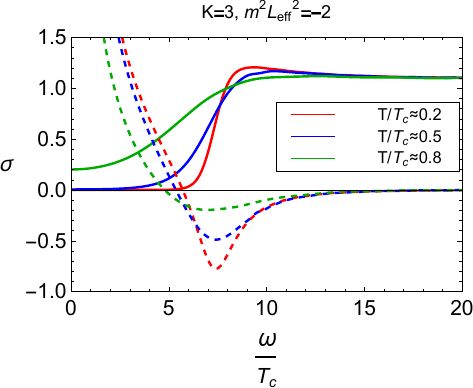}
		\hskip .1 cm
		\epsfxsize = 4 cm
		\includegraphics[keepaspectratio=true,scale=0.72]{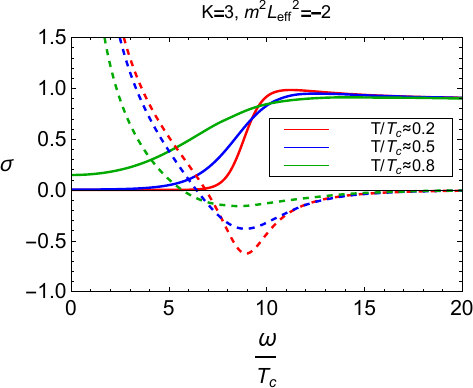}
		\hskip .1 cm
		\epsfxsize = 4 cm
		\includegraphics[keepaspectratio=true,scale=0.72]{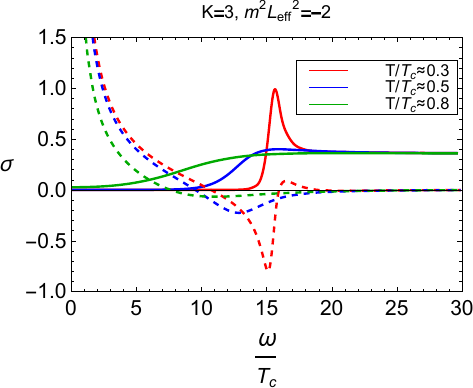}
		\vskip .1 cm
		\epsfxsize = 4 cm
		\includegraphics[keepaspectratio=true,scale=0.72]{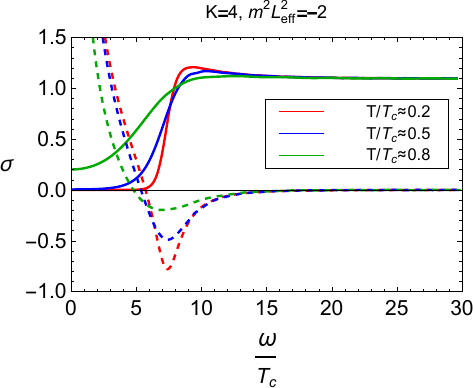}
		\hskip .1 cm
		\epsfxsize = 4 cm
		\includegraphics[keepaspectratio=true,scale=0.72]{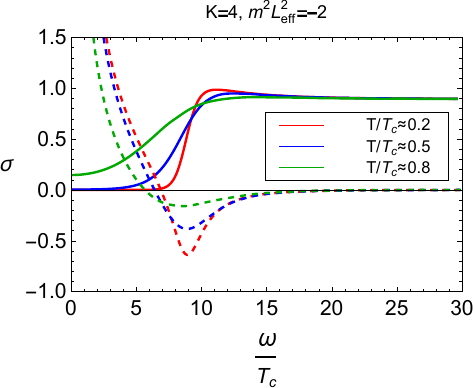}
		\hskip .1 cm
		\epsfxsize = 4 cm
		\includegraphics[keepaspectratio=true,scale=0.72]{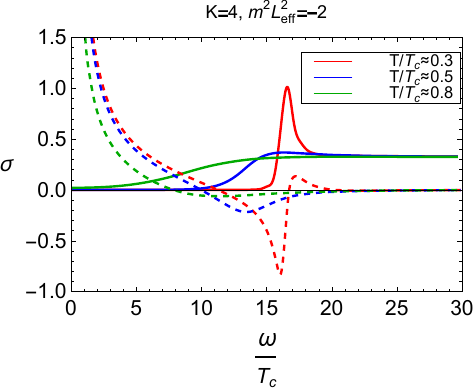}
	\end{center}
	\caption{
		\textit{Conductivity of the $ s $-wave holographic superconductors versus $\frac{\omega }{{{T_c}}}$ in $4D$ Einstein-Lovelock gravity theories (labeled by $K=2, 3$, and $4$), at three different values of $\frac{T}{{{T_c}}} \approx 0.2,0.5$ and $0.8$. The solid and dashed curves display the real and imaginary parts of the conductivity, respectively. The left and middle panels are plotted for $\alpha=-0.1$ and $\alpha=0.1$, respectively. The right panels are plotted for $\alpha=0.25, 1/3$ and $3/8$ which are the upper limits of $\alpha$ (the Chern-Simons limits), i.e. $\alpha = \frac{{(K - 1){L^2}}}{{2K}}$, resulting in $\alpha=0.25, 1/3$, and $3/8$ for $K=2,3$, and $4$, respectively. The mass of the scalar field is also fixed by ${m^2}L_{{\rm{eff}}}^2 =  - 2$.}}
	    \label{fig:conductivity:Swave_K_Leff}
\end{figure}

Regarding the black branes from the regularized $4D$ Einstein-Lovelock gravity as the gravitational background for building $s$-wave holographic superconductors, we plot the AC conductivity, $\sigma(\omega)$, versus frequency (normalized by $T_c$) obtained by solving Eq. (\ref{purturbed_EM_Swave}) numerically with
the boundary conditions explained above. In our considerations, we focus on the conductivity associated with the $\left\langle {{{\cal O}_ +}} \right\rangle$ condensate ($\propto$ the fast falloff), not only for $m^2 L_\text{eff}^2 =\textit{fixed}$ that is common in the literature but also for the alternative mass fixing, $m^2 L^2 =\textit{fixed}$.\footnote{As far as we know, all the studies on the conductivity of holographic superconductors within the framework of $4D$ Einstein-Gauss-Bonnet gravity have been performed by fixing the mass of the scalar field relative to the effective AdS scale, $L_{\rm{eff}}$; see refs. \cite{HS-4GB-2020JHEP,HS-4GB-2021PLBb}. Calculations for the cases with $m^2 L_\text{eff}^2 =\textit{fixed}$ are relatively more straightforward than the ones with $m^2 L^2 =\textit{fixed}$. However, it seems fixing the scalar mass with respect to the black hole/brane, i.e. $m^2 L^2 =\textit{fixed}$, is the correct physical choice, as first indicated in ref. \cite{HS-GB-2009Gregory}. The reason is that the physical measurables of black holes including the ADM mass ($M$), the temperature ($T$), and (if any exists) the angular momentum ($J$) and the $U(1)$ charge ($Q$) vary as we vary $\alpha$ and $K$. Subsequently, if we assume $m^2 L_\text{eff}^2 = \textit{fixed}$, the scalar mass varies with respect the conserved charges of the black hole (like $M$ and $T$) whenever we vary $\alpha$ or $K$. Our findings may provide new insights into the $4D$ Einstein-Lovelock gravity theories including the Einstein-Gauss-Bonnet class.} Assuming $m^2 L^2 =-2$, the plots in Figs. \ref{fig:conductivity:Swave_K2} and \ref{fig:conductivity:Swave_K} show the results for various $K$th orders of the theory as well as different variations of the fine-tuned coupling constant $\alpha$. The solid lines represent Re[$\sigma(\omega)$] while the dashed lines represent Im[$\sigma(\omega)$]. By paying attention to the plots, we see that the real parts of the optical conductivities exhibit the existence of a frequency gap, $\omega_g$. The frequency gap $\omega_g$ ensures the existence of the energy gap and the Re[$\sigma(\omega)$] almost vanishes for $\omega < \omega_g$. Both the real and imaginary parts of the AC conductivity approach constant values for large $\omega$ that is typical of holographic superconductors in $(2+1)$-dimensions, in agreement with Re[$\sigma(\omega \to \infty)$] $\propto \omega^{D-4}$ where $D$ is the spacetime dimensions of the bulk theory. Interestingly, the asymptotic value of Re[$\sigma(\omega)$] for large frequencies depends on both $\alpha$ and $K$,\footnote{This is not a trivial result; for example, considering 2+1 holographic superconductors in Einsteinian cubic gravity, the asymptotic value of Re[$\sigma(\omega)$] for large-$\omega$ is independent of the cubic coupling constant \cite{Edelstein2022}.} while Im[$\sigma(\omega)$] is independent of them (approaches to zero, as expected). As $\alpha$ increases, the asymptotic value of Re[$\sigma(\omega \to \infty)$] decreases. Its value is always smaller than that of Einstein-Maxwell-scalar system in the positive-$\alpha$ regime but larger for the regime with $\alpha <0$. However, the effect of adding higher curvature corrections is indirect: assuming a fixed value for $\alpha$, one finds that adding higher-order curvature corrections (via increasing $K$) does not alter the asymptotic value of Re[$\sigma(\omega \to \infty)$], compared to the Gauss-Bonnet ($K = 2$) case. But, the upper bound of $\alpha$ increases as the maximum order of curvature corrections increases; see Eq. (\ref{restriction}). So, for the holographic system with more higher-order curvature corrections, $\alpha$ can take larger values which consequently leads to smaller asymptotic values of Re[$\sigma(\omega \to \infty)$], as is obvious from the right panels of Figs. \ref{fig:conductivity:Swave_K2} and \ref{fig:conductivity:Swave_K}. This latter result may change for the $4D$ Einstein-Lovelock theory in the general case where the Lovelock coupling constants vary independently. In this case, the maximum order of the model might have a direct effect on the asymptotic value of Re[$\sigma(\omega)$]. We have also repeated the above calculations for the alternative mass fixing. Considering $m^2 L_\text{eff}^2 =-2$, i.e. fixing the mass of the scalar field with respect to the effective AdS scale (\ref{Leff}), the plots in Fig. \ref{fig:conductivity:Swave_K_Leff} show the detailed information for the theory up to quadratic ($K=2$), cubic ($K=3$), and quartic ($K=4$) curvature corrections. As seen, the results are qualitatively the same as the previous case where $m^2 L^2 =-2$. Quantitatively, we observe that the results for $\omega_g/T_c$ and Re[$\sigma(\omega \to \infty)$] almost coincide with those from the fixing the scalar mass with respect to the AdS scale $L$, that is an interesting feature of the holographic model (not reported before for Gauss-Bonnet and Lovelock gravity theories, to our knowledge). \vspace{1mm}


Considering both options for fixing the scalar mass (relative to $L$ and $L_{\rm{eff}}$), from Figs. \ref{fig:conductivity:Swave_K2}, \ref{fig:conductivity:Swave_K}, and \ref{fig:conductivity:Swave_K_Leff}, it is seen that the ratio of the gap frequency over the critical temperature, $\omega_g/T_c$, deviates from the Einsteinian universal value ($\approx 8$) and depends on both the fine-tuned coupling constant $\alpha$ and the maximum order of higher curvature corrections, $K$. The effect of the Gauss-Bonnet term for the case of $m^2 L^2_{\rm{eff}}=\textit{fixed}$ has already been studied in ref. \cite{HS-4GB-2020JHEP} that is a suitable case to compare the effects of higher-order curvature corrections with respect to it. For the theory up to the Gauss-Bonnet curvature corrections for $m^2 L^2_{\rm{eff}}=\textit{fixed}$, it was found that the increase of the Gauss-Bonnet parameter ($\alpha_2=\alpha$, in our notation) results in larger gap frequencies ($\omega_g$) \cite{HS-4GB-2020JHEP}; we also confirm this result for the case with $m^2 L^2=\textit{fixed}$, see Fig. \ref{fig:conductivity:Swave_K2} for more details. Returning to our model with higher curvature corrections beyond Gauss-Bonnet, from Figs. \ref{fig:conductivity:Swave_K}, and \ref{fig:conductivity:Swave_K_Leff}, we observe that the gap frequency (and therefore $\omega_g/T_c$) becomes larger as $\alpha$ increases.\footnote{This result has also been confirmed for various modified gravity theories with higher (quadratic and cubic) curvature corrections \cite{HS-GB-2009Gregory,HS-GB-2010Gregory,HS-HCC-2010Siani,HS-4GB-2020JHEP,Edelstein2022}.
} Compared to the Gauss-Bonnet case, adding higher-order curvature corrections has also a dramatic effect on the ratio of $\omega_g/T_c$: it pushes the frequency gap to higher values (it is assumed $\alpha$ is kept constant while adding higher-order curvature corrections.). Of course, the impact of increasing $\alpha$ on $\omega_g/T_c$ is much stronger than adding $K$. In all examples, we also see that lowering the temperature pushes the frequency gap to larger values. The values of both the real and the imaginary parts of the conductivity decrease as either $\alpha$ or $K$ increases. Furthermore, for our holographic system, we generally observe that the ratio of $\omega_g/T_c$ is greater than the universal value of Einstein gravity ($\approx 8$) for positive fine-tuned coupling constants. However, in the negative fine-tuned coupling regime, the ratio becomes smaller than its Einstein-gravity counterpart. It is also worth emphasizing that the results are qualitatively the same when the scalar mass is fixed relative to either the usual AdS or the effective AdS scales ($L$ or $L_{\rm{eff}}$), as explicitly shown through the plots in Figs. \ref{fig:conductivity:Swave_K2}, \ref{fig:conductivity:Swave_K}, and \ref{fig:conductivity:Swave_K_Leff} for the theory up to quadratic ($K=2$), cubic ($K=3$), and quartic ($K=4$) curvature corrections. \vspace{1mm}

Finally, regardless of whether we choose $m^2 L^2 =\textit{fixed}$ or $m^2 L_\text{eff}^2 =\textit{fixed}$ for fixing the scalar mass, taking the $\omega \to 0$ limit of Eq. (\ref{purturbed_EM_Swave}) also reveals the $1/\omega$ pole in Im[$\sigma(\omega)$] which subsequently implies the real part of the conductivity contains a delta function, Re[$\sigma(\omega \to 0)]$ $\propto \delta(\omega)$, confirming a diverging DC conductivity. The $1/\omega$ behavior can also be understood from the imaginary parts of the conductivities in the plots of Figs. \ref{fig:conductivity:Swave_K2}, \ref{fig:conductivity:Swave_K}, and \ref{fig:conductivity:Swave_K_Leff} which exhibit a pole close to $\omega = 0$. All these capture the essential physics of conductivity in the superconducting state.

\section{Holographic $p$-wave superconductors} \label{sect4:pWave}
 In the previous section, using AdS/CMT duality, we discussed the phenomenon of $s$-wave superconductivity that is more common and well-understood in the literature. However, the phenomenon of $p$-wave superconductivity, for which the conductivities are strongly anisotropic \cite{Pwave1963} and BCS theory fails to explains them \cite{PwaveF1981,PwaveF1984,PwaveF2000}, is relatively rare in nature\footnote{For example, $\text{Sr}_2\text{Ru}\text{O}_4$ is a possible $p$-wave superconductor below the critical temperature $T_c=1.5 \text{K}$ \cite{Annett2004Book}. $\text{UBe}_{13}$ is another candidate for $p$-wave Superconductivity \cite{PwaveF1984}.} and still is an active area of research using the tools of quantum field theory \cite{PwaveNew2001,PwaveNew2007,PwaveNew2013,PwaveNew2018,PwaveNew2019,PwaveNew2021,PwaveNew2022,PwaveNew2024} as well as gauge/gravity duality \cite{Gubser2008pWave,Aprile2011pWave,Cai2013pWave,Cai2014pWave,HS2020Nam,HS-GB-2010Cai,HS-HCC-2021,PwaveH2013}. In order for mimicking the physics of a $p$-wave superconductor with an anisotropic order parameter, it was shown that a complex vector field $\rho_\mu$, with mass $m$ and charge $q$, should be introduced into the Einstein-Maxwell-AdS gravity \cite{Cai2014pWave}. Following this proposal, the appropriate matter action reads \cite{Cai2014pWave}
\begin{equation}
{S_{matter}} =  - \frac{1}{{16\pi {G_N}}}\int {{d^4}x\sqrt { - g} \left[ {\frac{1}{4}{F_{\mu \nu }}{F^{\mu \nu }} + \frac{1}{2}\rho _{\mu \nu }^\dag {\rho ^{\mu \nu }} + {m^2}\rho _\mu ^\dag {\rho ^\mu } - iq\gamma {\rho _\mu }\rho _\nu ^\dag {F^{\mu \nu }}} \right]},
\end{equation}
where ${\rho _{\mu \nu }}$ is defined by ${\rho _{\mu \nu }} = {D _\mu }{\rho _\nu } - {D _\nu }{\rho _\mu }$ with the covariant derivative ${D_\mu } \equiv ({\nabla _\mu } - iq{A_\mu })$ and ${A_\mu }$ is the $U(1)$ gauge field again. In the absence of external magnetic field, the last term do not contribute to the field equations \cite{Cai2013pWave}. Adopting the following ansatz \cite{Cai2014pWave}
\begin{equation} \label{ansatz_pWave}
{A_\mu } = \Phi (r)\delta _\mu ^t\,, \quad {\rho _\mu } = {\rho _x}(r)\delta _\mu ^x,
\end{equation}
in the black brane background of the regularized $4D$ model under consideration, the field equations of motion are obtained in the probe limit as
\begin{equation}\label{scalarEq_p-wave}
\Phi '' + \frac{2}{r}\Phi ' - \frac{{2{q^2}\rho _x^2}}{{{r^2}f}}\Phi  = 0,
\end{equation}
\begin{equation}\label{vectorEq_p-wave}
{\rho ''_x} + \frac{{f'}}{f}{\rho '_x} + \left( {\frac{{{q^2}{\Phi ^2}}}{{{f^2}}} - \frac{{{m^2}}}{f}} \right){\rho _x}=0,
\end{equation}
where, again, the prime denotes the derivative with respect to $r$ and $f \equiv f(r)$. The ansatz for the complex vector field $\rho_\mu$ in Eq. (\ref{ansatz_pWave}) yields the vacuum expectation value of the $x$-component of the vector operator in the dual gauge field theory. So, one direction is special and the conductivities will be anisotropic, as expected for $p$-wave superconductors. \vspace{1mm}

Parallel to the steps we did in the previous section to study $s$-wave superconductivity using the tools of AdS/CMT duality, in the continuation of this section, we first examine the regularity conditions of the fields, then obtain solutions to the matter field equations and, consequently, present a holographic description of the $p$-wave superconductivity phenomenon (with an anisotropic order parameter) through investigating the vector condensate, the critical temperature, and the (super)conductivity of dual boundary gauge theory. We again consider matter fields as probes. We find that holographic $p$-wave superconductors can be built in the presence of any number of higher curvature corrections and both $\alpha$ and $K$ have dramatic effects on them compared to the ones in Einstein gravity and Einstein-Gauss-Bonnet gravity theories.

\subsection{Regularity at the horizon and the AdS boundary}

In order to solve the matter field equations (\ref{scalarEq_p-wave}) and (\ref{vectorEq_p-wave}), we impose the regularity conditions at both the horizon and the AdS boundary. At the event horizon, since the norm of the gauge field ($A_\mu A^\mu$) must not diverge, one should assume $A_t(r=r_+)=0$, implying
\begin{equation}
\Phi ({r_ + }) = 0 \, , \quad {\rho _x}({r_ + }) = \frac{{f'({r_ + }){\rho '_x}({r_ + })}}{{{m^2}}}.
\end{equation}
Regularity at the AdS boundary ($r \to \infty$) also determines the asymptotic forms of the solutions to (\ref{scalarEq_p-wave}) and (\ref{vectorEq_p-wave}), yielding
\begin{equation} \label{fields behavior at AdS_pWave}
{\left. {{A_t}} \right|_{r \to \infty }} = \mu  - \frac{\rho }{r} \,, \quad {\left. {{\rho _x}} \right|_{r \to \infty }} = \frac{{{\rho^-_{x }}}}{{{r^{{\Delta _ - }}}}} + \frac{{{\rho^+_{x }}}}{{{r^{{\Delta _ + }}}}},
\end{equation}
where the conformal dimension $\Delta_\pm$ at strong coupling in the boundary is given by
\begin{eqnarray} \label{lamdapm_Pwave}
{\Delta _ \pm } &=& \frac{1}{2} \pm \frac{1}{2}\sqrt {1 + 4{m^2}L_{{\text{eff}}}^2}  \nonumber \\
&=& \frac{1}{2} \pm \frac{1}{2}\sqrt {1 + \frac{{8{m^2}\alpha }}{{K - 1}}{{\left( {1 - {{\left[ {1 - \frac{{2\alpha K}}{{(K - 1){L^2}}}} \right]}^{1/K}}} \right)}^{ - 1}}} ,
\end{eqnarray}
in terms of $L_\text{eff}$ or explicitly in terms of $L$, $\alpha$, and $K$. Again, $\mu $ and $\rho$ are interpreted as the chemical potential and the charge density of the boundary gauge field theory, respectively. From Eq. (\ref{lamdapm_Pwave}), it is inferred that the squared mass of complex vector field $\rho_\mu$ must obey the following BF bound
\begin{equation}\label{BF bound_Pwave}
{m^2} \geqslant m_{{\text{BF}}}^2 =  - \frac{1}{{4L_{{\text{eff}}}^2}},
\end{equation}
where ${L_{{\text{eff}}}^2}$ is given by (\ref{Leff}) or, equivalently, it can be understood from (\ref{lamdapm_Pwave}). We will again examine both options for fixing the squared mass of the complex vector field as $\rho_x$, i.e., $m^2 {L^2}=\textit{fixed}$ and $m^2 {L_{{\rm{eff}}}^2} = \textit{fixed}$ so that the BF bound (\ref{BF bound_Pwave}) is satisfied. Again, either $\rho^+_{x}$ or $\rho^-_{x}$ in Eq. (\ref{fields behavior at AdS_pWave}) can be dual to the expectation value of the vector operator ${\cal O}_x$ with the conformal dimension $\Delta_\pm$ presented in Eq. (\ref{lamdapm_Pwave}). 

\subsection{Vector condensate}

In the normal phase where the system is not superconducting, the complex vector field $\rho_\mu$ vanishes identically, so the hairless black brane solution, specified by the line element (\ref{metric}) and the metric function (\ref{metric function}), corresponds to the normal high-temperature state ($T > T_c$). Below the critical temperature ($T<T_c$), the vector field $\rho_x$ must condense to spontaneously have the $U(1)$ symmetry breaking when the source is off (the spatial rotational symmetry is also broken since the condensate will pick out one direction as special) \cite{Cai2013pWave,Cai2014pWave,Cai-2015Review}. In this region, we have black brane solution with the vector hair. However, the matter fields including the non-vanishing vector field $\rho_x$ are considered as probes in the black brane background. Since the physics of holographic superconductivity can be described by taking either $\rho^+_{x}$ or $\rho^-_{x}$ to be dual to the expectation value of the vector (condensation) operator ${\cal O}_x$, we will mainly choose to work in the standard picture where the \textit{fast falloff} sector with the conformal dimension $\Delta_+$ corresponds to the \textit{response} of the system and the slow falloff play the role of the source in the spontaneous vector condensation by setting $\rho^-_{x}=0$. In addition, we will examine both options for fixing the squared mass of the vector field $\rho_x$, i.e., $m^2 {L^2}=\textit{fixed}$ and $m^2 {L_{{\rm{eff}}}^2} = \textit{fixed}$. \vspace{1mm}

In what follows, we first choose the mass of the vector field to obey once $m^2 {L^2}=3/4$ and another time $m^2 {L_{{\rm{eff}}}^2} = 3/4$ to make contact with the previous study of $4D$ Einstein-Gauss-Bonnet gravity in \cite{HS-4GB-2020JHEP}. Note that for both cases, we must only work in the standard quantization picture, because the conformal dimension $\Delta_-$ for the vector operator $\left\langle {{{\cal O}^-_ x}} \right\rangle$ is negative which makes this operator nonrenormalizable; so, one should set $\rho^-_{x}=0$ as the source to ensure the stability of AdS as well as a normalizable solution. For this reason, we will use the convention $\left\langle {{{\cal O}_ x }} \right\rangle \equiv \left\langle {{\cal O}^+_ x } \right\rangle$ for convenience that is common in the literature. Furthermore, for the sake of completeness, we also consider the mass of the scalar field to obey once $m^2 {L^2}=- 3/16$ and another time $m^2 {L_{{\rm{eff}}}^2} = -3/16$ in order for examining the vector condensates in each of the possible quantizations, i.e., $\left\langle {{\cal O}^-_ x } \right\rangle$ and $\left\langle {{\cal O}^+_x } \right\rangle$, especially since the condensate associated with $\left\langle {{{\cal O}^-_ x }} \right\rangle$ has not been explicitly studied even for the cases of lower- and higher-dimensional Gauss-Bonnet gravity theories.

\subsubsection{Vector condensate with $m^2 {L^2} = \textit{fixed}$}

\begin{figure}[!htbp]
	\begin{center}
		\epsfxsize = 4 cm
		\includegraphics[keepaspectratio=true,scale=0.78]{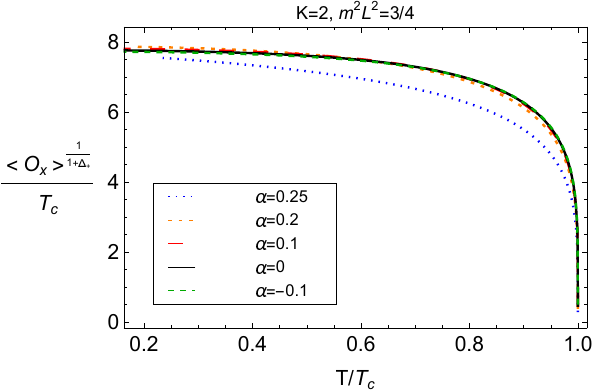}
		\hskip .1 cm
		\epsfxsize = 4 cm
		\includegraphics[keepaspectratio=true,scale=0.78]{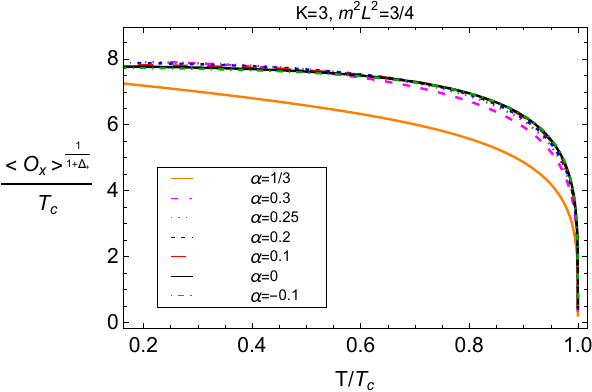}
		\vskip .1 cm
		\epsfxsize = 4 cm
		\includegraphics[keepaspectratio=true,scale=0.78]{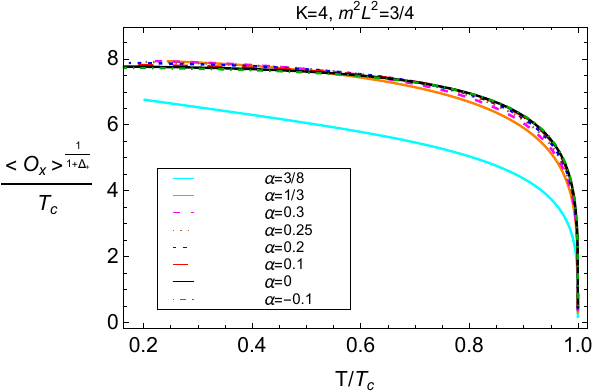}
		\hskip .1 cm
		\epsfxsize = 4 cm
		\includegraphics[keepaspectratio=true,scale=0.78]{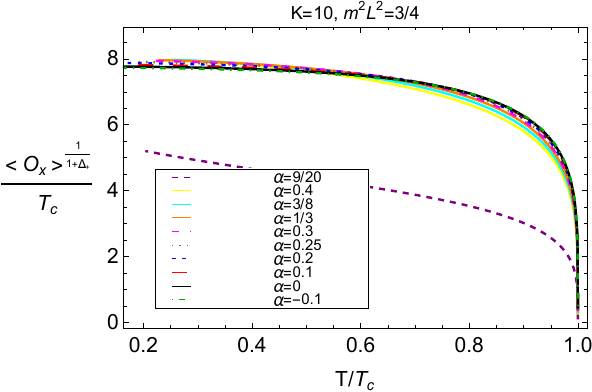}
		\vskip .1 cm
		\epsfxsize = 4 cm
		\includegraphics[keepaspectratio=true,scale=0.78]{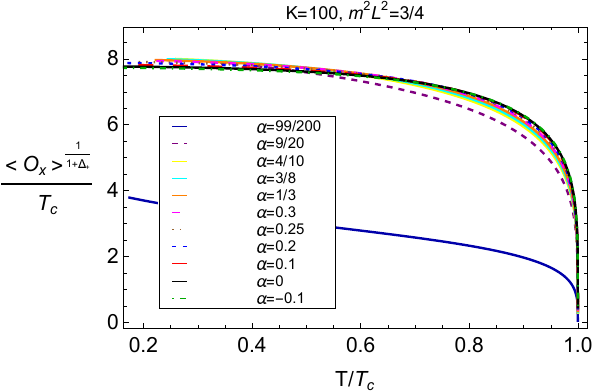}
	\end{center}
	\caption{\textit{Condensation of the vector operator ${\cal O}_ x $ as a function of temperature for five different orders of the theory, $K=2, 3,4,10$ and $100$, with the mass of the vector field fixed by $m^2 L^2 =  3/4$. For each $K$th order of the theory, different values of $\alpha$ in the allowed ranges are considered. The cases with $\alpha =0 $ correspond to Einstein gravity (they are actually the same in all panels).}}
	\label{fig:Ox_vector}
\end{figure}

Imposing $\rho^-_{x}=0$ as the dual source and $\rho^+_{x}$ as the spontaneous
condensation of the operator ${\cal O}_x$, we compute the condensation associated with this case, $\rho^+_{x}=\left\langle {{{\cal O}_ x }} \right\rangle$. Considering $m^2 {L^2} = 3/4$, in Fig. \ref{fig:Ox_vector}, condensation of the operator ${\cal O}_x$ as a function of temperature has been depicted for a variety of values of $K$ and $\alpha$. For all the cases, we observe that, the vector operator $\left\langle {{{\cal O}_ x }} \right\rangle $ has condensed for $T<T_c$ which goes to a constant value as $T \to 0$. Although it is difficult to be seen in this figure, we confirm that,
generally, increasing values of $\alpha$ leads to larger condensates at low temperatures , $T<T_c$. More specifically, all the condensation curves cross with the one belong to the Einstein gravity (the case with $\alpha = 0$). Assuming the curves with $\alpha > 0$, the condensate is larger than that of Einstein gravity before crossing, while its value becomes smaller after crossing. However, for the range $\alpha < 0$, the reverse behavior is observed. So, for both positive and negative fine-tuned Lovelock couplings, the condensate can be larger or smaller compared to the one in Einstein gravity. Increasing $K$ also leads to a larger condensate provided that $\alpha$ is kept fixed. We also observed an abnormal behavior of the condensate near the upper bound of $\alpha$ since they do not follow the pattern mentioned above and exhibit an unusual (in fact, stronger) variation of the condensate with $T$. But, as we already pointed out for the $s$-wave superconductors, the causality constraint from the dual gauge field theory will restrict the parameter space more than Eq. (\ref{restriction}) \cite{KSS2008violation,deBoer2010,causality4dGH2020}, meaning that such curves should be excluded. \vspace{1mm}


\begin{figure}[!htbp]
	\begin{center}
		\epsfxsize = 4 cm
		\includegraphics[keepaspectratio=true,scale=0.78]{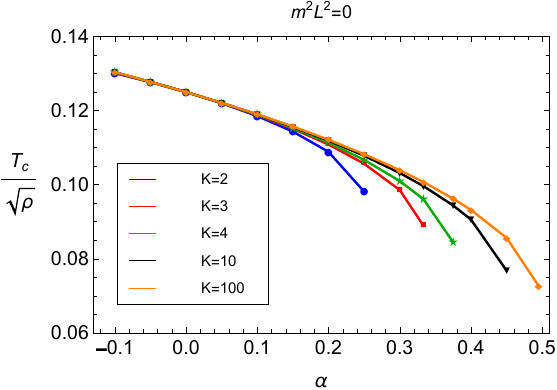}
		\hskip .1 cm
		\epsfxsize = 4 cm
		\includegraphics[keepaspectratio=true,scale=0.78]{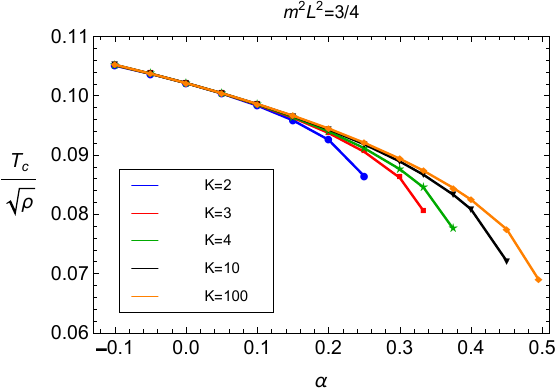}
		\vskip .1 cm
		\epsfxsize = 4 cm
		\includegraphics[keepaspectratio=true,scale=0.78]{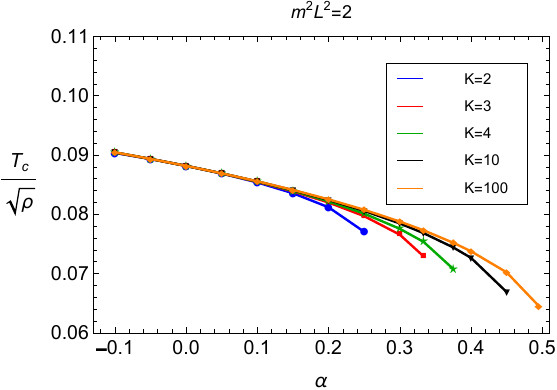}
	\end{center}
	\caption{\textit{Critical temperature of the superconducting transition, associated with the vector operator ${\cal O}_x$, as a function of $\alpha$ for five different orders of higher curvature corrections as $K=2,3,4,10$ and $100$. Different fixing of $m^2 L^2$ has been examined and the points with $\alpha =0$ correspond to the results of Einstein gravity where $L_{\rm{eff}} \to L$. Note that the panel with $m^2 L^2 =3/4$ corresponds to Fig. \ref{fig:Ox_vector}.}}
	\label{fig:Tc_pWave_forOx}
\end{figure}

We now analyze critical temperature of the superconducting transition. In Fig. \ref{fig:Tc_pWave_forOx}, the critical temperature with respect to $\alpha$ has been plotted for different $K$th orders of the theory and, in all panels of this figure, the point with $\alpha =0$ corresponds to the result of Einstein gravity; e.g., assuming $m^2 {L^2} = 3/4$, one obtains $T_c \approx 0.102 \rho^{1/2}$. We observe the same qualitative behavior for the three different fixing of the mass as $m^2 L^2 =0$, $3/4$ and $2$. It is inferred that the critical temperature $T_c$ decreases with increasing $\alpha$. On the other hand, the effect of of adding higher curvature corrections (via $K$) is to increase the critical temperature provided that $\alpha$ is held fixed. Generally, compared to the Einstein gravity, high-$T_c$ $p$-wave superconductors can always be found in the region with negative fine-tuned coupling constants, $\alpha<0$. Furthermore, in all cases, near the critical temperature, we observe that the order parameter behaves as $\left\langle {{{\cal O}_ x }} \right\rangle  \propto {\left( {{T_C} - T} \right)^{1/2}}$, reflecting a second-order phase transition exactly the same as the Ginzburg-Landau mean-field theory. \vspace{1mm}


\begin{figure}[!htbp]
	\begin{center}
		\epsfxsize = 4 cm
		\includegraphics[keepaspectratio=true,scale=0.78]{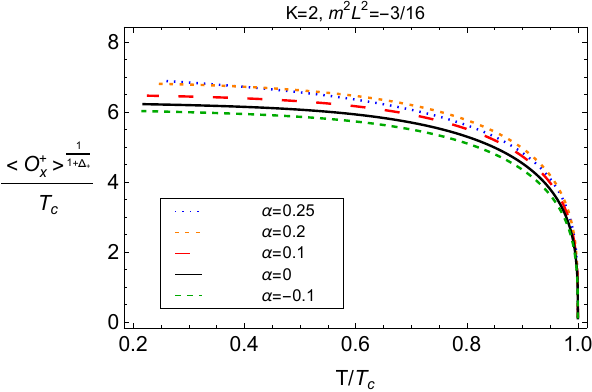}
		\hskip .1 cm
		\epsfxsize = 4 cm
		\includegraphics[keepaspectratio=true,scale=0.78]{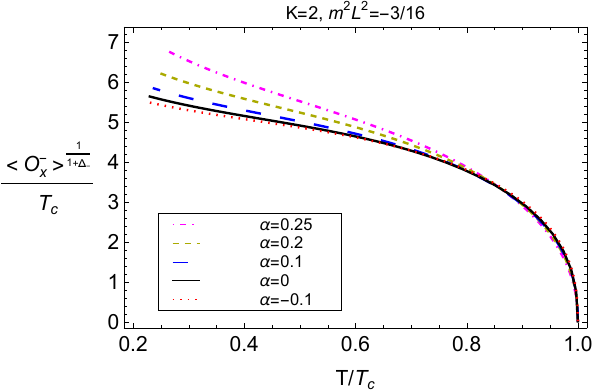}
		\vskip .1 cm
		\epsfxsize = 4 cm
		\includegraphics[keepaspectratio=true,scale=0.78]{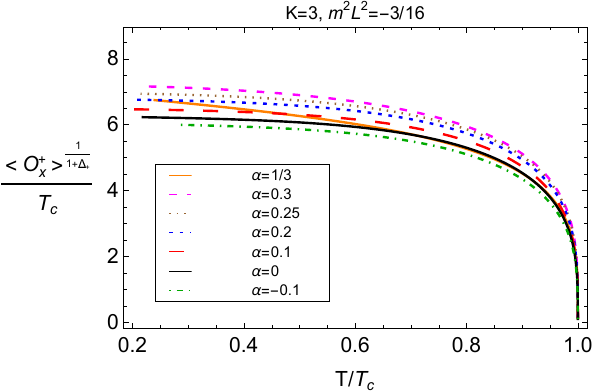}
		\hskip .1 cm
		\epsfxsize = 4 cm
		\includegraphics[keepaspectratio=true,scale=0.78]{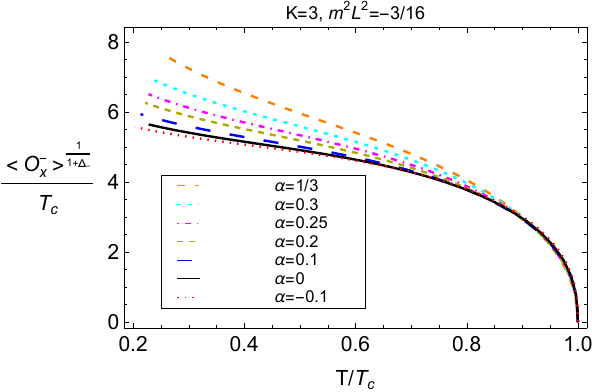}
	\end{center}
	\caption{\textit{Condensation of the vector operators  ${\cal O}_ x^+ $ and ${\cal O}_x^- $  as a function of temperature for the theory up to quadratic ($K=2$) and cubic ($K=3$) curvature corrections with the mass of the vector field fixed by $m^2 L^2 =  -3/16$. For each case, different values of $\alpha$ in the allowed ranges are considered.}}
	\label{fig:Ox_vector_pm}
\end{figure}

For the sake of completeness, we now fix the mass of the vector field as $m^2 {L^2}=- 3/16$ to see the differences compared to the previous ones presented in Figs. \ref{fig:Ox_vector} and \ref{fig:Tc_pWave_forOx}. (We will examine the other case, $m^2 {L^2_{\rm{eff}}}=- 3/16$, in the next subsection.) We only present the detailed information for the regularized $4D$ theory up to quadratic ($K=2$) and cubic ($K=3$) curvature corrections, but the general situation is commented. In Fig. \ref{fig:Ox_vector_pm}, the vector condensates associated with both possible (the standard and the alternative) pictures, i.e. $\rho^+_{x}=\left\langle {{\cal O}^+_x } \right\rangle$ with the dual source as $\rho^-_{x}=0$ and $\rho^-_{x}=\left\langle {{\cal O}^-_ x } \right\rangle$ with the dual source as $\rho^+_{x}=0$, have been depicted with respect to temperature for different values of $\alpha$. As seen, the expectation values of the anisotropic operators, $\left\langle {{\cal O}^+_ x } \right\rangle$ and $\left\langle {{\cal O}^-_x } \right\rangle$, qualitatively behave the same as those isotropic order parameters, $\left\langle {{\cal O}_ + } \right\rangle$ and $\left\langle {{\cal O}_- } \right\rangle$, in $s$-wave superconductors (see Figs. \ref{fig:O+_scalar} and \ref{fig:O-_scalar}). Except for the abnormal behavior of the curves near to the upper bound of $\alpha$, we observe increasing $\alpha$ results in an increase of both $\left\langle {{\cal O}^+_ x } \right\rangle$ and $\left\langle {{\cal O}^-_x } \right\rangle$. While adding higher-order curvature corrections through increasing $K$ leads to smaller condensates associated with $\left\langle {{\cal O}^+_ x } \right\rangle$ but larger values for the $\left\langle {{\cal O}^-_ x } \right\rangle$ condensate. Interestingly, leaving aside the abnormal behavior for the curves near the upper bound of $\alpha$, we see that the condensate curves from the regularized $4D$ quadratic and cubic theories (as well as beyond cubic curvature corrections) do not intersect the one belong to Einstein gravity (the curves with $\alpha=0$). Therefore, compared to Einstein gravity, the values of condensate are larger for $\alpha > 0$ but smaller for $\alpha < 0$. In addition, we confirm for both condensates the mean-field behavior close to $T_c$, namely $\left\langle {{{\cal O}^{\pm}_ x }} \right\rangle  \propto {\left( {{T_C} - T} \right)^{1/2}}$. \vspace{1mm}

\begin{figure}[!htbp]
	\begin{center}
		\epsfxsize = 4 cm
		\includegraphics[keepaspectratio=true,scale=0.8]{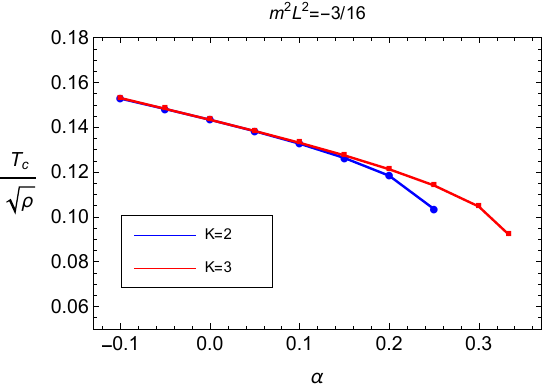}
		\hskip .1 cm
		\epsfxsize = 4 cm
		\includegraphics[keepaspectratio=true,scale=0.8]{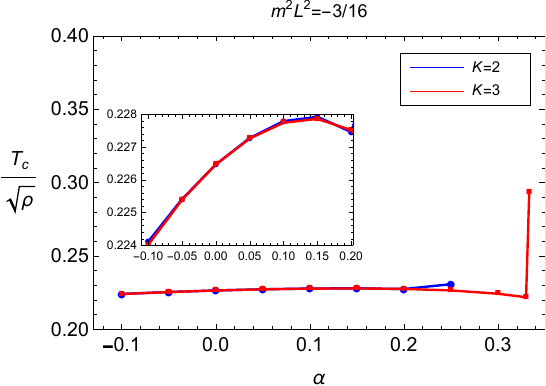}
	\end{center}
	\caption{\textit{Critical temperature of the superconducting transition for the vector operator ${\cal O}_x^ +$ (left panel) and  ${\cal O}_x^ -$ (right panel) as a function of $\alpha$ for the theory up to quadratic ($K=2$) and cubic ($K=3$) curvature corrections with the mass of the vector field fixed by $m^2 L^2=-3/16$. The points with $\alpha =0$ correspond to the results of Einstein gravity.}}
	\label{fig:Tc_pWave_forOx_pm}
\end{figure}

The behavior of the critical temperature $T_c$ has also been presented in Fig. \ref{fig:Tc_pWave_forOx_pm}. For critical temperature of the superconducting transition associated with the $\left\langle {{\cal O}^+_ x } \right\rangle$ condensate (the left panel in Fig. \ref{fig:Tc_pWave_forOx_pm}), we see that increasing $\alpha$ lowers $T_c$ while adding higher (cubic) curvature corrections raise it up provided that $\alpha$ is kept fixed. Regarding the $\left\langle {{\cal O}^-_ x } \right\rangle$ condensate, the right panel of Fig. \ref{fig:Tc_pWave_forOx_pm} shows that increasing $\alpha$ initially increases $T_c$, then lowers it, and again increases when approaching the upper bound of $\alpha$ (Chern-Simons limit). Also, in the limit $\alpha \to 0$, one finds $T_c \approx 0.143 \rho^{1/2}$ associated with the $\left\langle {{\cal O}^+_ x } \right\rangle$ condensate while $T_c \approx 0.226 \rho^{1/2}$ for the $\left\langle {{\cal O}^-_ x } \right\rangle$ condensate. We should emphasize that these features persist for any $K$th order of the regularized $4D$ theory. This example clearly shows that, in the holographic superconductivity phenomenon,  the sign of fixing $m^2 L^2$ matters and the results including variations of basic quantities with the theory's parameters ($\alpha$ and $K$) could be different.

\subsubsection{Vector condensate with $m^2 {L_{{\rm{eff}}}^2} = \textit{fixed}$}

\begin{figure}[!htbp]
	\begin{center}
		\epsfxsize = 4 cm
		\includegraphics[keepaspectratio=true,scale=0.78]{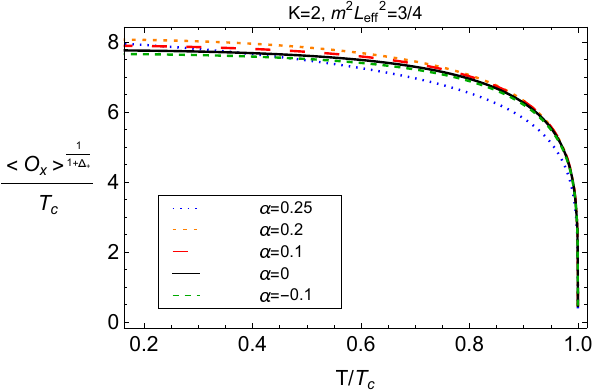}
		\hskip .1 cm
		\epsfxsize = 4 cm
		\includegraphics[keepaspectratio=true,scale=0.78]{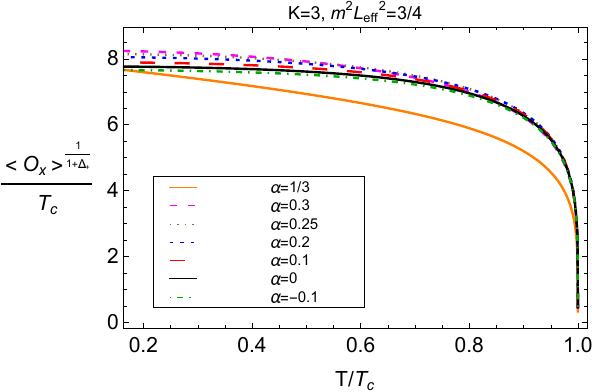}
		\vskip .1 cm
		\epsfxsize = 4 cm
		\includegraphics[keepaspectratio=true,scale=0.78]{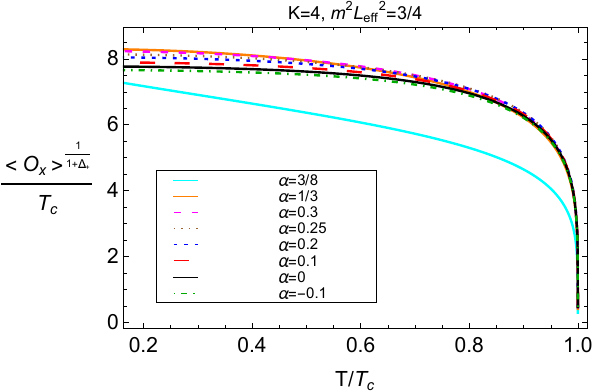}
		\hskip .1 cm
		\epsfxsize = 4 cm
		\includegraphics[keepaspectratio=true,scale=0.78]{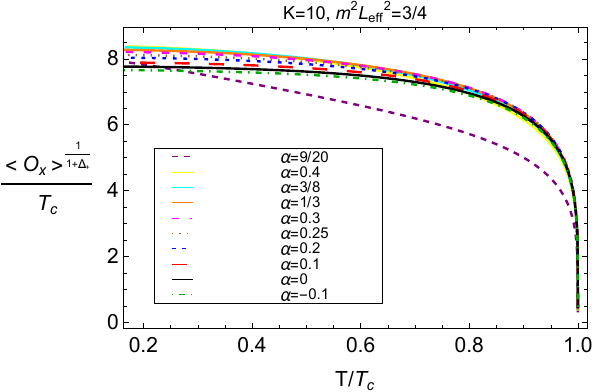}
		\vskip .1 cm
		\epsfxsize = 4 cm
		\includegraphics[keepaspectratio=true,scale=0.78]{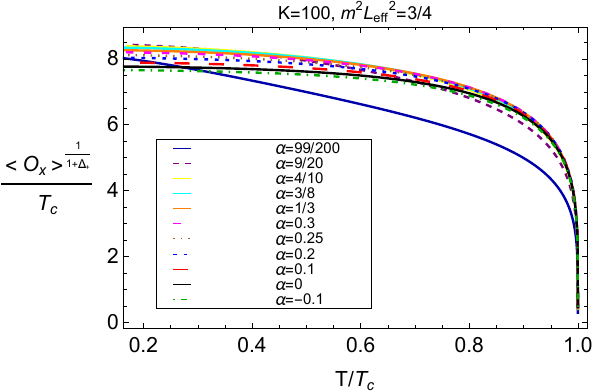}
	\end{center}
	\caption{\textit{Condensation of the vector operator ${\cal O}_ x $ as a function of temperature for five different values of $K=2, 3,4,10$ and $100$ with the mass of the vector field fixed by  ${m^2}L_{{\rm{eff}}}^2 = 3/4$. For each $K$th order of the theory, different values of $\alpha$ in the allowed ranges are considered. The cases with $\alpha =0 $ correspond to Einstein gravity where $L_\text{eff} \to L$ (they are actually the same in all panels).}}
	\label{fig:Ox_vector_Leff}
\end{figure}

We now consider the mass of the complex vector field ($\rho_\mu$) to obey $m^2{L_{{\rm{eff}}}^2}= 3/4$, in order for making contact with the results of Einstein gravity as well as $4D$ Einstein-Gauss-Bonnet gravity theories \cite{HS-4GB-2020JHEP}. $m^2{L_{{\rm{eff}}}^2}=3/4$ implies the standard picture automatically since the conformal dimension for the slow falloff is negative and we need to set it zero as the dual source, so $\rho^-_x=0$ and $\rho^+_x=\left\langle {{{\cal O}_ x }} \right\rangle$. Regarding this case, the spontaneous condensation of the operator ${\cal O}_x$ has been presented in Fig. \ref{fig:Ox_vector_Leff}. From this figure, it is inferred that increasing the fine-tuned coupling constant $\alpha$ results in larger condensates for any $K$th order Einstein-Lovelock gravity theory. While, adding higher curvature corrections (via increasing $K$) to the theory makes it easier for the scalar hair ($\propto$ the condensate) to form provided that $\alpha$ is held fixed. We generally observed that the vector condensates are larger for $\alpha > 0$ but smaller for $\alpha <0$, compared to the result of Einstein gravity. Near the critical temperature, these curves also behave as $\left\langle {{{\cal O}_ x }} \right\rangle  \propto {\left( {{T_C} - T} \right)^{1/2}}$, as expected. Furthermore, the previous abnormal behavior associated with the standard quantization picture (\textit{fast falloff} $\to$ \textit{response}) is again observed near the upper bound of the fine-tuned Lovelock coupling constant, i.e. for values near $\alpha=\frac{{(K - 1){L^2}}}{{2K}}$. \vspace{1mm}


\begin{figure}[!htbp]
	\begin{center}
		\epsfxsize = 4 cm
	\includegraphics[keepaspectratio=true,scale=0.8]{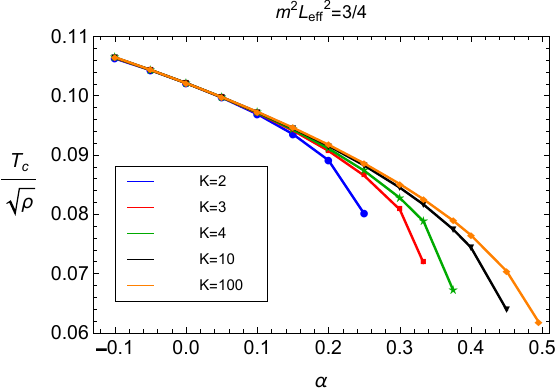}
	\end{center}
	\caption{\textit{Critical temperature of the superconducting transition for the vector operator ${\cal O}_x$ as a function of $\alpha$ for five different values of $K=2,3,4,10$ and $100$ with the mass of the vector field fixed by ${m^2}L_{{\rm{eff}}}^2 = 3/4$. The point with $\alpha =0$ corresponds to the result of Einstein gravity.}}
	\label{fig:Tc_pWave_forOx_Leff}
\end{figure}

Fig. \ref{fig:Tc_pWave_forOx_Leff} shows critical temperature of the superconducting transition for the vector operator ${\cal O}_x$, corresponding to Fig. \ref{fig:Ox_vector_Leff}, as a function of $\alpha$. As seen, increasing $\alpha$ reduces the critical temperature of the system. However, assuming $\alpha$ is held fixed, adding higher curvature corrections ($K$) leads the critical temperature to increase. Compared to Einstein gravity, we also observe that $T_c$ is larger in the range with $\alpha < 0$ but smaller for $\alpha > 0$. \vspace{1mm}


\begin{figure}[!htbp]
	\begin{center}
		\epsfxsize = 4 cm
		\includegraphics[keepaspectratio=true,scale=0.78]{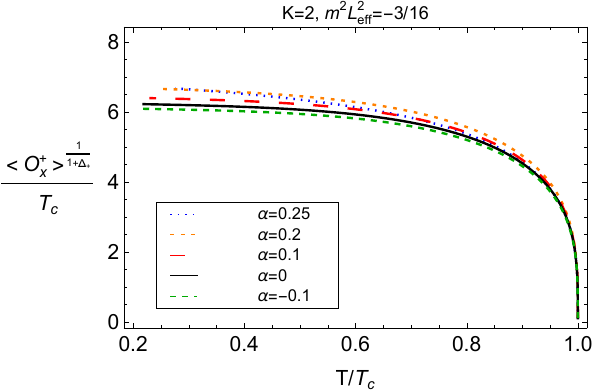}
		\hskip .1 cm
		\epsfxsize = 4 cm
		\includegraphics[keepaspectratio=true,scale=0.78]{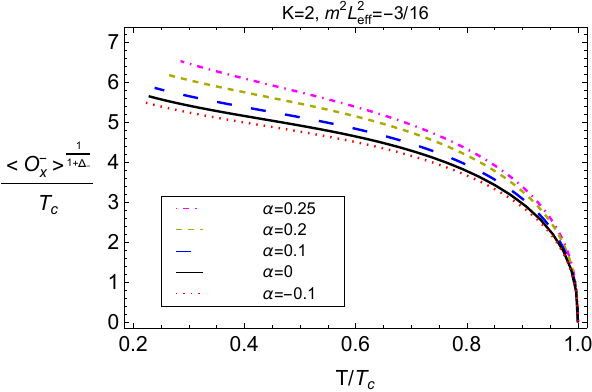}
		\vskip .1 cm
		\epsfxsize = 4 cm
		\includegraphics[keepaspectratio=true,scale=0.78]{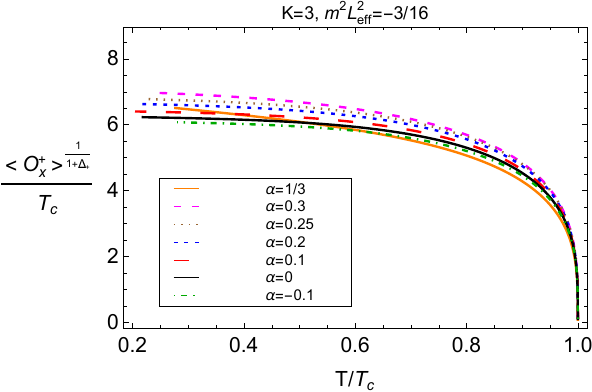}
		\hskip .1 cm
		\epsfxsize = 4 cm
		\includegraphics[keepaspectratio=true,scale=0.78]{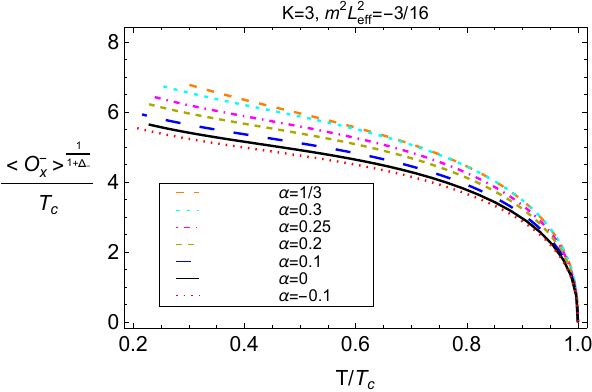}
	\end{center}
	\caption{\textit{Condensation of the vector operator $O_ x^+ $ and $O_x^- $  as a function of temperature for the theory up to quadratic ($K=2$) and cubic ($K=3$) curvature corrections with the mass of the vector field fixed by $m^2  L_{{\rm{eff}}}^2   =  -3/16$. For each case, different values of $\alpha$ in the allowed ranges are considered.}}
	\label{fig:Ox_vector_pm_Leff}
\end{figure}

We now study vector condensation for the case where $m^2 {L^2_{\rm{eff}}}=- 3/16$, in order for examining both the $\left\langle {{\cal O}^+_ x } \right\rangle$ and the  $\left\langle {{\cal O}^-_x } \right\rangle$ condensates and comparing the results with the previous case $m^2 L^2=- 3/16$ that was presented in the previous subsection. We again present the detailed information for the theory up to quadratic ($K=2$) and cubic ($K=3$) curvature corrections, although our statements are valid for the general case. The vector condensates for both possible quantizations, $\left\langle {{\cal O}^+_ x } \right\rangle$ and $\left\langle {{\cal O}^-_x } \right\rangle$, are presented in Fig. \ref{fig:Ox_vector_pm_Leff} for these two particular subclasses of the theory. Generally, we observe that increasing $\alpha$ results in an increase of the $\left\langle {{\cal O}^+_ x } \right\rangle$ condensate, while adding higher curvature corrections ($K$) has the opposite effect. For the alternative picture, the $\left\langle {{\cal O}^-_ x } \right\rangle$ condensate is getting larger as $\alpha$ increases but smaller as we add higher curvature corrections (via $K$) to the model. For both cases operators, the values of condensates are greater than the one in Einstein gravity background for $\alpha >0$ but smaller for $\alpha <0$. Near the critical temperature $T_c$, we again confirm $\left\langle {{{\cal O}^\pm_ x }} \right\rangle  \propto {\left( {{T_C} - T} \right)^{1/2}}$. \vspace{1mm}

\begin{figure}[!htbp]
	\begin{center}
		\epsfxsize = 4 cm
		\includegraphics[keepaspectratio=true,scale=0.8]{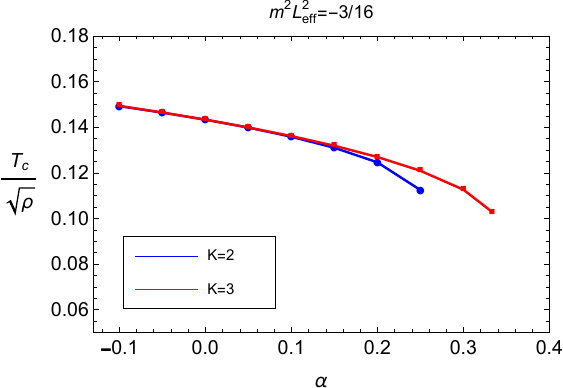}
		\hskip .1 cm
		\epsfxsize = 4 cm
		\includegraphics[keepaspectratio=true,scale=0.8]{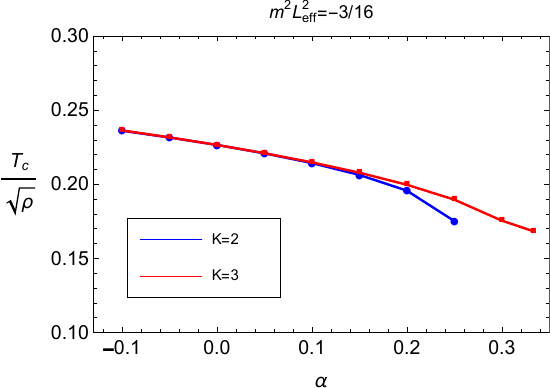}
	\end{center}
	\caption{\textit{Critical temperature of the superconducting transition for the vector operator ${\cal O}_x^ +$ (left panel) and  ${\cal O}_x^ -$ (right panel) as a function of $\alpha$ for the theory up to quadratic ($K=2$) and cubic ($K=3$) curvature corrections with the mass of the vector field fixed by $m^2 L_{{\rm{eff}}}^2  =-3/16$. The points with $\alpha =0$ correspond to the results of Einstein gravity.}}
	\label{fig:Tc_pWave_forOx_pm_Leff}
\end{figure}

The critical temperature ($T_c$) of the superconducting transition corresponding to the two vector operators has also been depicted in Fig. \ref{fig:Tc_pWave_forOx_pm_Leff}. We generally observe that increasing values of $\alpha$ leads to smaller critical temperatures. For allowed fixed values of $\alpha$, the effect of adding more higher curvature corrections ($K$) is to increase the value of critical temperature. Again, for both cases, high-$T_c$ superconductors are found in the range with negative fine-tuned coupling constant ($\alpha < 0$), compared to the one in the context of Einstein gravity.


\subsection{Electric conductivity}

We now investigate the conductivity of the boundary gauge field theory in the superconducting phase (the phase where the condensate appears below the critical temperature) by considering perturbation of the $U(1)$ gauge field in the bulk. We choose the perturbed Maxwell field along the $y$-direction as $\delta {A_y}(t,r) = {A_y}(r){e^{ - i\omega t}}dy$, thus resulting in an anisotropic superconductivity when $T<T_c$ \cite{Cai2013pWave,Cai2014pWave,Cai-2015Review}. The linearized equation of motion of $A_y(r)$ is then obtained as
\begin{equation}\label{purturbed_EM_Pwave}
{A''_y}(r) + \frac{{f'(r)}}{{f(r)}}{A'_y}(r) + \left( {\frac{{{\omega ^2}}}{{{f^2}(r)}} - 2{q^2}\frac{{\rho _x^2(r)}}{{{r^2}f(r)}}} \right){A_y}(r) = 0,
\end{equation}
which, in the asymptotic region, behaves as 
\begin{equation}
{A''_y}(r) + \frac{2}{r}{A'_y}(r) + \frac{{{\omega ^2}L_{{\text{eff}}}^4}}{{{r^4}}}{A_y}(r) = 0\, .
\end{equation}
At the black brane's horizon, we again impose the ingoing wave boundary condition, ${A_y}(r) \sim f{(r)^{ - i\omega /4\pi T}}$. At the AdS boundary, the asymptotic behavior of the Maxwell field reads
\begin{equation}
{A_y}(r) = {A_y^{(0)}} + \frac{{{A_y^{(1)}}}}{r} + ... ,
\end{equation}
where ${A_y^{(0)}}$ and ${{A_y^{(1)}}}$ are the dual source and the expectaion value of the dual current $\left\langle {{J_y}} \right\rangle  = A_y^{(1)}$, respectively. The optical conductivity is then obtained via the AdS/CFT dictionary as
\begin{equation}
{\sigma _y} = \frac{{A_y^{(1)}}}{{i\omega A_y^{(0)}}}.
\end{equation}

\begin{figure}[!htbp]
	\begin{center}
		\epsfxsize = 4 cm
		\includegraphics[keepaspectratio=true,scale=0.8]{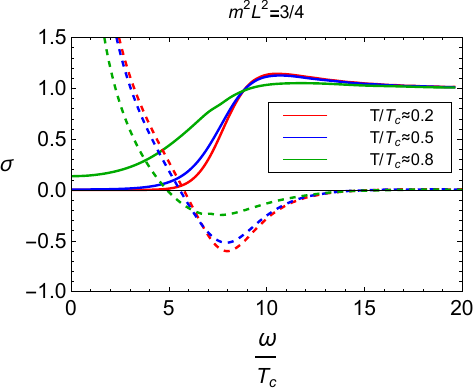}
	\end{center}
	\caption{\textit{Conductivity of the p-wave holographic superconductor versus $\frac{\omega }{{{T_c}}}$  for the Einstein limit (where $\alpha=0$ and $L_{{\rm{eff}}} \to L$) at three different values of $\frac{T}{{{T_c}}} \approx 0.2,0.5$ and $0.8$. The solid and dashed curves show the real and imaginary parts of the conductivity, respectively.}}
	\label{fig:conductivity:Einstein_PW}
\end{figure}

\begin{figure}[!htbp]
	\begin{center}
		\epsfxsize = 4 cm
		\includegraphics[keepaspectratio=true,scale=0.72]{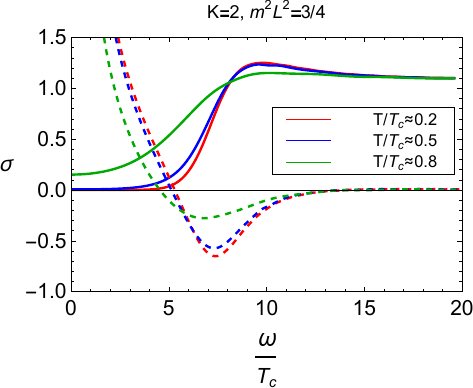}
		\hskip .1 cm
		\epsfxsize = 4 cm
		\includegraphics[keepaspectratio=true,scale=0.72]{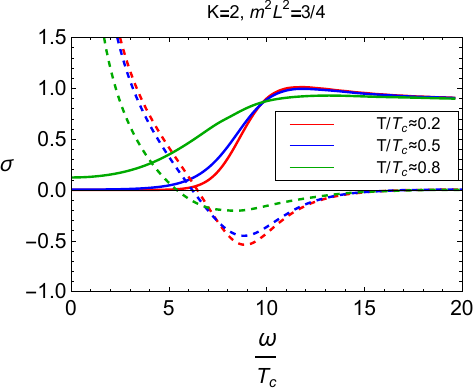}
		\hskip .1 cm
		\epsfxsize = 4 cm
		\includegraphics[keepaspectratio=true,scale=0.72]{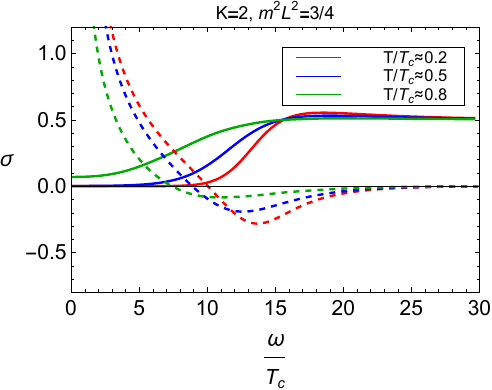}
	\end{center}
	\caption{
		\textit{Conductivity of the p-wave holographic superconductor versus $\frac{\omega }{{{T_c}}}$  for the $4D$ Einstein-Gauss-Bonnet gravity ($K=2$) at three different values of $\frac{T}{{{T_c}}} \approx 0.2,0.5$ and $0.8$. The solid and dashed curves display the real and imaginary parts of the conductivity, respectively. The left, middle and right panels are plotted for $\alpha=-0.1,0.1$ and $0.25$ (the upper bound), respectively and the mass of the scalar field is fixed by $m^2 L^2=3/4$. }}
	\label{fig:conductivity:Pwave_K2}
\end{figure}

\begin{figure}[!htbp]
	\begin{center}
		\epsfxsize = 4 cm
		\includegraphics[keepaspectratio=true,scale=0.72]{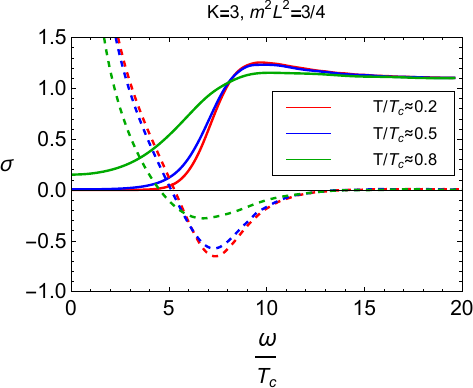}
		\hskip .1 cm
		\epsfxsize = 4 cm
		\includegraphics[keepaspectratio=true,scale=0.72]{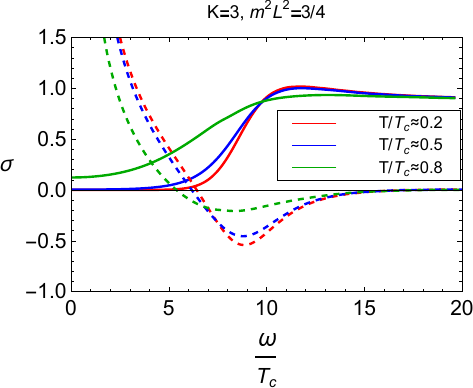}
		\hskip .1 cm
		\epsfxsize = 4 cm
		\includegraphics[keepaspectratio=true,scale=0.71]{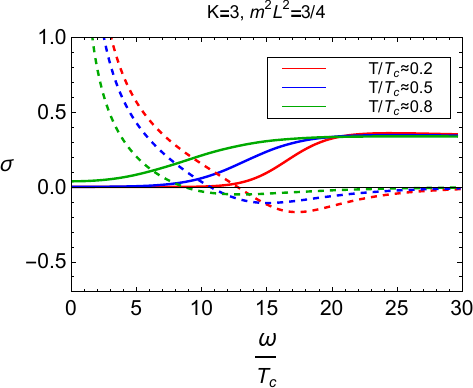}
		\vskip .1 cm
		\epsfxsize = 4 cm
		\includegraphics[keepaspectratio=true,scale=0.72]{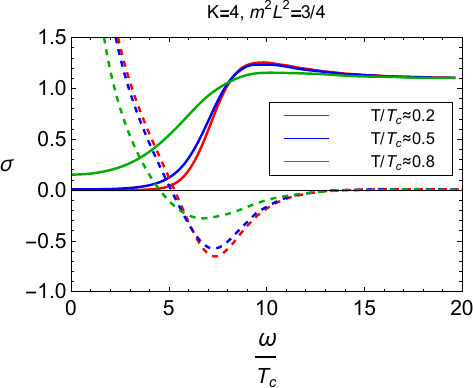}
		\hskip .1 cm
		\epsfxsize = 4 cm
		\includegraphics[keepaspectratio=true,scale=0.72]{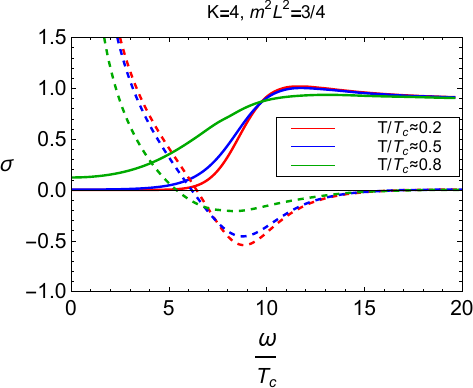}
		\hskip .1 cm
		\epsfxsize = 4 cm
		\includegraphics[keepaspectratio=true,scale=0.72]{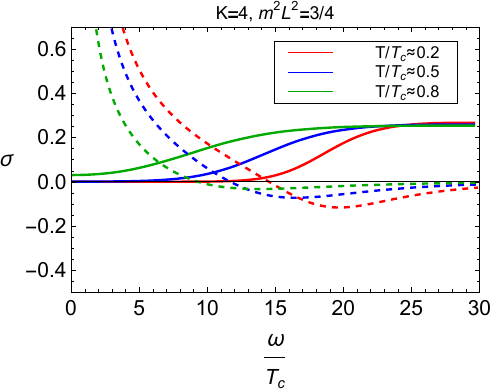}
		\vskip .1 cm
		\epsfxsize = 4 cm
		\includegraphics[keepaspectratio=true,scale=0.72]{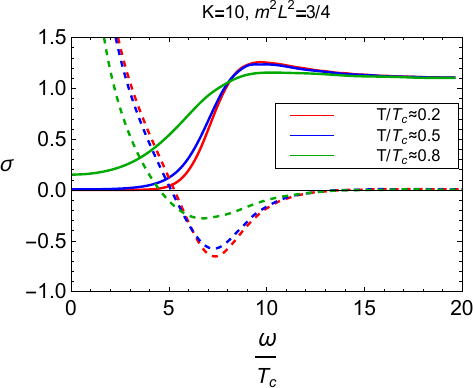}
		\hskip .1 cm
		\epsfxsize = 4 cm
		\includegraphics[keepaspectratio=true,scale=0.72]{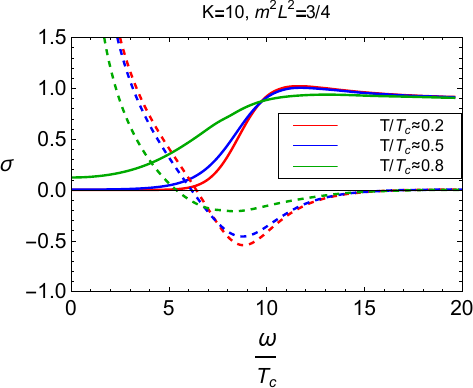}
		\hskip .1 cm
		\epsfxsize = 4 cm
		\includegraphics[keepaspectratio=true,scale=0.72]{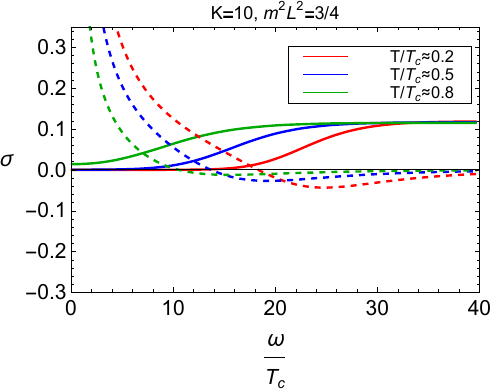}
		\vskip .1 cm
		\epsfxsize = 4 cm
		\includegraphics[keepaspectratio=true,scale=0.72]{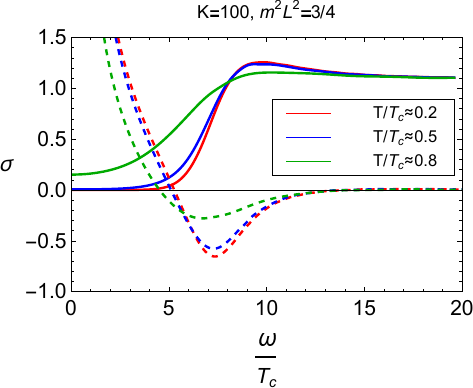}
		\hskip .1 cm
		\epsfxsize = 4 cm
		\includegraphics[keepaspectratio=true,scale=0.72]{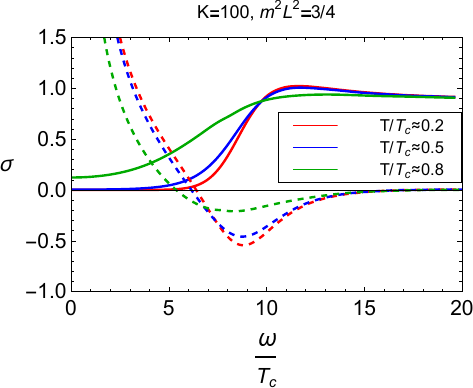}
		\hskip .1 cm
		\epsfxsize = 4 cm
		\includegraphics[keepaspectratio=true,scale=0.7]{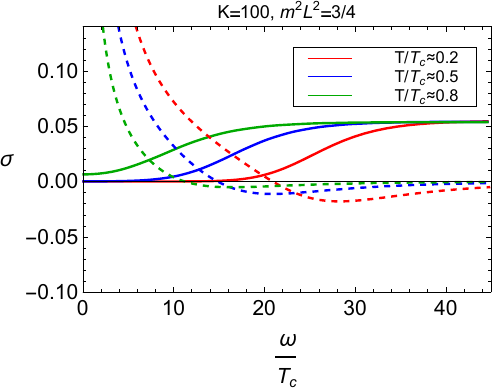}
	\end{center}
	\caption{
		\textit{Conductivity of the $ p $-wave holographic superconductors versus $\frac{\omega }{{{T_c}}}$ in $4D$ Einstein-Lovelock gravity theories (labeled by $K=3, 4, 10$ and $100$), at three different values of $\frac{T}{{{T_c}}} \approx 0.2,0.5$ and $0.8$. The solid and dashed curves display the real and imaginary parts of the conductivity, respectively. The left and middle panels are plotted for $\alpha=-0.1$ and $\alpha=0.1$, respectively. The right panels are plotted for the upper bound of $\alpha$ (the Chern-Simons limits), i.e. $\alpha = \frac{{(K - 1){L^2}}}{{2K}}$, resulting in $\alpha=1/3, 3/8, 9/20$ and $99/200$ for $K=3,4,10$ and $100$, respectively. The mass of the scalar field is also fixed by $m^2 L^2=3/4$.}}
	   \label{fig:conductivity:Pwave_K}
\end{figure}

\begin{figure}[!htbp]
	\begin{center}
		\epsfxsize = 4 cm
		\includegraphics[keepaspectratio=true,scale=0.72]{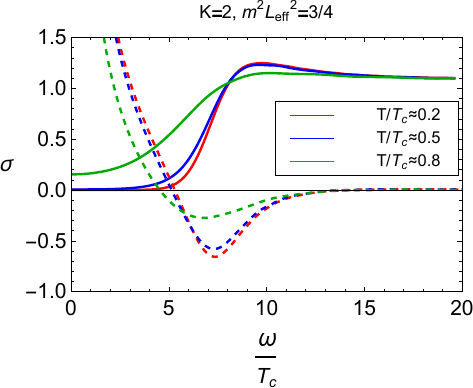}
		\hskip .1 cm
		\epsfxsize = 4 cm
		\includegraphics[keepaspectratio=true,scale=0.72]{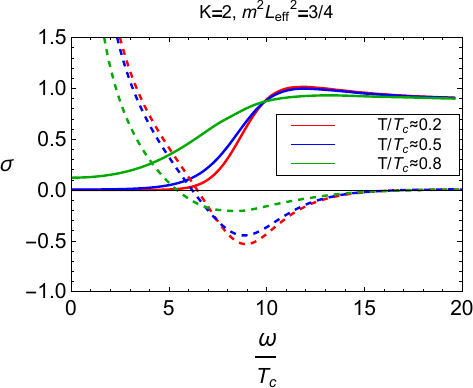}
		\hskip .1 cm
		\epsfxsize = 4 cm
		\includegraphics[keepaspectratio=true,scale=0.72]{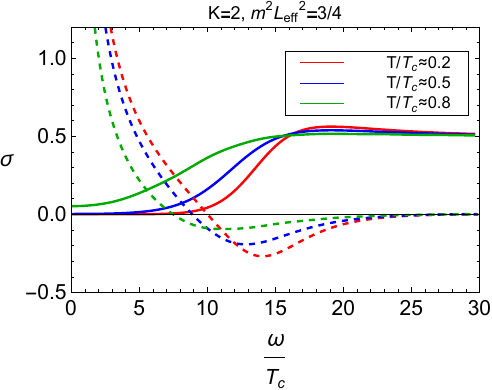}
		\vskip .1 cm
		\epsfxsize = 4 cm
		\includegraphics[keepaspectratio=true,scale=0.72]{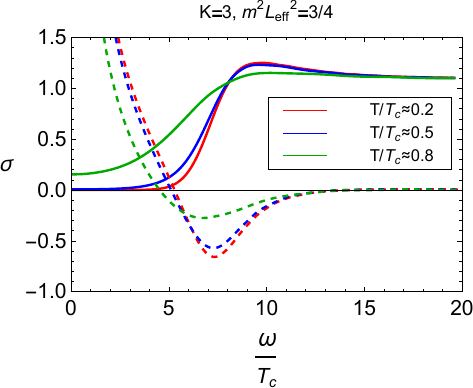}
		\hskip .1 cm
		\epsfxsize = 4 cm
		\includegraphics[keepaspectratio=true,scale=0.72]{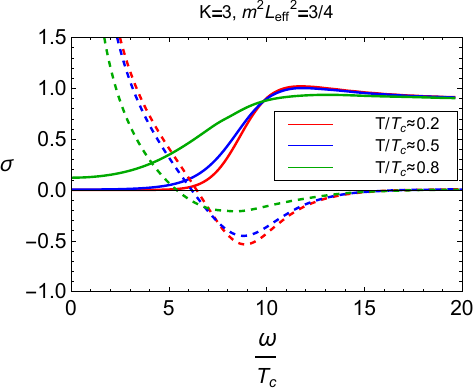}
		\hskip .1 cm
		\epsfxsize = 4 cm
		\includegraphics[keepaspectratio=true,scale=0.72]{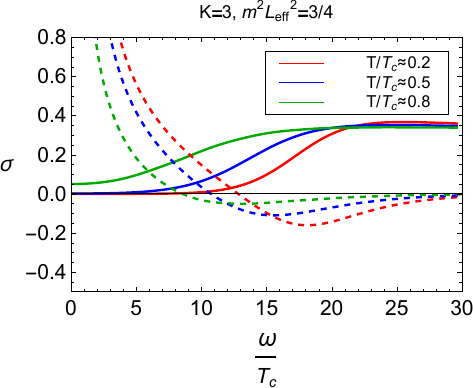}
		\vskip .1 cm
		\epsfxsize = 4 cm
		\includegraphics[keepaspectratio=true,scale=0.72]{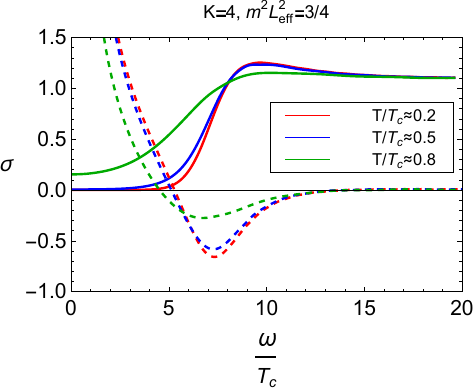}
		\hskip .1 cm
		\epsfxsize = 4 cm
		\includegraphics[keepaspectratio=true,scale=0.72]{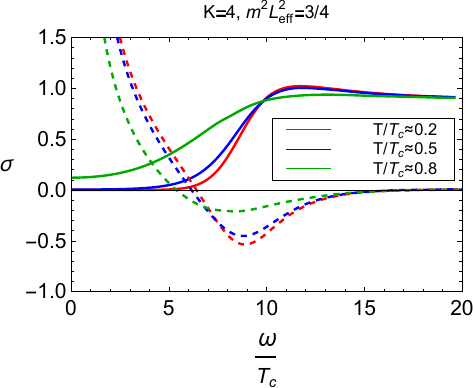}
		\hskip .1 cm
		\epsfxsize = 4 cm
		\includegraphics[keepaspectratio=true,scale=0.72]{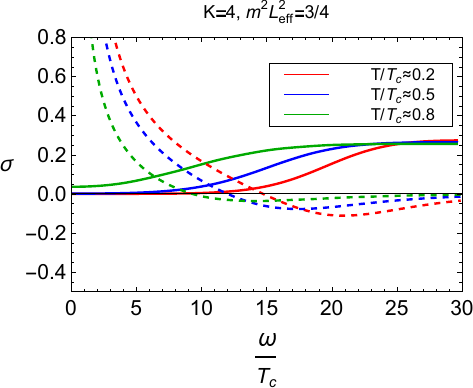}
	\end{center}
	\caption{
		\textit{Conductivity of the $ p $-wave holographic superconductors versus $\frac{\omega }{{{T_c}}}$ in $4D$ Einstein-Lovelock gravity theories (labeled by $K=2, 3$, and $4$), at three different values of $\frac{T}{{{T_c}}} \approx 0.2,0.5$ and $0.8$. The solid and dashed curves display the real and imaginary parts of the conductivity, respectively. The left and middle panels are plotted for $\alpha=-0.1$ and $\alpha=0.1$, respectively. The right panels are plotted for $\alpha=0.25, 1/3$ and $3/8$ which are the upper limits of $\alpha$ (the Chern-Simons limits), i.e. $\alpha = \frac{{(K - 1){L^2}}}{{2K}}$, resulting in $\alpha=0.25, 1/3$, and $3/8$ for $K=2,3$, and $4$, respectively. The mass of the scalar field is also fixed by ${m^2}L_{{\rm{eff}}}^2 =3/4$.}}
	    \label{fig:conductivity:Pwave_K_Leff}
\end{figure}

We again concentrate on the AC conductivities associated with the fast falloff condensate (the $\left\langle {{{\cal O}^+_ x}} \right\rangle$ condensate), once for $m^2 L^2 =\textit{fixed}$ (which also has not yet been investigated for the Gauss-Bonnet limit) and another time for $m^2 L_\text{eff}^2 =\textit{fixed}$. The findings might provide us with new insights into the $4D$ Einstein-Lovelock gravity theories including the Einstein-Gauss-Bonnet class. The plots in Figs. \ref{fig:conductivity:Einstein_PW} and \ref{fig:conductivity:Pwave_K2} show the results of frequency dependent ($p$-wave) conductivities for the Einsteinian and the Gauss-Bonnet limits for the case of $m^2 L^2 =3/4$, respectively, where the solid lines represent Re[$\sigma(\omega)$] while the dashed lines represent Im[$\sigma(\omega)$]. For the Einstein-Maxwell-vector $p$-wave system, the universal relation ${\omega_g}/{T_c} \approx 8$ is again observed, although a greater deviation form this value (with respect to the $s$-wave superconductor, see Fig. \ref{fig:conductivity:Einstein}) occurs for the subcritical temperatures near $T_c$. When the Gauss-Bonnet correction term is added to gravity, sensible deviation from the universal value ($\approx 8$) for the ratio of gap frequency over the critical temperature is observed as shown in Fig. \ref{fig:conductivity:Pwave_K2}; the gap frequency becomes larger as $\alpha$ increases. Going beyond the Einstein-Gauss-Bonnet class with the assumption of $m^2 L^2 =3/4$, the plots in Fig. \ref{fig:conductivity:Pwave_K} show the results for various $K$th orders of the theory as well as different variations of the fine-tuned coupling constant $\alpha$, indicating the gap in the real part of the optical conductivities. For all cases considered, we observe that Re[$\sigma(\omega)$] almost vanishes for $\omega < \omega_g$ and the gap frequency is getting larger as either $\alpha$ or $K$ grow. Generally the value of $\omega_g/T_c$ is greater than the one in the Einstein-Maxwell-vector $p$-wave system provided that $\alpha > 0$. However, it takes smaller values for the negative-$\alpha$ regime. In addition, like the $s$-wave superconductors, lowering the temperature pushes the frequency gap to larger values. Interestingly, all the aforementioned general results are not affected by choosing the alternative fixing for the vector mass, $m^2 L_\text{eff}^2 =\textit{fixed}$, as this fact is evident from the plots in Fig. \ref{fig:conductivity:Pwave_K_Leff}. \vspace{1mm}

For all the cases, the real parts of the AC conductivities approach constant values for large-$\omega$, like the holographic $s$-wave superconductors studied before, confirming the behavior Re[$\sigma(\omega \to \infty)$] $\propto \omega^{D-4}$ ($D$ is the spacetime dimension of the bulk theory). While, Im[$\sigma(\omega \to \infty)$] approaches to zero, as expected. We observe that the asymptotic value of Re[$\sigma(\omega \to  \infty)$] for large frequencies decreases as $\alpha$ increases. Comparing with the Einstein-Maxwell-vector $p$-wave superconductors, this asymptotic value is always smaller for $\alpha >0$ but larger for $\alpha <0$. Although adding higher-order curvature corrections (via increasing $K$) beyond Gauss-Bonnet does not alter the asymptotic value of Re[$\sigma(\omega \to \infty)$], but the holographic system with more higher-order curvature corrections can take larger values of $\alpha$ (according to Eq. \ref{rescalings}), thereby leading to smaller asymptotic values of Re[$\sigma(\omega \to \infty)$]. This can be understood from Figs. \ref{fig:conductivity:Pwave_K2}, \ref{fig:conductivity:Pwave_K} and \ref{fig:conductivity:Pwave_K_Leff}. However, this result might change and the maximum order of the model has a direct effect if the Lovelock coupling constants can vary independently. The diverging DC conductivity is again confirmed by taking the $\omega \to 0$ limit of Eq. (\ref{purturbed_EM_Pwave}) which shows a $1/\omega$ pole in Im[$\sigma(\omega)$], thereby implying ${\sigma _{{\text{DC}}}} = \mathop {\lim }\limits_{\omega  \to 0} \operatorname{Re} [\sigma (\omega )] \propto \delta (\omega )$ by implementing the Kramers-Kronig relation. We should emphasize these results are valid for both choices of fixing the vector mass (relative to either $L$ or $L_{\rm{eff}}$), as obvious from the plots in Figs. \ref{fig:conductivity:Pwave_K} and \ref{fig:conductivity:Pwave_K_Leff}.\\

\section{Summary and discussions} \label{sect5:final}

We have studied holographically dual descriptions of $s$- and $p$-wave superconductors in $2+1$ dimensions by including the ghost-free combinations of stringy higher curvature corrections beyond Gauss-Bonnet using the Einstein-Lovelock Lagrangian. In order to incorporate higher curvature corrections up to any desired order, the four-dimensional ($4D$) regularization of Einstein-Lovelock gravity theory is considered, in which the Lovelock coupling constants are finely tuned according to (\ref{fine tuning}), thereby leading to finding the exact $4D$ black brane solutions. In this special class of regularized $4D$ Einstein-Lovelock gravity theories, when the first coupling constant ($\alpha_2 = \alpha$) is known, the rest of Lovelock coupling constants are determined accordingly, and, for this reason, we called $\alpha$ the fine-tuned (Lovelock) coupling constant. So, this class of black brane solutions, defined by the line element (\ref{metric}) with the emblackening factor (\ref{metric function}) or (\ref{metric_function_newform}), is distinguished from the one in Einstein gravity by the fine-tuned coupling constant ($\alpha$) as well as the positive integer $K$ that is the highest power of curvature in the Lagrangian (representing the maximum order of higher curvature corrections). In both the $\alpha \to 0$ and the $K \to 1$ limits, we recover the planar Schwarzschild black branes in four-dimensions. The $K \to 2$ and $K \to 3$ limits of this class also reduce to the black brane solutions in $4D$ Einstein-Gauss-Bonnet gravity and $4D$ cubic Einstein-Lovelock gravity theories \cite{Glavan2020,Zerbini2013,Tomozawa2011,Zhidenko2020PRDa,Casalino2021}, respectively, which both can also be obtained through different regularization procedures \cite{EGB4Dreview,Alkac2022}. More generally, the planar reduced background considered here also admits an independent realization through conformal regularization, providing further theoretical support for its use as a four-dimensional holographic background beyond the $K=3$ sector. In our considerations, we have also paid special attention to the model in the negative-$\alpha$ regime.\footnote{This is motivated by the fact that a bound on negative values of $\alpha$ can be found by demanding that atomic nuclei should not be shielded by a horizon \cite{Charmousis2022}. On the other hand, the bulk causal structure analysis further constrain the Lovelock coupling constants more than Eq. (\ref{restriction}), and for the $4D$ Einstein-Gauss-Bonnet gravity, it has been shown that the Gauss-Bonnet coupling constant ($\alpha_2 =\alpha$, in our notation) is restricted to the negative-$\alpha$ regime \cite{causality4dGH2020}.} \vspace{1mm}

In order for studying holographic construction of superconductors, we have then clarified the nomenclature used in this paper. Since a holographic renormalization is necessary for driving the field/operator correspondence for any bulk field, the essential aspects of the superconductivity phenomenon including the condensate, critical temperature, and infinite DC conductivity can be explored in each possible quantization schemes. 
The prevailing consensus, in line with physical intuition, posits that the the slow falloff of bulk field serves as the external source of the corresponding boundary operator, with the fast falloff representing its response \cite{Witten1998b,Rey2001Yee,Policastro2000PRL,Policastro2002a,Policastro2002b,Hartnoll2009Lectures,Zaanen2015Book,ErdmengerBook2015,Nastase2015Book,NatsuumeBookAdS/CFT}. We referred to this, ``\textit{fast falloff} $ \leftrightarrow$ \textit{response}'', as the standard picture while addressing the other possible quantization, ``\textit{slow falloff} $ \leftrightarrow$ \textit{response}'', as the alternative picture. We have also conducted our analysis for superconducting systems relative to both the standard and the effective AdS scales, $L$ and $L_{\rm{eff}}$. In fact, the Einstein-Lovelock black branes as the holographic background introduce the effective AdS scale $L_{\rm{eff}}$ in the asymptotic region ($r \to \infty$); see Eq. (\ref{Leff}). For holography applications, it is important to distinguish between $L_{\rm{eff}}$ (coming from higher curvature gravity models) and $L$ (in Einstein's general relativity), since the mass of matter fields in the bulk, which are dual to some boundary operators in the dual CFT, can be fixed relative to either $L$ or $L_{\rm{eff}}$, thereby leading to different results \cite{HS-GB-2009Gregory,HS-GB-2010Gregory,Natsuume2026}. \vspace{1mm}

The main objective of this work is to understand how the strength of higher curvature interactions, controlled by the coupling $\alpha$, and the inclusion of increasingly higher curvature terms, encoded in $K$, affect the physics of holographic superconductors. We first focused on the $s$-wave model, whose isotropic order parameter is described by the scalar operator $\mathcal{O}$ dual to the bulk scalar field $\Psi$. Considering both the standard and alternative quantization schemes, and allowing the scalar mass to be fixed relative to either $L$ or $L_{\rm eff}$, we performed a detailed investigation of the condensate formation and the corresponding critical temperature as functions of $\alpha$ and $K$. A summary of our results for the $s$-wave sector is presented in Table \ref{tab:s-wave}. As shown, both the condensate and the critical temperature exhibit a systematic dependence on the fine-tuned coupling $\alpha$ and the order of theory $K$, although the detailed behavior depends on the choice of quantization scheme and the prescription for fixing the scalar mass. As seen, the results for the condensate and the critical temperature associated with the fast falloff of the scalar field exhibit the same qualitative behavior for both choices of the scalar mass fixing. However, for the alternative picture, we observe opposite variations of $\left\langle {{{\cal O}-}} \right\rangle$ and $T_c$ in two choices of the scalar field's mass. We have disclosed that in the standard picture, positive higher curvature couplings make the condensation harder, thereby delaying the superconducting phase transition (i.e., leading to lower critical temperatures). In contrast, negative values of $\alpha$ enhance the condensation and facilitate the formation of scalar hair. Also in this picture,  increasing $K$ results in a decrease of the condensate provided that $\alpha$ is held fixed. However, the combined effect of increasing both $\alpha$ and $K$ leads to an overall enhancement of $\left\langle {{{\cal O}+}} \right\rangle$. More importantly, when compared with the Einstein-Maxwell-scalar $s$-wave system \cite{HHH2008PRL}, the condensate becomes larger than in GR within the positive-$\alpha$ regime, but smaller in the negative-$\alpha$ regime. Therefore, positive fine-tuned couplings make the condensation more difficult, whereas negative $\alpha$ values make it easier for scalar hair to form in any subclass of 4$D$ Einstein-Lovelock gravity theories. Every result discussed in this paragraph for the standard picture also holds for the alternative picture when $m^2 L^2_{\rm{eff}}=-2$, whereas the case $m^2 L^2 = -2$ displays precisely the opposite behavior. These results indicate that, a common feature emerging from all cases is that negative values of the fine-tuned coupling $\alpha$ facilitate the formation of scalar hair and typically lead to higher critical temperatures compared to Einstein gravity, whereas positive $\alpha$ generally has the opposite effect. This conclusion remains robust across both quantization schemes, even though the detailed behavior of the condensate may vary with the scalar mass prescription.\vspace{1mm}

\begin{table}[]
\caption{\textit{$s$-wave superconductors.}
This table summarizes variations of the condensate
($\langle \mathcal{O}\rangle$) and the critical temperature
for the superconducting transition ($T_c$) with respect to the
fine-tuned coupling ($\alpha$) and the highest order of curvature
corrections ($K$). As seen, the results may depend on the choice
of scalar-mass fixing. By ``increasing $K$'' we mean increasing
$K$ while keeping $\alpha$ fixed in order to isolate the pure
effect of $K$. The symbol ``$^*$'' indicates deviations from the
general pattern near the Chern-Simons (CS) limit. GR is a shorthand for the relevant quantity in Einstein's General Relativity.}
\label{tab:s-wave}

\begin{tabular}{|c|c|c|c|c|c|c|}
\cline{2-7}
\multicolumn{1}{l|}{} &
Quantity &
Increasing $\alpha$ &
Increasing $K$ &
\begin{tabular}[c]{@{}c@{}}
Negative-$\alpha$\\
regime
\end{tabular} &
\begin{tabular}[c]{@{}c@{}}
Positive-$\alpha$\\
regime
\end{tabular} &
\begin{tabular}[c]{@{}c@{}}
Scalar mass\\
fixing
\end{tabular}
\\ \hline

\multicolumn{1}{|c|}{\multirow{2}{*}{
\begin{tabular}[c]{@{}c@{}}
Standard\\
picture
\end{tabular}}}
&
$\langle \mathcal{O}_{+}\rangle$
&
increases$^*$
&
decreases
&
smaller than GR
&
larger than GR$^*$
&
\multirow{4}{*}{
\begin{tabular}[c]{@{}c@{}}
$m^2L^2=-2$\\
(The same behavior\\
is seen for the\\
other choices.)
\end{tabular}}
\\ \cline{2-6}

&
$T_c$
&
decreases
&
increases
&
higher than GR
&
lower than GR
&
\\ \cline{1-6}

\multicolumn{1}{|c|}{\multirow{2}{*}{
\begin{tabular}[c]{@{}c@{}}
Alternative\\
picture
\end{tabular}}}
&
$\langle \mathcal{O}_{-}\rangle$
&
decreases
&
increases
&
larger than GR
&
smaller than GR
&
\\ \cline{2-6}

&
$T_c$
&
increases
&
decreases
&
lower than GR
&
higher than GR
&
\\ \hline

\multicolumn{1}{|c|}{\multirow{2}{*}{
\begin{tabular}[c]{@{}c@{}}
Standard\\
picture
\end{tabular}}}
&
$\langle \mathcal{O}_{+}\rangle$
&
increases$^*$%
\footnote{When $\alpha$ approaches the CS limit, the condensation curves intersect each other as well as the Einstein-gravity curve.}
&
decreases
&
smaller than GR
&
larger than GR$^*$%
\footnote{Near the CS limit, the condensate becomes smaller than the GR value after a characteristic temperature $T_i$.}
&
\multirow{4}{*}{
\begin{tabular}[c]{@{}c@{}}
$m^2L_{\rm eff}^2=-2$\\
(The same behavior\\
is seen for the\\
other choices.)
\end{tabular}}
\\ \cline{2-6}

&
$T_c$
&
mostly decreases$^*$%
\footnote{Very close to the CS limit, $T_c$ starts increasing.}
&
increases
&
higher than GR
&
lower than GR$^*$%
\footnote{For $K\ge 3$, the condensate becomes larger than the GR value near the CS limit.}
&
\\ \cline{1-6}

\multicolumn{1}{|c|}{\multirow{2}{*}{
\begin{tabular}[c]{@{}c@{}}
Alternative\\
picture
\end{tabular}}}
&
$\langle \mathcal{O}_{-}\rangle$
&
increases
&
decreases
&
smaller than GR
&
larger than GR
&
\\ \cline{2-6}

&
$T_c$
&
mostly decreases$^*$%
\footnote{Very close to the CS limit, $T_c$ starts increasing.}
&
mostly increases$^*$%
\footnote{Very close to the CS limit, $T_c$ starts decreasing.}
&
higher than GR
&
lower than GR
&
\\ \hline

\end{tabular}
\end{table}

We then studied the AC conductivity of s-wave holographic superconductors in regularized 4$D$ Einstein-Lovelock black branes. We numerically analyzed the frequency-dependent optical conductivity and the effects of higher curvature corrections characterized by $\alpha$ and $K$. We found a clear gap frequency $\omega_g$, where Re[$\sigma(\omega)$] is suppressed for $\omega<\omega_g$, while both components approach constant asymptotic values consistent with the expected scaling behavior. The asymptotic Re[$\sigma(\omega)$] decreases with increasing $\alpha$, while showing no direct dependence on $K$, which instead acts indirectly through the allowed range of $\alpha$. The ratio $\omega_g/T_c$ deviates from Einstein gravity and increases with both $\alpha$ and higher-order curvature corrections. At low frequencies, Im[$\sigma(\omega)$] exhibits a $1/\omega$ pole, implying a delta function in Re[$\sigma(\omega)$] and confirming infinite DC conductivity. Overall, higher curvature corrections significantly affect both the optical conductivity and the superconducting gap structure.\vspace{1mm}

For the sake of completeness, we have also investigated the condensation behavior and critical temperature of holographic $p$-wave superconductors for different choices of vector-field mass and quantization schemes. The key features of the condensate and critical temperature, together with their dependence on the model parameters, are summarized in Tables \ref{tab:p-wave-positive} and \ref{tab:p-wave-negative}. In all cases, the vector order parameter develops below a critical temperature and exhibits the mean-field behavior $\langle O_x\rangle \propto (T_c-T)^{1/2}$, indicating a second-order phase transition.
Our results show that the fine-tuned Lovelock coupling $\alpha$ and the order $K$ of curvature corrections significantly affect both the condensate and the critical temperature. For positive vector masses, increasing $\alpha$ generally enhances the condensate while suppressing the critical temperature, whereas higher-order curvature corrections tend to increase $T_c$ (the condensation behavior is not universal and depends sensitively on the choice of mass-fixing prescription, exhibiting qualitatively different patterns for $m^2L_{\rm eff}^2=\mathrm{fixed}$ and $m^2L^2=\mathrm{fixed}$.). For negative vector masses, the behavior becomes quantization dependent and the effects of higher curvature corrections on the condensate may reverse depending on the choice of the dual operator. Nevertheless, a common feature of all considered models is that negative values of the fine-tuned coupling constant lead to larger critical temperatures compared to Einstein gravity, implying that the superconducting phase is easier to form in this regime. We also observed that the sign and prescription used for fixing the vector-field mass, either with respect to the AdS scale ($L$) or the effective AdS scale ($L_{\rm{eff}}$), can qualitatively modify the response of the condensate to the Lovelock parameters. This demonstrates that holographic $p$-wave superconductors in regularized 4$D$ Einstein-Lovelock gravity possess a richer phase structure than their Einstein and Gauss--Bonnet counterparts. Overall, our analysis reveals that higher curvature corrections provide an efficient mechanism for controlling both the condensation process and the superconducting transition temperature in holographic $p$-wave systems. Taken together, unlike the $s$-wave case, the $p$-wave condensates exhibit a stronger sensitivity to the prescription used for fixing the vector-field mass. In particular, the response of the condensate to higher curvature corrections may qualitatively change depending on whether the mass is fixed relative to the AdS scale or the effective AdS scale.\vspace{1mm}

\begin{table}[]
\caption{\textit{$p$-wave superconductors with positive vector mass.}
Summary of the effects of the fine-tuned Lovelock coupling ($\alpha$) and the highest order of curvature corrections ($K$) on the condensate and critical temperature. The symbol ``$^*$'' indicates deviations near the Chern-Simons (CS) limit. GR is a shorthand for the relevant quantity in Einstein's General Relativity.}
\label{tab:p-wave-positive}

\begin{tabular}{|c|c|c|c|c|c|c|}
\cline{2-7}
\multicolumn{1}{l|}{} &
Quantity &
Increasing $\alpha$ &
Increasing $K$ &
\begin{tabular}[c]{@{}c@{}}
Negative-$\alpha$\\ regime
\end{tabular} &
\begin{tabular}[c]{@{}c@{}}
Positive-$\alpha$\\ regime
\end{tabular} &
\begin{tabular}[c]{@{}c@{}}
Scalar mass\\ fixing
\end{tabular}
\\ \hline

\multirow{6}{*}{
\begin{tabular}[c]{@{}c@{}}
Standard\\ picture
\end{tabular}}
&
$\langle {\cal O}_x\rangle$
&
increases$^*$
&
increases
&
\begin{tabular}[c]{@{}c@{}}
changes from smaller\\
to larger than GR\\
across crossing
\end{tabular}
&
\begin{tabular}[c]{@{}c@{}}
changes from larger\\
to smaller than GR\\
across crossing$^*$
\end{tabular}
&

\begin{tabular}[c]{@{}c@{}}
$m^2L^2=3/4$\\
(The same behavior is\\
seen for the other\\
choices.)
\end{tabular}
\\ \cline{2-6}

&
$T_c$
&
decreases
&
increases
&
higher than GR
&
lower than GR
&
\\ \cline{2-7}

&
$\langle {\cal O}_x\rangle$
&
increases$^*$
&
decreases
&
smaller than GR
&
larger than GR$^*$
&
\begin{tabular}[c]{@{}c@{}}
$m^2L_{\rm eff}^2=3/4$\\
(The same behavior is\\
seen for the other\\
choices.)
\end{tabular}
\\ \cline{2-6}

&
$T_c$
&
decreases
&
increases
&
higher than GR
&
lower than GR
&
\\ \hline

\end{tabular}
\end{table}


\begin{table}[]
\caption{\textit{$p$-wave superconductors with negative vector mass.}
Summary of the effects of the fine-tuned Lovelock coupling ($\alpha$) and the highest order of curvature corrections ($K$) on the condensates and critical temperature. The symbol ``$^*$'' indicates deviations from the
general pattern near the Chern-Simons (CS) limit. GR is a shorthand for the relevant quantity in Einstein's General Relativity.}
\label{tab:p-wave-negative}
\begin{tabular}{|c|c|c|c|c|c|c|}
\cline{2-7}
\multicolumn{1}{l|}{} &
Quantity &
Increasing $\alpha$ &
Increasing $K$ &
\begin{tabular}[c]{@{}c@{}}
Negative-$\alpha$\\
regime
\end{tabular} &
\begin{tabular}[c]{@{}c@{}}
Positive-$\alpha$\\
regime
\end{tabular} &
\begin{tabular}[c]{@{}c@{}}
Scalar mass\\
fixing
\end{tabular}
\\ \hline

\multicolumn{1}{|c|}{\multirow{2}{*}{
\begin{tabular}[c]{@{}c@{}}
Standard\\
picture
\end{tabular}}}
&
$\langle \mathcal{O}_{x}^{+}\rangle$
&
increases$^*$
&
decreases
&
smaller than GR
&
larger than GR$^*$
&
\multirow{6}{*}{
\begin{tabular}[c]{@{}c@{}}
$m^2L^2=-3/16$\\
(The same behavior\\
is seen for the\\
other choices.)
\end{tabular}}
\\ \cline{2-6}

&
$T_c$
&
decreases
&
increases
&
higher than GR
&
lower than GR
&
\\ \cline{1-6}

\multicolumn{1}{|c|}{\multirow{4}{*}{
\begin{tabular}[c]{@{}c@{}}
Alternative\\
picture
\end{tabular}}}
&
$\langle \mathcal{O}_{x}^{-}\rangle$
&
increases
&
increases
&
smaller than GR
&
larger than GR
&
\\ \cline{2-6}

&
$T_c$
&
\begin{tabular}[c]{@{}c@{}}
increases,\\ then decreases,\\ then increases$^*$
\end{tabular}
&
no clear trend
&
lower than GR
&
no clear trend
&
\\ \hline

\multicolumn{1}{|c|}{\multirow{2}{*}{
\begin{tabular}[c]{@{}c@{}}
Standard\\
picture
\end{tabular}}}
&
$\langle \mathcal{O}_{x}^{+}\rangle$
&
increases$^*$%
&
decreases
&
smaller than GR
&
larger than GR$^*$%
&
\multirow{4}{*}{
\begin{tabular}[c]{@{}c@{}}
$m^2L_{\rm eff}^2=-3/16$\\
(The same behavior\\
is seen for the\\
other choices.)
\end{tabular}}
\\ \cline{2-6}

&
$T_c$
&
decreases
&
increases
&
higher than GR
&
lower than GR
&
\\ \cline{1-6}

\multicolumn{1}{|c|}{\multirow{2}{*}{
\begin{tabular}[c]{@{}c@{}}
Alternative\\
picture
\end{tabular}}}
&
$\langle \mathcal{O}_{x}^{-}\rangle$
&
increases
&
decreases
&
smaller than GR
&
larger than GR
&
\\ \cline{2-6}

&
$T_c$
&
decreases
&
increases
&
higher than GR
&
lower than GR
&
\\ \hline

\end{tabular}
\end{table}

Finally, we explored the AC conductivity of holographic $p$-wave superconductors with fast falloff condensates for two choices of vector mass fixing, $m^2L^2=\mathrm{const.}$ and $m^2L_{\rm eff}^2=\mathrm{const.}$. In the Einstein-Maxwell-vector model, the standard result $\omega_g/T_c \simeq 8$ is recovered, while Gauss-Bonnet and higher-order Lovelock corrections lead to significant deviations. The gap frequency increases with both $\alpha$ and $K$, and the ratio $\omega_g/T_c$ is larger (smaller) than the Einstein value for positive (negative) $\alpha$. Lower temperatures shift the gap to higher frequencies. At high frequencies, $\mathrm{Re}[\sigma(\omega)]$ shows the expected asymptotic behavior, while $\mathrm{Im}[\sigma(\omega)] \to 0$. Increasing $\alpha$ reduces the asymptotic value of the real part, while $K$ affects it only indirectly through the allowed range of $\alpha$. We confirmed that the system exhibits divergent DC conductivity via a $1/\omega$ pole in $\mathrm{Im}[\sigma(\omega)]$ and a delta function in $\mathrm{Re}[\sigma(\omega)]$. These features remain unchanged under both mass-fixing prescriptions, indicating robustness of the conductivity behavior and its primary dependence on the Lovelock coupling $\alpha$ and higher curvature corrections.\vspace{1mm}

Overall, our results reveal that higher curvature corrections play a significant role in shaping the superconducting phase, enhancing the sensitivity of both the condensate and the critical temperature to the underlying model parameters and consequently amplifying the imprint of the gravitational sector on the superconducting observables. Furthermore, we find that the regularized 4D Einstein-Lovelock theory naturally accommodates high-$T_c$ superconductors relative to Einstein gravity in the regime of negative fine-tuned coupling constants and sufficiently high orders of curvature corrections. Interestingly, this increase in the critical temperature is generally accompanied by a suppression of the condensate at low temperatures.\vspace{1mm}

The present work also opens several promising directions for future investigations. A natural extension is to consider the $K \to \infty$ limit, which effectively resums the entire tower of higher curvature corrections and may provide a deeper understanding of their non-perturbative gravitational effects on holographic superconductivity, as we already suggested in ref. \cite{conference2022}. A complete analytical treatment of the conformal-regularization realization of these backgrounds (along with the finite-$K$ black brane backgrounds in this paper), including the connections among the different regularization procedures and the geometric, thermodynamic, and holographic properties of the resulting solutions, will be presented in a separate forthcoming work \cite{ZD2026}. Another interesting avenue is to couple nonlinear $U(1)$ gauge-invariant extensions of Maxwell electrodynamics \cite{DSZ2022,Sorokin2022} to the four-dimensional Einstein--Lovelock theory. Since these models incorporate string-inspired higher-derivative corrections in both the gravitational and matter sectors, they offer a more comprehensive effective description of the dual holographic system and may unveil novel features of the superconducting phase. Finally, it would be important to investigate the causality constraints arising from the dual boundary CFT. Such constraints may further restrict the admissible parameter space beyond the range considered in Eq.~(\ref{restriction}), in close analogy with the corresponding analyses in higher-dimensional Gauss--Bonnet and Lovelock gravities \cite{deBoer2010,KSS2008violation}. To the best of our knowledge, a systematic study of these constraints has not yet been carried out for the regularized four-dimensional Einstein--Lovelock theory. Incorporating them could potentially exclude the region near the upper bound of the fine-tuned coupling constant, where the anomalous behavior observed in the present work typically emerges. Exploring these issues would therefore provide a more complete understanding of the consistency and physical viability of holographic superconductors in regularized 4$D$ Lovelock gravity theories.

\section{Declarations}

\noindent\textbf{Acknowledgements} S.Z. gratefully acknowledges the support of the Research Council of the University of Sistan and Baluchestan.

\noindent\textbf{Funding} The authors received no specific funding for this work.

\noindent\textbf{Availability of data and material} No external datasets were used in this study. The analytical and numerical results generated during the current study are available from the corresponding author upon reasonable request.

\noindent\textbf{Competing interests} The authors declare that they have no competing interests.

\noindent\textbf{Authors' contributions} Both authors contributed equally to all aspects of this work, including the conception of the study, analytical and numerical calculations, interpretation of the results, and preparation of the manuscript. Both authors read and approved the final manuscript.

\end{document}